\documentclass[a4paper]{article}

\usepackage{graphicx}
\usepackage{atlasphysics}
\usepackage{subfigure}
\usepackage{lineno}
\usepackage{amsmath}
\usepackage{upgreek}

\def\mubinv{\ensuremath{\upmu \mathrm b^{-1}}}%
\def\mum{\ensuremath{\upmu \mathrm m}}%
\def\dedx{\ensuremath{{\mathrm d}E/{\mathrm d}x}}%

\begin{document}

\title{Performance of the ATLAS Detector using First Collision Data}
\author{The ATLAS Collaboration}
\maketitle

\begin{abstract}
  More than half a million minimum-bias events of LHC collision data 
  were collected by the ATLAS experiment in December 2009 at
  centre-of-mass energies of 
  $0.9$~\TeV\ and $2.36$~\TeV.
  This paper reports on studies of the initial performance of the ATLAS
  detector from these data. Comparisons between data and Monte Carlo
  predictions are shown for
  distributions of several track- and calorimeter-based
  quantities. 
The good performance
of the ATLAS detector in these first data gives confidence for
successful running at higher energies.

\end{abstract}


\section{Introduction}

In December 2009, the ATLAS detector~\cite{DetectorPaper}
 recorded data from a first series of  LHC runs at
centre-of-mass energies of $0.9$~\TeV\ and $2.36$~\TeV.
When the beams were colliding and declared to be stable by 
the LHC  operators, all main detector components
were fully operational and all levels
of the trigger and data acquisition system performed as expected,
assuring smooth and well monitored data taking.
The data
sample at 0.9~\TeV\ contains nearly 400\,000 events
recorded with high-quality calorimeter and tracking  information,
corresponding  to an integrated luminosity of approximately 
$9$~\mubinv~\cite{Collaboration:2010rd}.
The data at 2.36~\TeV, 36\,000 events, which are only used here for
calorimeter studies,   correspond to approximately $0.7$~\mubinv.
These data sets do not contain very many high-\pt\ objects,
and 
therefore do not correspond to the environment for which ATLAS was
designed. 


The ATLAS detector
was thoroughly commissioned and initial calibration and performance
studies were done using cosmic ray data recorded during 2008 and
2009.
Performance close to design goals was obtained for
the different detector components, details can be found in
Refs.~\mbox{\cite{readiness1,muonpaper,id-paper,readiness2}}.

This paper presents performance established with data taken in first
collisions in 2009.
The detector components are outlined in
Section~\ref{secDet}.
The trigger and data acquisition performance together with the initial
event selection are discussed in
Section~\ref{secPresel}; the simulation to which the data are
compared is explained in Section~\ref{sec:mc}.
The performance of the inner tracking system is reviewed in
Section~\ref{sec:tracking}, and the  combined analysis of calorimeter data and tracking information to
study electrons and photons is discussed  in Section~\ref{e-gamma}.
Studies of jets and missing transverse energy, \MET,
using the calorimeter cells
are presented
in Sections~\ref{jets} and~\ref{met}. Finally, kinematic distributions of the first reconstructed muon
candidates are shown in Section~\ref{sec:muons}.


%
%
%
%
%
%

\section{The ATLAS Detector}
\label{secDet}

The
ATLAS detector~\cite{DetectorPaper} covers almost the entire solid
angle around the nominal interaction point and comprises the following
sub-components:
\begin{itemize}
%
  \item An inner tracking system: operating inside an axial magnetic field of
 2~T, it is based on three types of tracking devices. These are an outer
 tracker using 
straw tubes with particle identification capabilities based on
transition radiation (Transition Radiation Tracker, TRT), a silicon
strip detector (SemiConductor Tracker, SCT) and an innermost silicon
pixel detector (Pixel).

\item A hybrid calorimeter system: for the
  electromagnetic portion (EM), the hadronic end-cap (HEC) and the
  forward calorimeter (FCal) a
  liquid argon (LAr) technology with different types of absorber materials is
  used. The central hadronic calorimeter (Tile) is a
  sampling calorimeter with steel as the absorber material and
  scintillator as the active medium.
  The electromagnetic sections use an accordion geometry to ensure
  fast and uniform response.
   A presampler detector, to correct for energy losses in the upstream
  material,  is installed in front of
  the EM calorimeter in the range $ | \eta | <$ 
1.8.\footnote{The origin of the coordinate system used to describe the ATLAS
detector is the nominal interaction point. The positive $x$ axis is defined as
pointing from the interaction point to the centre of the LHC ring,
the positive $y$ axis is defined as pointing upwards
and the beam direction defines the $z$ axis of a right-handed
coordinate system.
Transverse momenta are measured in the $x$-$y$ plane with radius $r$.
Polar ($\theta$)
and azimuthal ($\phi$) angles are measured with respect to this
reference system. The pseudorapidity is defined as $\eta = -\ln
\tan(\theta/2)$.}

\item
  A large muon spectrometer: an air-core toroid system
  generates an average field of 0.5~T (1~T), in the barrel (end-cap) region
  of this spectrometer, resulting in a bending power between 2.0 and
  7.5~Tm. Over most of the $\eta$-range, tracks are measured by Monitored Drift
  Tubes (MDT);  in the high $\eta$-regime the closest  of four wheels
  to the interaction region
   is instrumented with Cathode Strip Chambers (CSC). 
  Trigger information is provided by Thin Gap Chambers (TGC) in the
  end-cap and Resistive Plate 
  Chambers (RPC) in the barrel.
  \item
  Specialized detectors in the forward region: two dedicated forward
  detectors, the LUCID Cherenkov counter and the Zero Degree
  Calorimeter (ZDC). In addition 
  the BPTX, an electrostatic beam-pickup which monitors the timing of
  the beam near ATLAS and two scintillator wheels (MBTS) were mounted in
  front of the 
  electromagnetic end-caps to provide trigger signals with minimum bias.
\end{itemize}

\section{Data-Taking Performance and Event Selection}
\label{secPresel}

\begin{table}[htp]
\caption{Luminosity-weighted fraction of the time during stable beam
  operation   for which  the different
  detectors were able to take data under nominal conditions.}
\begin{center}
\hspace{-0.2cm}
\begin{tabular}{ccccccccc}
\hline\hline
               & Pixel & SCT  & TRT & LAr  & Tile & MDT  & RPC & TGC \\ \hline
Efficiency [\%] &  80.9 & 86.2 & 100 & 99.0 & 100  & 87.4 & 88.6 & 84.4 \\
\hline\hline
\end{tabular}
\end{center}
\label{ta:det-eff}
\end{table}%

The ATLAS operating procedure in 2009 maintained the
calorimeters and TRT in standard operating conditions, but
the silicon
trackers and muon chambers were at a reduced or `standby' voltage until
after stable beams were declared by the LHC.
Most of the studies in this paper required the tracking detectors to be
in operating conditions and used approximately 400\,000 events
while the \MET\ studies  only required the calorimeters and used
some 600\,000  and 36\,000  events at 0.9~\TeV\ and 2.36~\TeV\ respectively.
 The luminosity-weighted
availability of the various sub-detectors (see Section~\ref{sec:lumi})
during stable beam operations
is summarized in Table~\ref{ta:det-eff}.

\subsection{Trigger/DAQ System}

%
%
The ATLAS trigger and data acquisition (TDAQ) is a multi-level system
with buffering at all levels~\cite{DetectorPaper}. Trigger 
decisions are 
based on calculations done at three consecutive trigger levels.
While  decisions at the first two levels are pending, the data
acquisition system buffers  the event data from each
sub-detector. Complete events are built after the second level decision.
The first level trigger (L1) is largely based on custom built
electronics. It incorporates timing from the BPTX  and 
coarse detector information from
the muon trigger chambers and the trigger towers of the calorimeters,
along with multiplicity information from the MBTS scintillators
and the ATLAS forward detectors, LUCID and ZDC.
The L1 system is designed to select events at a rate not exceeding 75
kHz from an input rate  of 40 MHz and identify regions-of-interest
(RoIs), needed by the high-level trigger system (HLT), for
potentially interesting physics objects. 
 HLT runs on a processor farm and comprises the
second level (L2) and the third level trigger or Event Filter (EF). The L2
system evaluates event characteristics by examining the RoIs using
more detector information and more complete algorithms.
The EF analyzes the
L2 selected events again looking at the RoIs' measurements. The
analysis of the complete event is also possible. The output rate is
reduced to approximately 200~Hz.

\begin{figure}[htp]
\begin{center}
\includegraphics[width=0.42\textwidth]{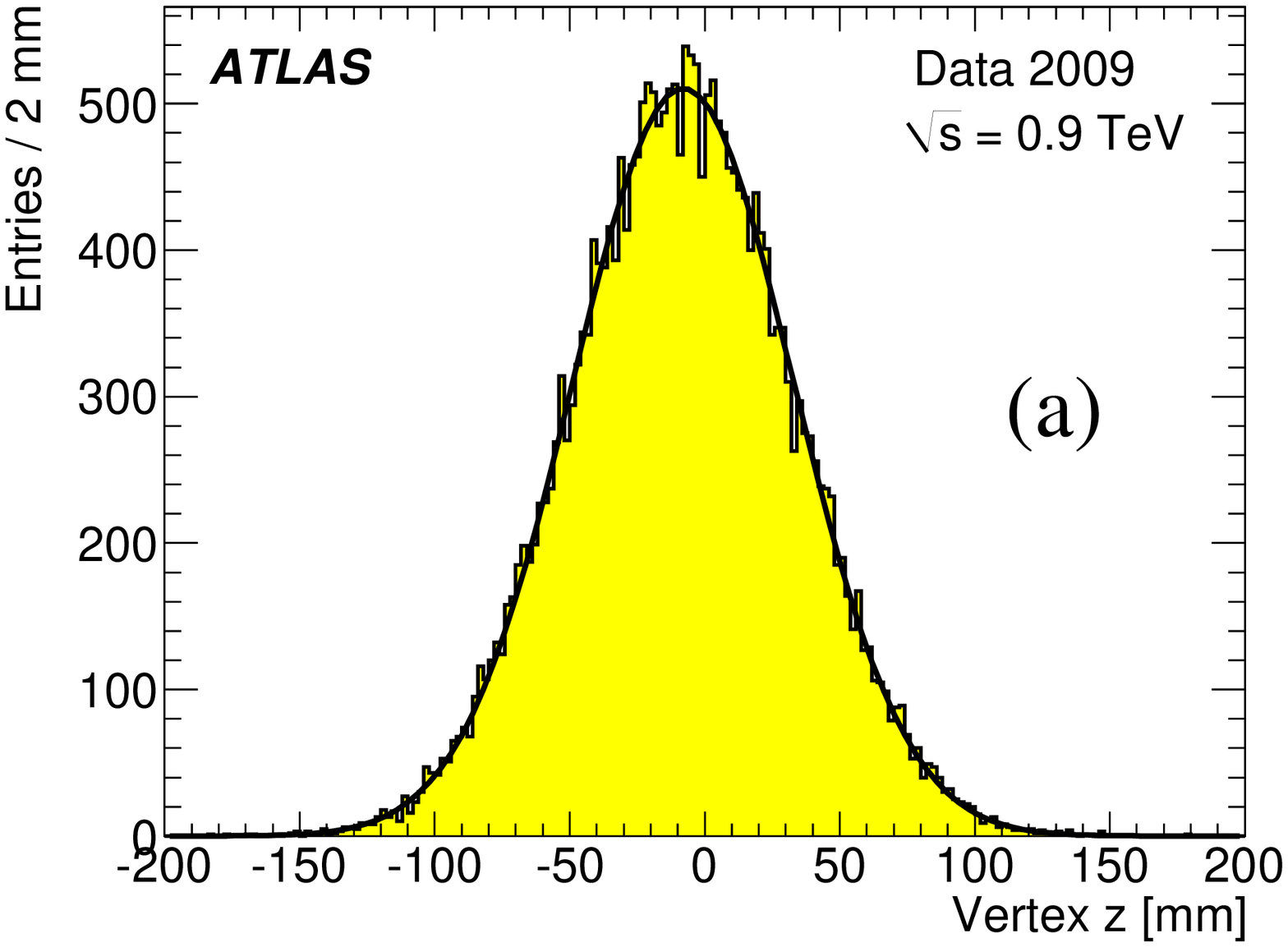}
\raisebox{-.25cm}{\includegraphics[width=0.57\textwidth,height=4.14cm]{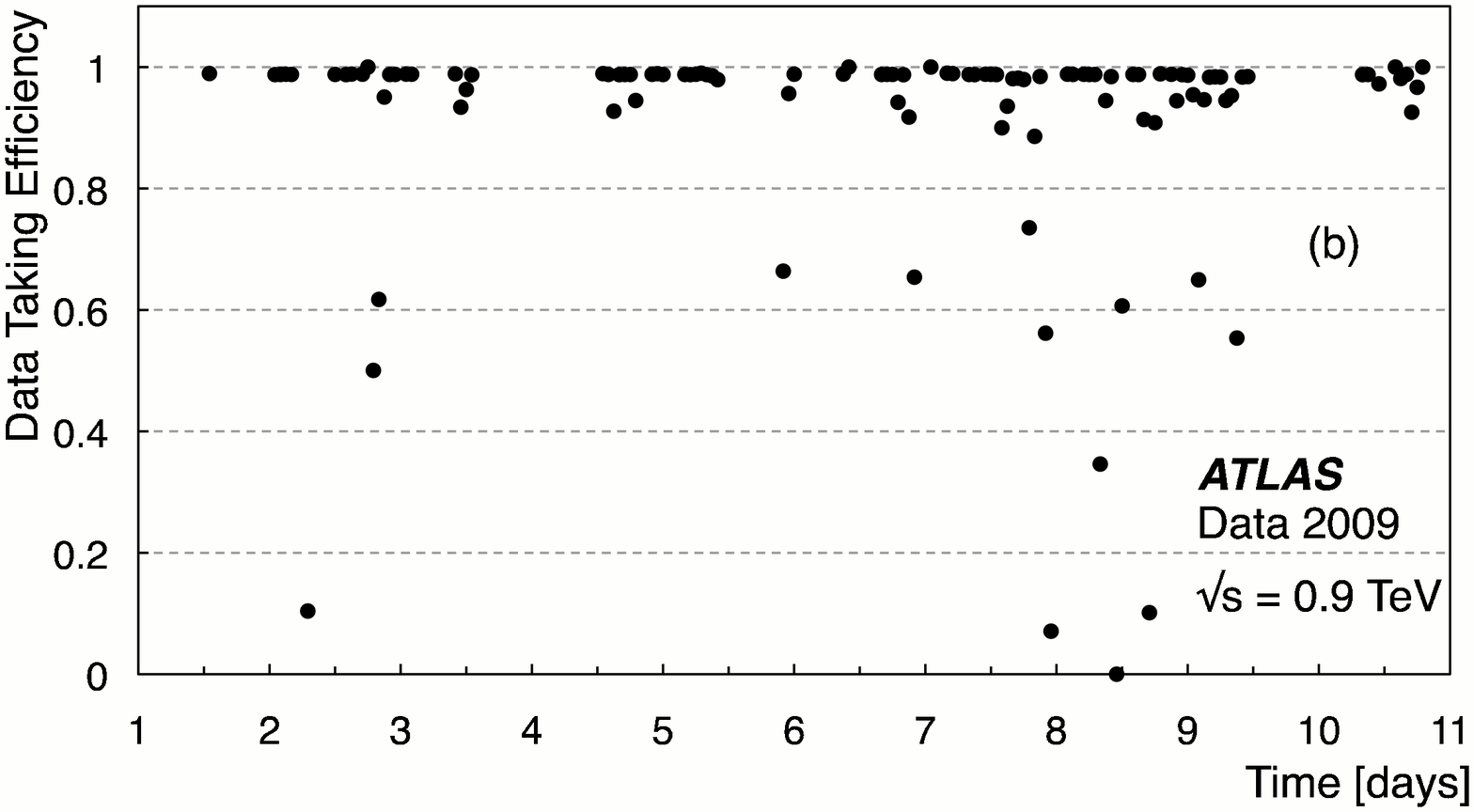}}

\caption{(a) Reconstructed $z$-vertex distribution calculated online by the higher level
trigger for monitoring purposes. The width includes a small
contribution from the experimental resolution. (b) Data-taking efficiency
for periods with  two circulating beams in December 2009. 
}
\label{HLT}
\end{center}
\end{figure}

All trigger decisions
used  were made by the L1 systems as the
luminosity was low.
Minimum-bias events were triggered by a coincidence between the the
BPTX and signals indicating hits in one or both of the MBTS
scintillator wheels.
However, the functionality of the main L2 and EF algorithms, for
example  those used for inner track reconstruction and jet finding,
were validated by 
running in `passthrough' mode, i.e.\ calculating relevant L2 and
EF decisions without rejecting events. Important
distributions like the vertex position (Fig.~\ref{HLT}(a)) were monitored
online and different data streams for fast analysis feedback based on
L1 trigger decisions were provided. For the December 2009 running
period
 the data-taking efficiency of the overall Trigger/DAQ system, as illustrated
in Fig.~\ref{HLT}(b), averaged to 90\%.


\subsection{Event Selection}


\begin{figure}[htp]
\begin{center}
\includegraphics[width=0.49\textwidth]{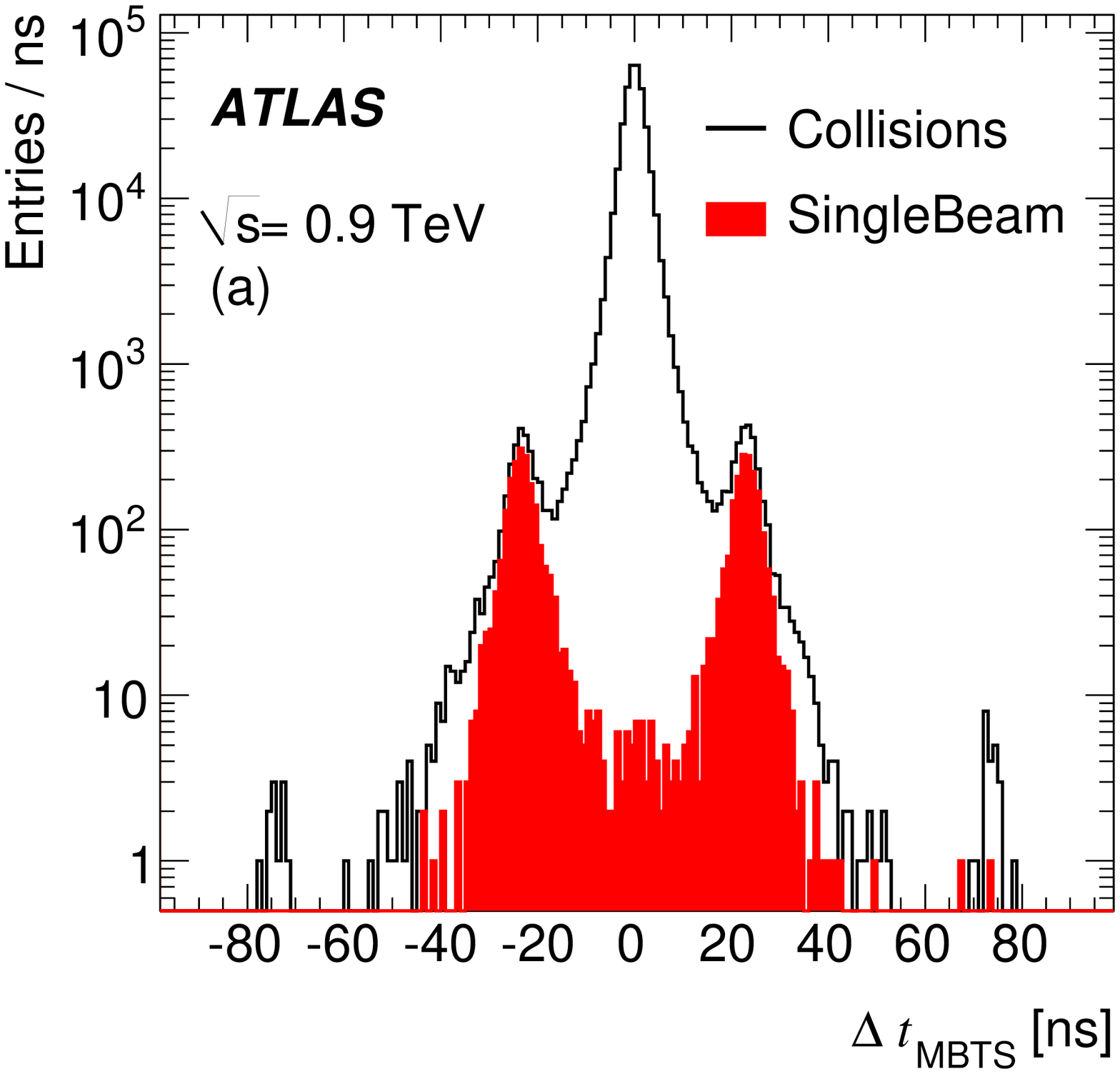}
\includegraphics[width=0.49\textwidth]{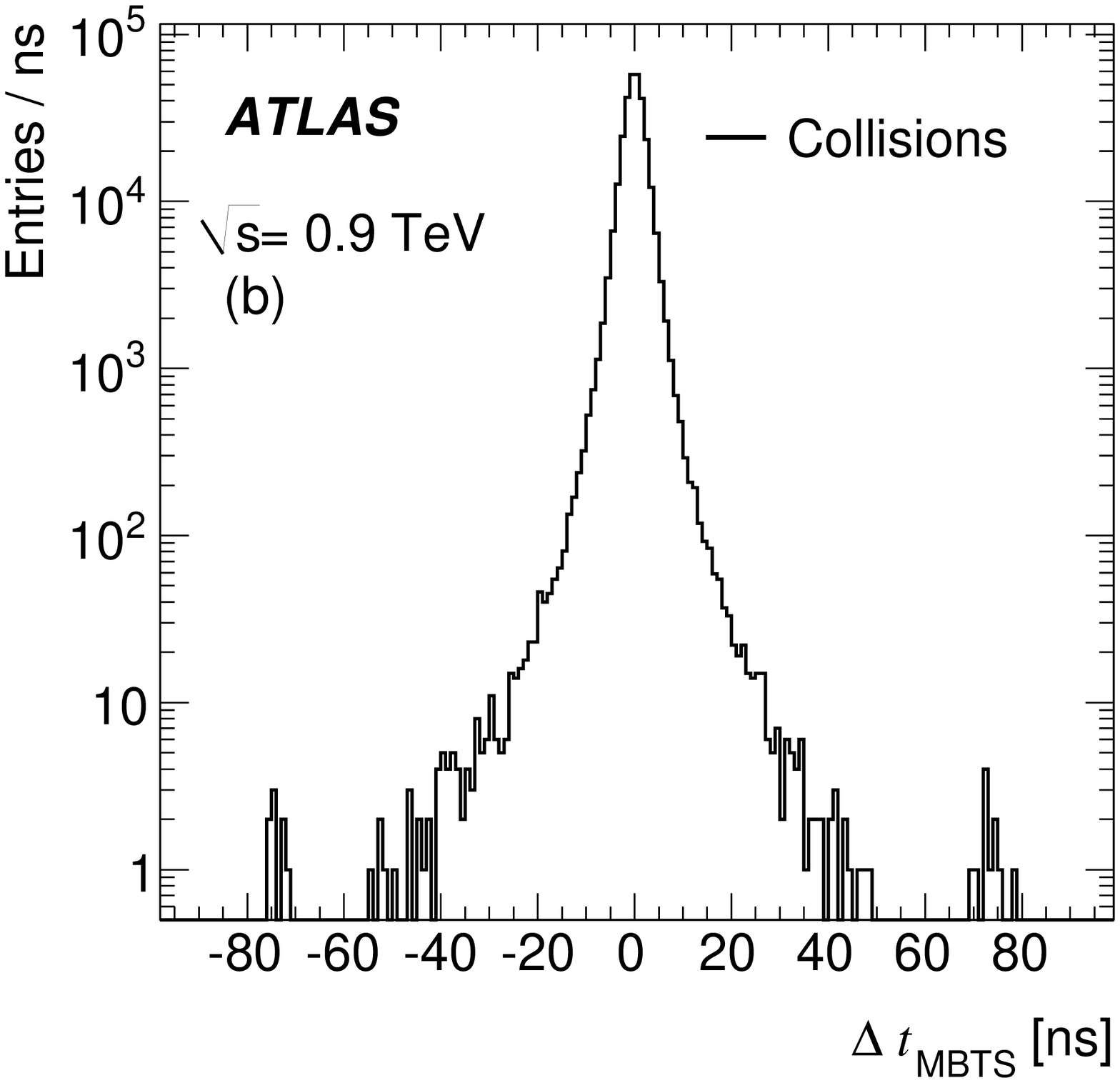}
\caption{Time difference $\Delta t_{\mathrm MBTS}$ of hits recorded by the two MBTS
  scintillator wheels
mounted in front of the electromagnetic end-cap wheels on both sides
of the ATLAS detector; (a) time 
difference without any selection and (b) requiring a
well-reconstructed vertex. 
}
\label{preselfig}
\end{center}
\end{figure}

To  select collision candidates and remove beam-related background
two different strategies were employed:
\begin{itemize}
\item For those studies  based mainly on track information, the
  presence of a primary vertex, reconstructed using at least three
  tracks with sufficient transverse momenta, typically $p_T>150$~MeV, and a transverse distance
  of closest approach compatible with the nominal interaction point
  are required. This selection strategy, which was used in
  Ref.~\cite{Collaboration:2010rd},
   uses events triggered by a
  single hit in one of the two MBTS scintillator wheels.
\item Alternatively, the selection is based on the
      timing difference of signals detected on both sides of the ATLAS detector.
      Coincident signals, within a time window of 5 or 10~ns from either the electromagnetic
      calorimeters (end-cap or FCal) or from
      the two MBTS wheels, respectively, are required. 
   The event must again be triggered by an MBTS signal. In case
    no timing coincidence is found, a two hit MBTS trigger with at least one hit per side is
    required.


\end{itemize}
The detailed track quality criteria used for the first strategy vary
slightly for the different studies presented and are described later
when appropriate.

Without any event selection the MBTS-triggered events contain
backgrounds from beam-related events as shown in
Fig.~\ref{preselfig}(a), where the time difference, $\Delta t_{\rm MBTS}$, of
MBTS signals recorded on both sides of the ATLAS detector is
depicted. For events coming from the interaction point $\Delta t_{\rm MBTS}$ is
small. Beam-related background produced upstream or downstream
should have $\Delta t_{\rm MBTS}$ around 25~ns, with the sign giving the
direction.  
 Eighty percent of single
beam events are missing timing information on one or
both sides and are therefore not shown.
Requiring a
well-reconstructed vertex with track quality requirements reduces
the beam-related background by more than three orders of magnitude
while retaining genuine collision events (Fig.~\ref{preselfig}(b)).
There are twelve single-beam events which meet this vertex
requirement, but all of them are missing timing information and are
not shown.
%
%
\subsection{Luminosity Measurement}
\label{sec:lumi}

The luminosity during the 2009 ATLAS data-taking period was estimated
 offline based upon the timing distributions measured by  the MBTS.
Events with signals
detected on opposite ends of the ATLAS detector  in
the MBTS are counted. After background subtraction the luminosity is calculated
using the number of events with a timing difference consistent with particles
originating from the interaction point (see Fig.~\ref{preselfig}), the
expected minimum-bias cross
section and the event selection efficiency determined from data and
Monte Carlo. The MBTS detector is
used for the absolute luminosity determination because of  its high
trigger efficiency for non-diffractive events~\cite{Collaboration:2010rd}.
The uncertainty on the luminosity, dominated by
the understanding of the   modelling  of inelastic pp interactions, 
is estimated to be around 20\%.

\begin{figure}[htp]
\begin{center}
\includegraphics[width=0.7\textwidth]{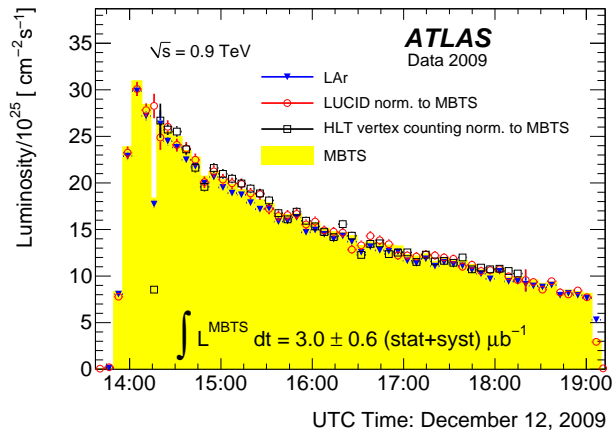}
\caption{
Instantaneous luminosity measured by the MBTS and LAr, with
superimposed the LUCID and HLT vertex counting estimates normalized in such
a way to give the same integrated luminosity as measured with the MBTS
system. All measurements  are
corrected for TDAQ dead-time, except LUCID which is free from dead
time effects.  The short luminosity
drop at 14:15 is due to inhibiting the trigger for ramping up the
silicon detectors after declaration of stable LHC beams.
%
}
\label{lumi} 
\end{center}
\end{figure}

Figure~\ref{lumi}
shows the luminosity as a
function of time, as measured using the MBTS system as well as with
three other techniques:
 timing in the LAr, the LUCID relative luminosity
monitor and particle vertices reconstructed online.
The LAr technique has a slightly larger systematic uncertainty in the
accepted cross section than that
from the MBTS and produces a result which agrees to 4\%.
The other methods are  normalized to the MBTS measurement.

\section{Monte Carlo Simulation}
\label{sec:mc}

Monte Carlo samples produced with the PYTHIA
6.4.21~\cite{Sjostrand:2006za} event generator are used
for comparison with the data.
ATLAS selected an optimized parameter
set~\cite{ATL-PHYS-PUB-2010-002},
using the \pt-ordered parton shower, tuned to describe
the underlying event and minimum bias data from Tevatron measurements
at $0.63$~\TeV\ and $1.8$~\TeV. 
The parton content of the proton is parameterized by
the MRST LO*
parton distribution functions~\cite{Martin:2004dh}.

Various samples of Monte
Carlo events were generated for single-diffractive, double-diffractive
and non-diffractive processes in pp collisions.
The different
contributions in the generated samples were mixed according to the
cross-sections calculated by the generator. 
There was no contribution from cosmic ray events in this simulation.
All the events were processed through the ATLAS detector
sim\-ula\-tion program~\cite{atlas-sim},
which is based on GEANT4~\cite{Agostinelli2003250}.
This simulation software has also  been systematically compared to
test-beam data over the past decade (see e.g. Ref.~\cite{tilecal}) and
it was constantly 
improved to describe these data.  
After the detector sim\-u\-la\-tion the events were reconstructed and
analyzed by the same software chain also used for data.

The  beam position and size, which did not correspond precisely to those
used in the simulation prepared beforehand, have  a
significant impact on some  distributions, particularly for
detectors close to the interaction region. The length of the
luminous region, as seen in Fig.~\ref{HLT}(a), is approximately half
that expected, and the simulated events were re-weighted to match
this. The  transverse offset in the simulation of about 2~mm cannot
be corrected for by this method.

The distributions presented in this paper always show  the
simulated sample normalized to the  number of data events in the figure.

\section{Tracking Performance}
\label{sec:tracking}

The inner tracking system measures charged particle tracks at all
$\phi$ and with 
pseudorapidity $|\eta| < 2.5$.  The pixel detector is closest to the
beam, covering radial distances of $50$ -- $150$~mm with three layers both
in the barrel region and in each end-cap.  The innermost Pixel layer
(known as the  B-layer) is located just outside the beam pipe at a
radius of $50$~mm. The pixels
are followed, at radii between $299\;$--$\;560$~mm, by the silicon
strip detector known as the SCT. This provides $4$ (barrel) or 9
(end-cap) double layers of detectors.  The Pixels are followed, for
radii between 
$563\;$--$\;1066$~mm, by the TRT. 
The TRT straw layout is designed so that charged particles with
transverse momentum \pt~$>$~0.5~\GeV\ and with pseudorapidity $|\eta|
< 2.0$  cross typically more than 30 straws. 
The intrinsic position resolutions
in $r\phi$ for the
Pixels,
 the SCT and the TRT are $10$, $17$ and $130$~\mum, respectively.
For the Pixels and the SCT the other space
coordinate is measured with $115$ and $580$~\mum\ accuracy,
where the SCT measurement derives from a 40~mrad stereo angle
between the two wafers in a layer.

\subsection{Hits on Tracks}

The sample of minimum-bias events  provides approximately two million
charged particles with \pt\ over 500~\MeV\ through the central
 detectors of ATLAS.
Their trajectories in the inner detector were reconstructed
using a pattern recognition algorithm that starts with the silicon
information  and adds TRT hits. This `inside-out' tracking procedure  selects
track candidates with transverse momenta above 500~\MeV~\cite{NEWT}.
  Two further pattern recognition steps were run, each
looking only at hits not previously used: one starts from the TRT and
works inwards adding silicon hits as it progresses and the other
repeats the first step, but with parameters adjusted to allow particle
transverse momenta down to 100~\MeV. The  multiple algorithms are necessary
partly because
a 100~\MeV\ \pt\ charged particle has a radius of
curvature of about 17~cm in the ATLAS magnetic field and will not
reach the TRT.


The track selection requirements vary slightly among the analyses
presented here. A typical set of selections
 is that charged particle tracks are required to have \pt$>$0.5~\GeV,
 $\ge$1 Pixel hit,  $\ge$6 SCT hits and impact parameters with respect to
 the primary vertex of $|d_0| <$ 1.5 mm and $|z_0 \sin\theta| <$ 1.5 mm.
The transverse impact parameter, $d_0$, of a track is its  distance from the
primary vertex at the point of closest approach when projecting into
the transverse plane, signed negative if the extrapolation inwards has the
primary vertex to the right, $z_0$ is the longitudinal distance at that point.

\begin{figure}[htp]
  \begin{center}
    \includegraphics[width=0.49\textwidth,angle=0]{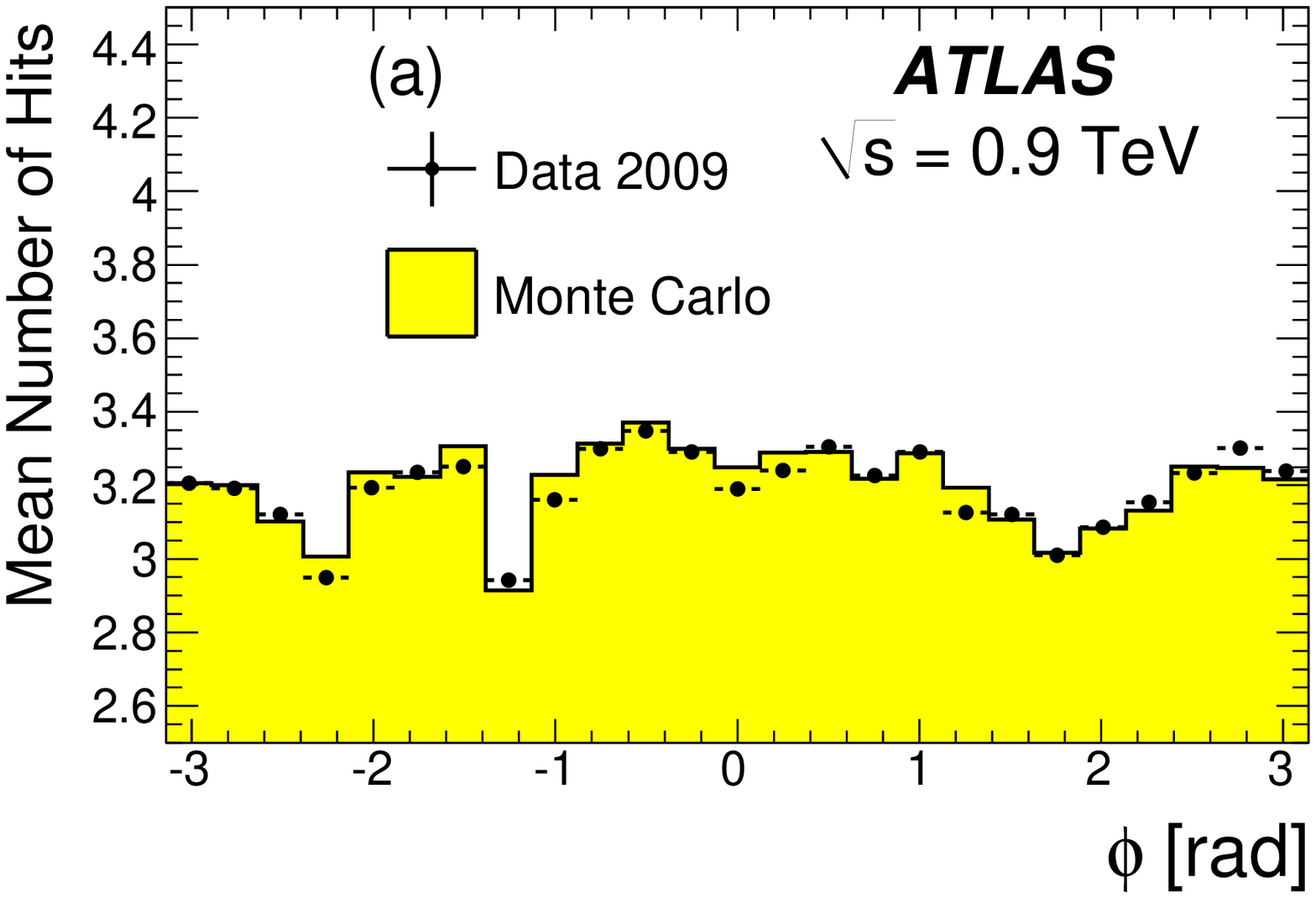}
    \includegraphics[width=0.49\textwidth,angle=0]{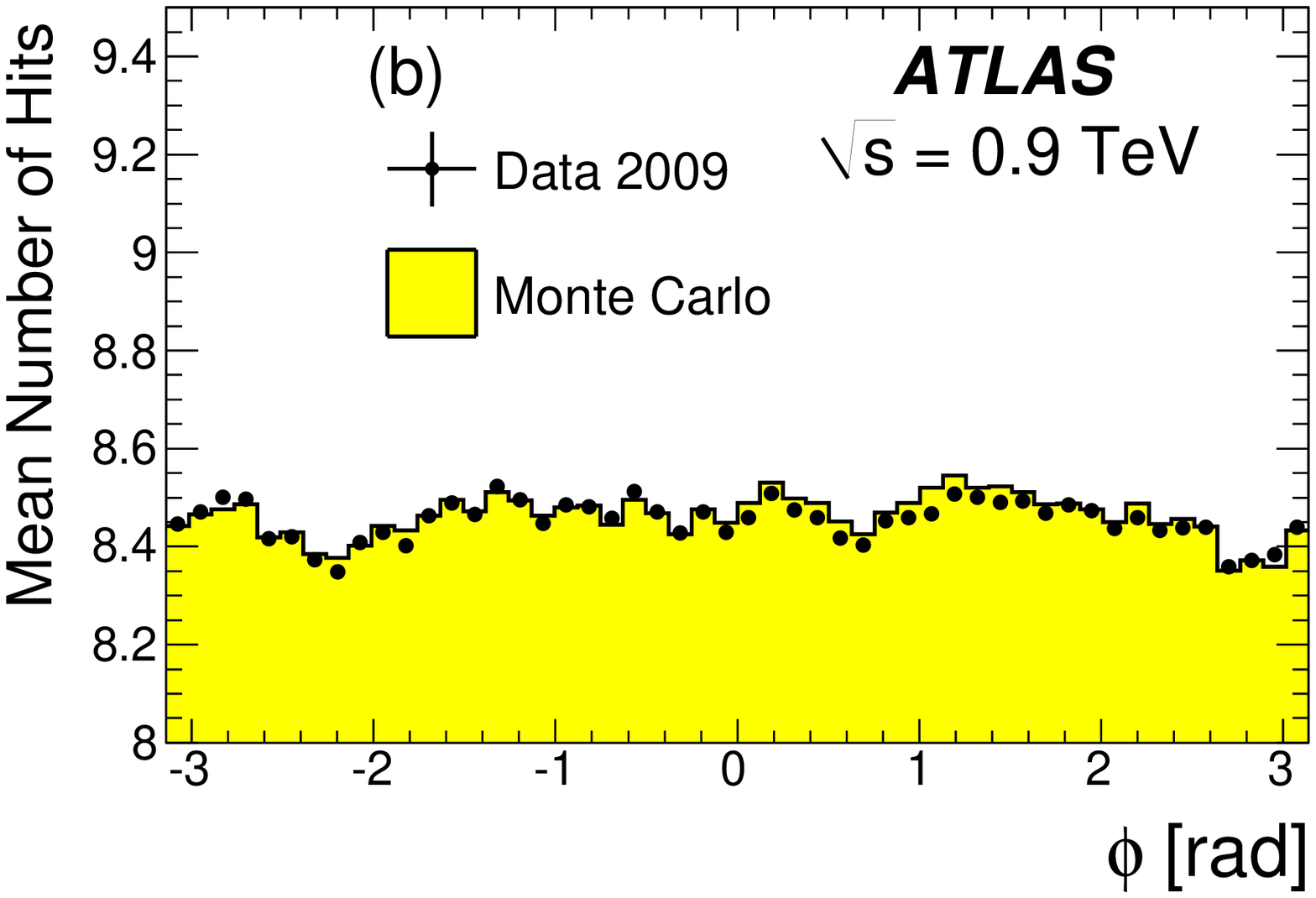}
   \vspace*{-0.5cm}
  \end{center}
  \caption{A comparison of data and simulation in the average number of hits
    in  (a) the Pixels and (b) the SCT
    versus $\phi$ on selected tracks. Comparable
    distributions versus $\eta$ can be found in
    Ref.~\cite{Collaboration:2010rd}.
\label{fig:nsihits}
}
\end{figure}

The  hit distributions in the silicon detectors as a function of
$\phi$ are shown in
Fig.~\ref{fig:nsihits} for tracks passing these requirements
 in data and simulation.
The fluctuations seen in $\phi$ correspond to non-responsive
detector modules which are modelled in the simulation. A small
mis-match between data and simulation arises  because the simulated
beam had a transverse displacement of about 2~mm from the true
position, as discussed in Section~\ref{sec:mc}.

\begin{figure}[htbp]
  \begin{center}
    \includegraphics[width=0.49\textwidth]{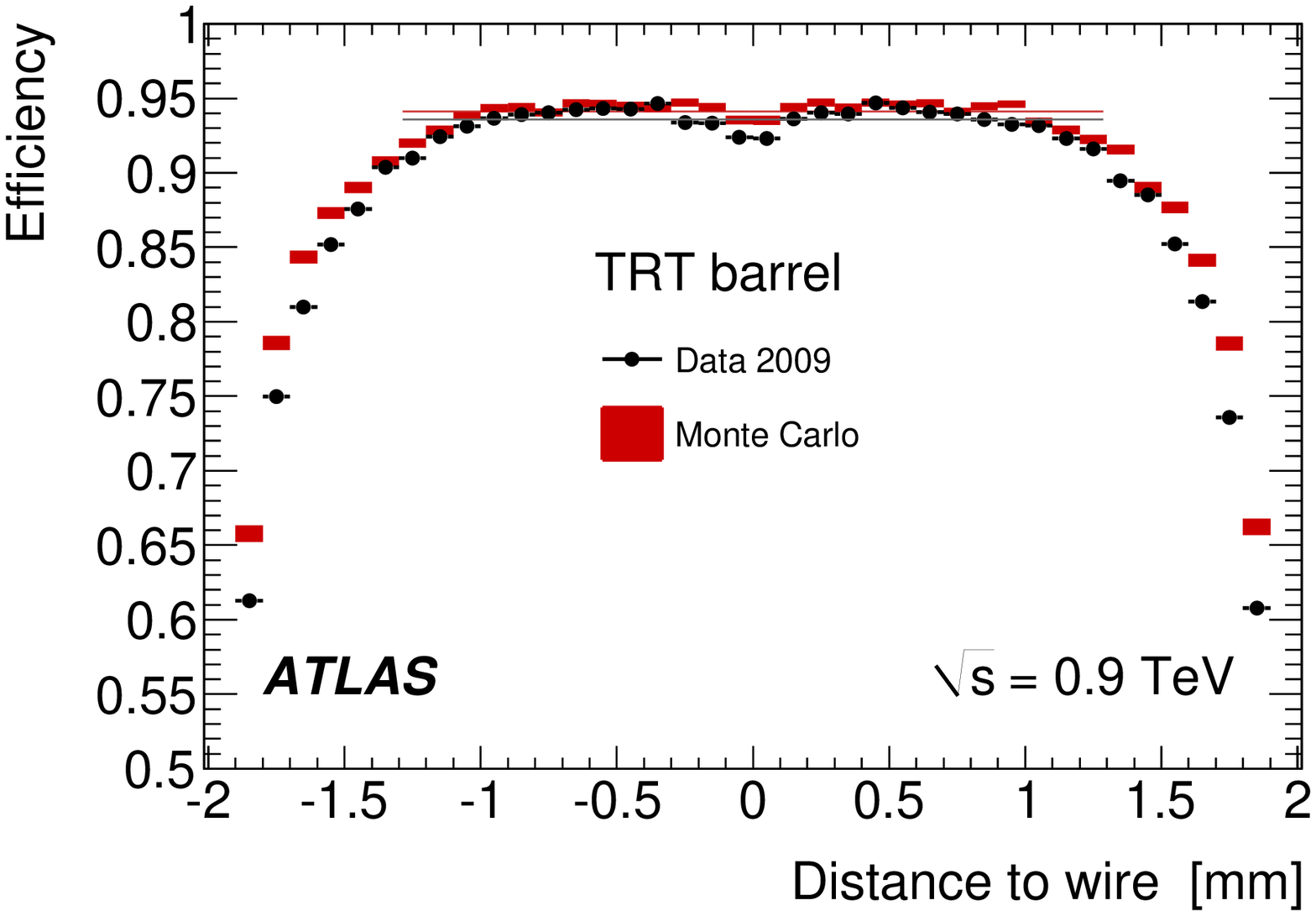}
    \includegraphics[width=0.49\textwidth]{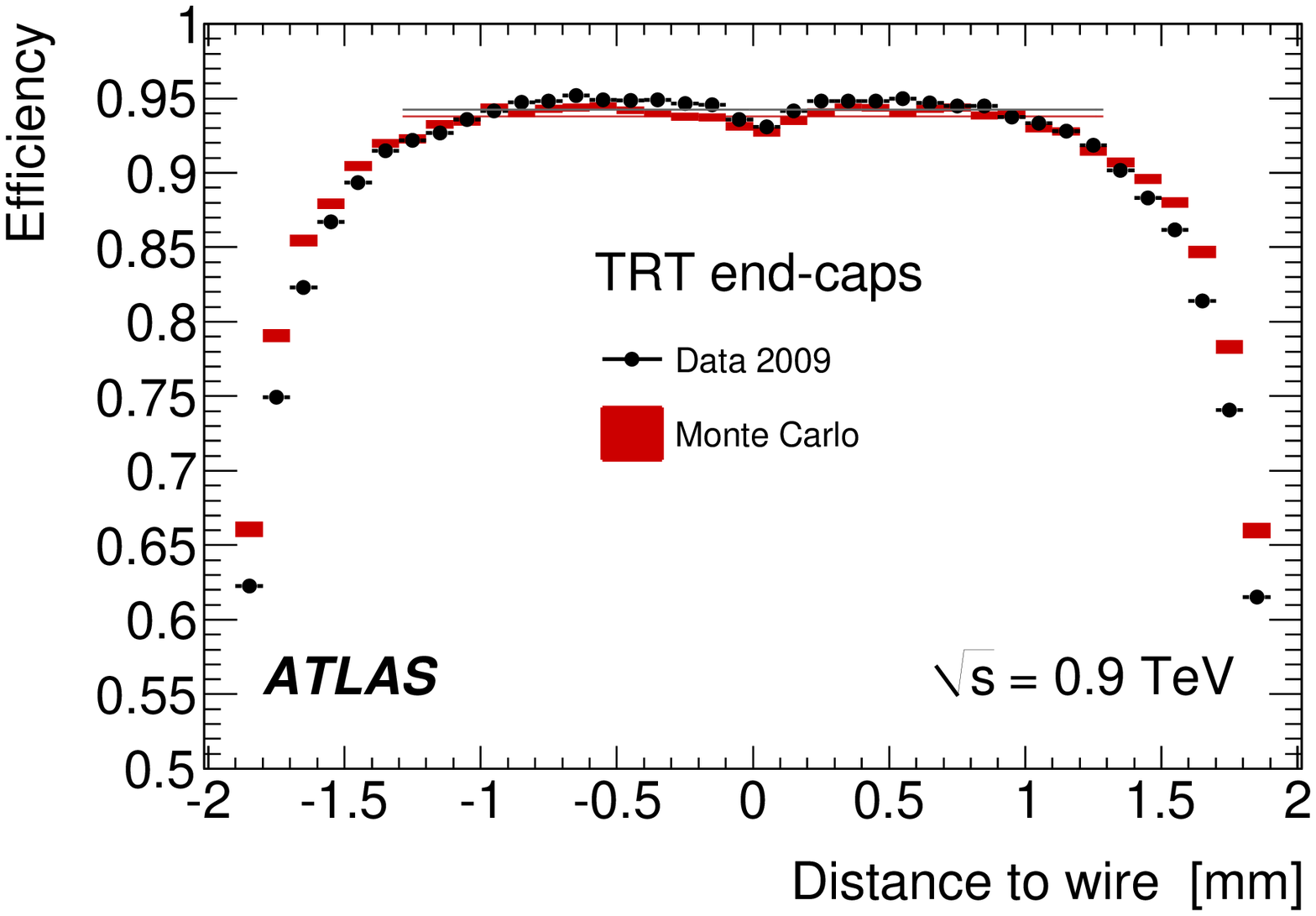}
     \put(-290, 75){\small\bf (a)}
     \put(-120, 75){\small\bf (b)}
   \vspace*{-0.5cm}
  \end{center}
  \caption{TRT hit efficiency as a function of the distance of the track
 from  the wire in the centre of the straw
in (a) the barrel and     (b) the end-caps. 
\label{fig:trteff}
}
\end{figure}

The efficiency of the individual TRT straws is displayed in
Fig.~\ref{fig:trteff} as a function of the distance of the test track
to the wire in the  centre of the straw. The efficiency for data and
simulation, barrel and end-cap, has a plateau close to 94\%.

The alignment of the tracking detectors benefited from the precision
construction and survey followed by an extended period of data taking
using cosmic ray muons~\cite{id-paper}.
The alignment was improved using the 0.9~\TeV\ collision data,
although the particles have rather low momentum and therefore their
tracks suffer from multiple scattering.  
The quality of the alignment can be checked by the study of the
residuals, which are defined as the measured hit position minus that expected
from the track extrapolation.

\begin{figure}[htbp]
  \begin{center}
    \includegraphics[width=0.49\textwidth]{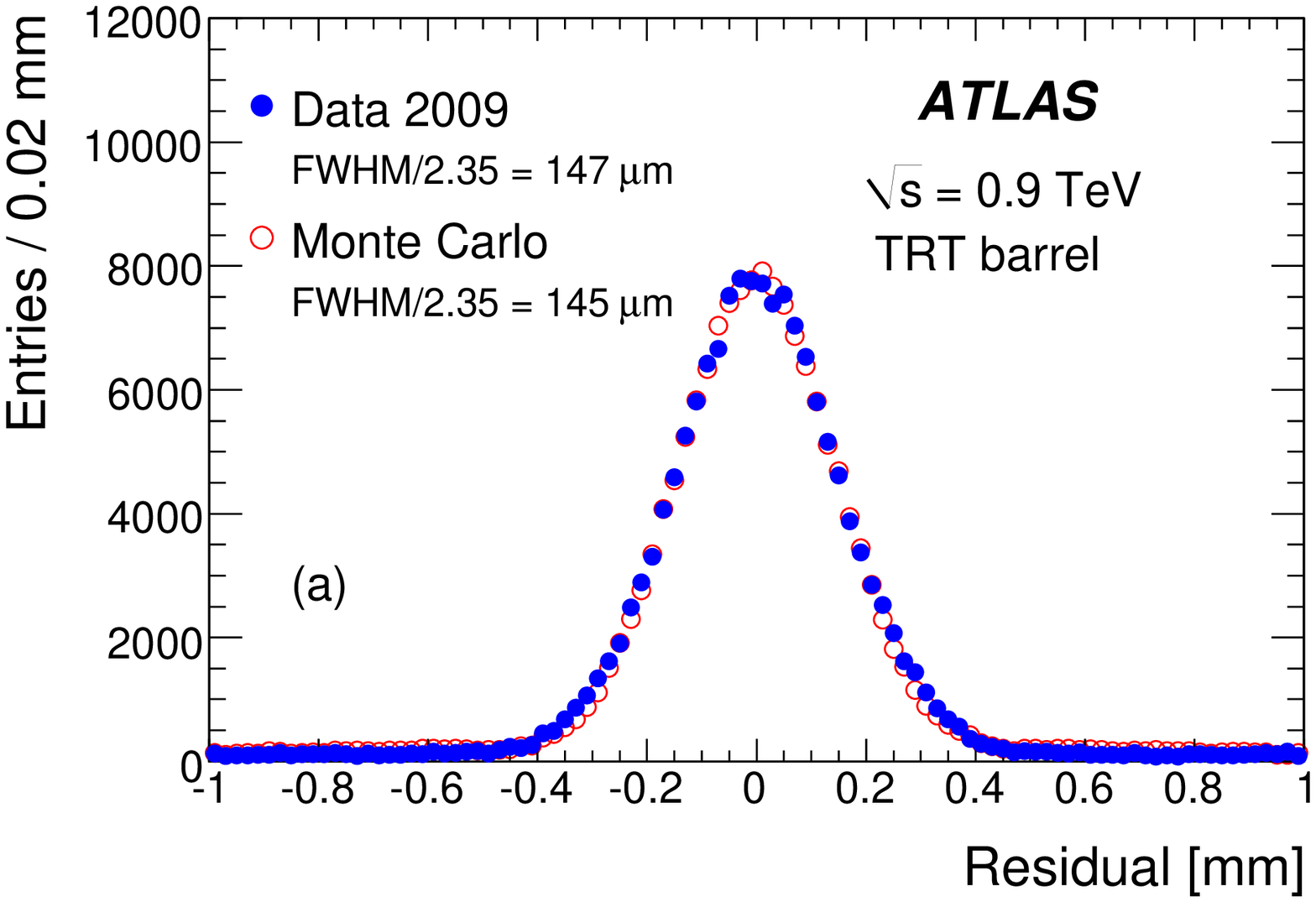}
    \includegraphics[width=0.49\textwidth]{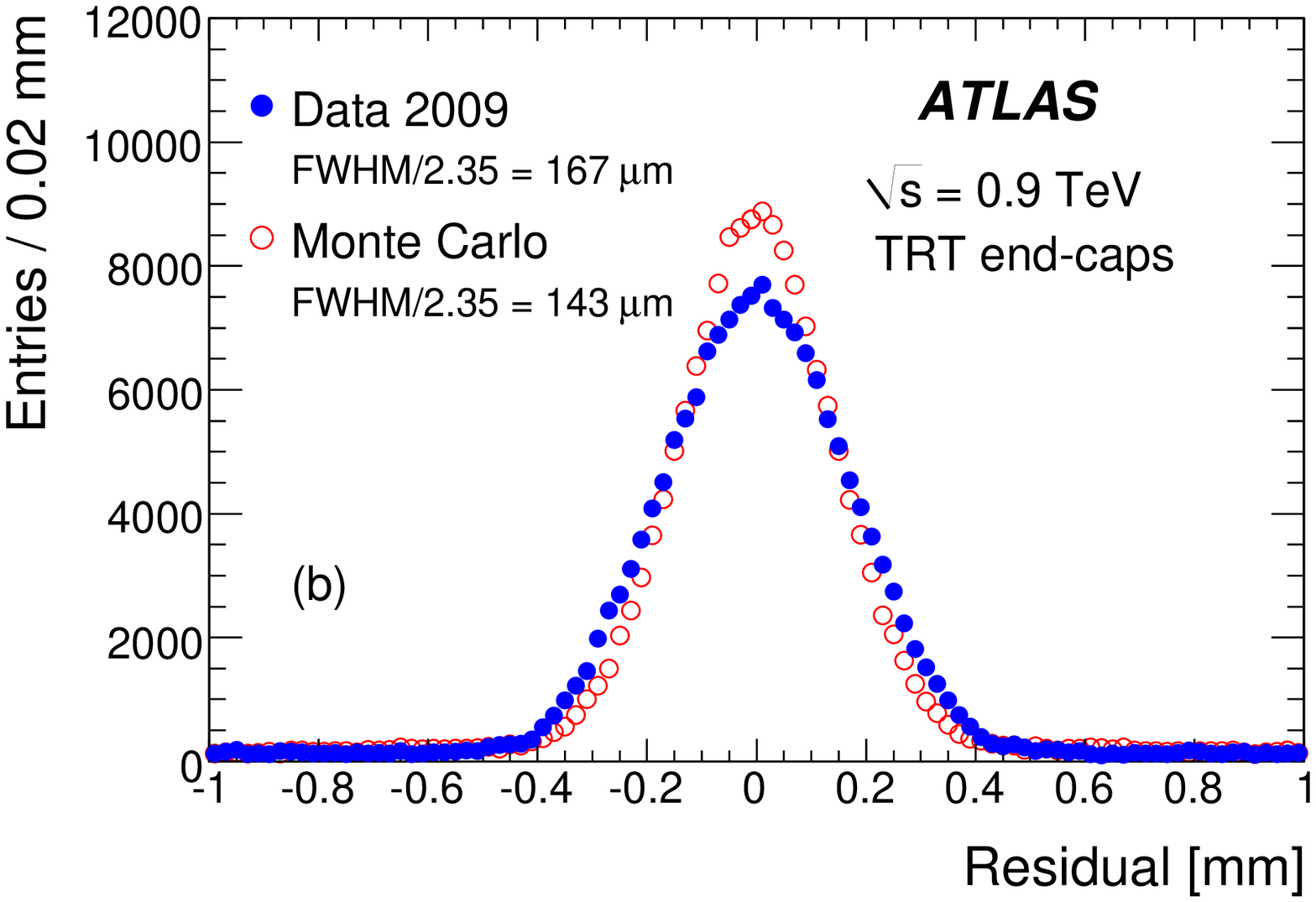}
   \vspace*{-0.7cm}
  \end{center}
  \caption{Unbiased residual distributions in the TRT  barrel (a) and
    end-caps (b). The data points are in filled circles and the simulation in
    empty ones.
\label{fig:trtres}}
\end{figure}

Unbiased residuals between  tracks and barrel TRT  hits
are plotted in Fig.~\ref{fig:trtres}.
This figure is made using charged particles  with \pt$>$1~\GeV\ with over 6 hits in
the SCT and at least 14 in the TRT.
The equivalent Gaussian width is extracted from the full-width at half maximum.
The end-cap shows a resolution somewhat
worse than simulation, while in the barrel part of the detector, where
a higher cosmic ray flux can be used for alignment, data and
simulation  are in close agreement. 

\begin{figure}[htbp]
  \begin{center}
    \includegraphics[width=0.49\textwidth]{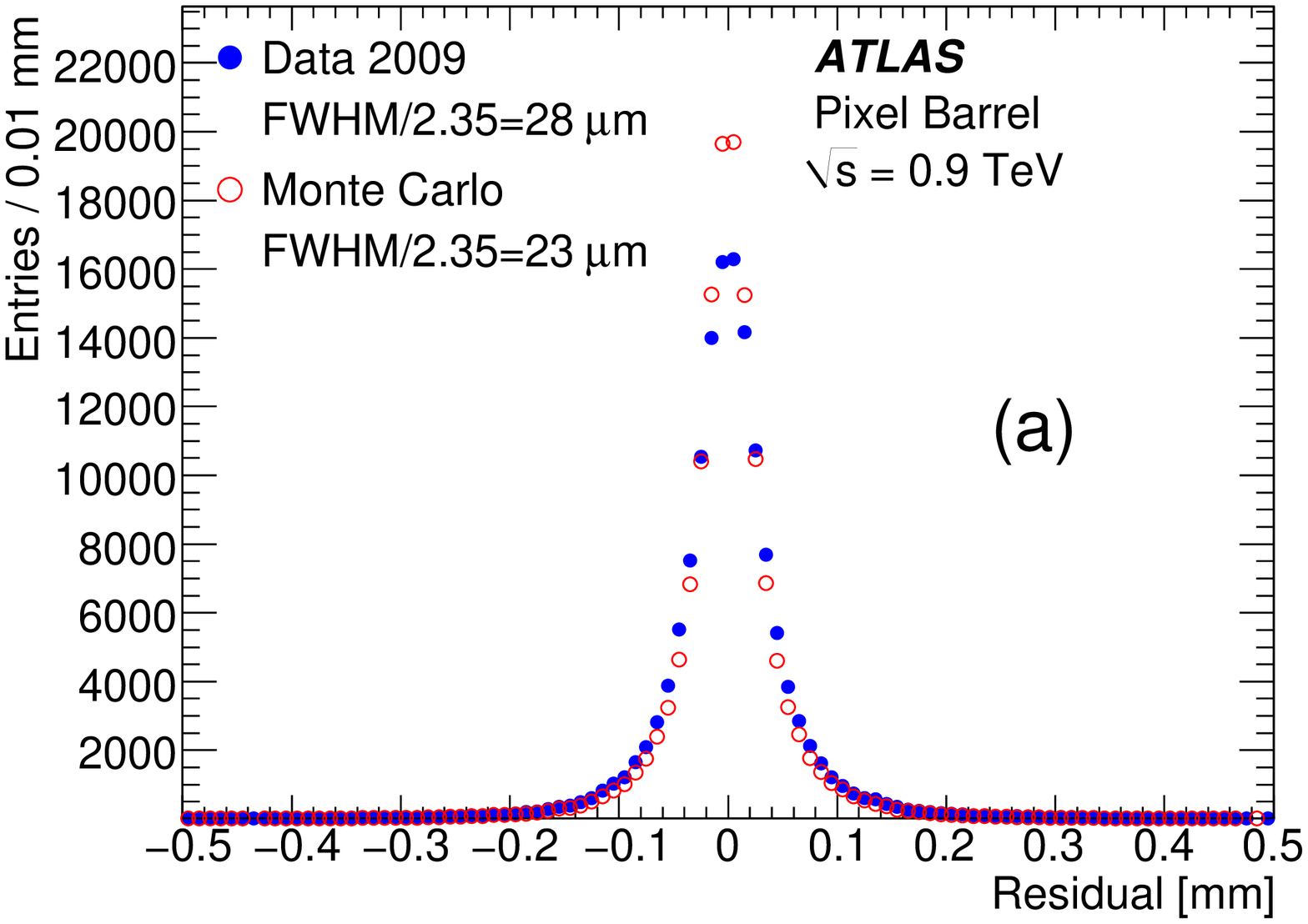}
    \includegraphics[width=0.49\textwidth]{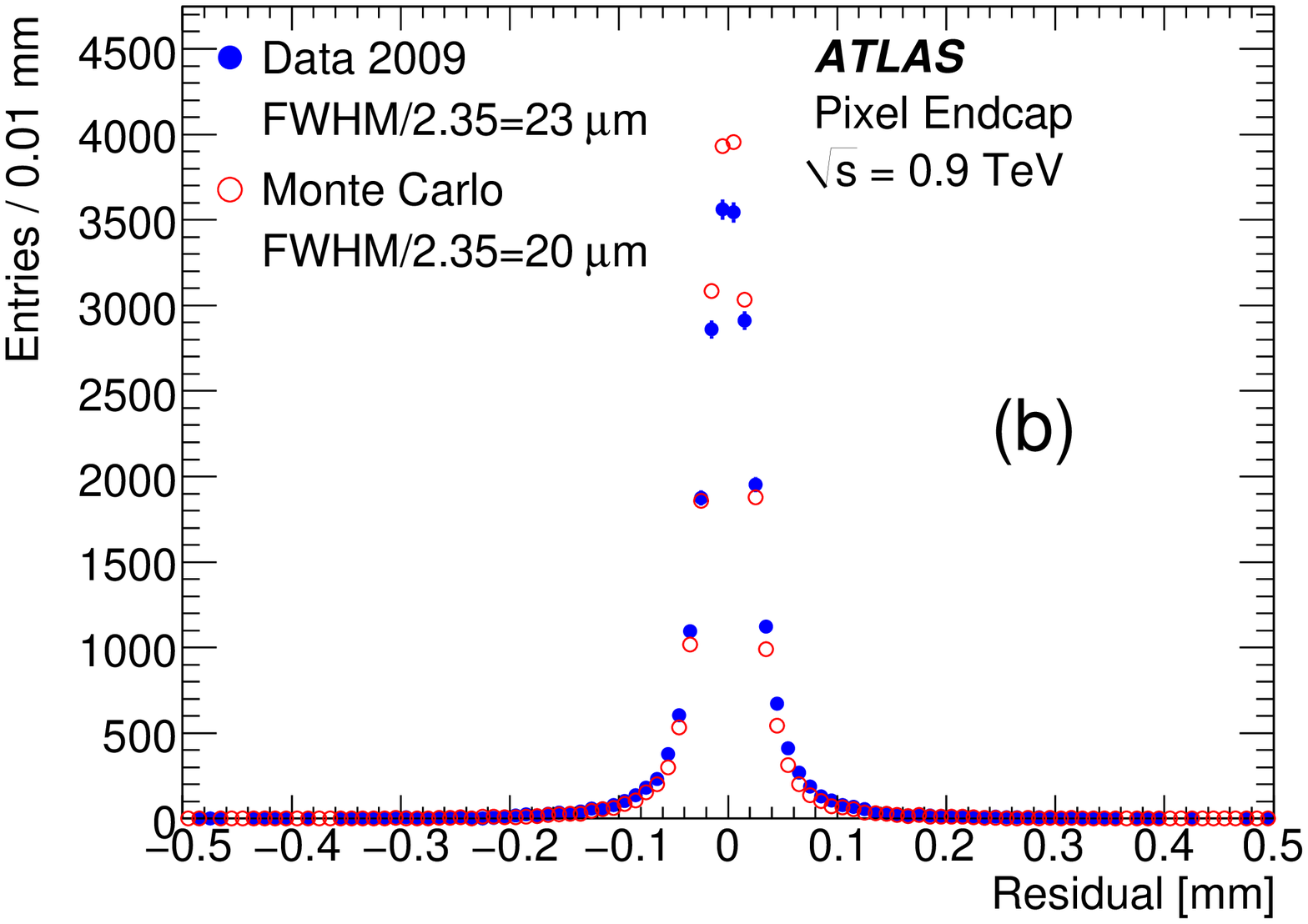}
    \includegraphics[width=0.49\textwidth]{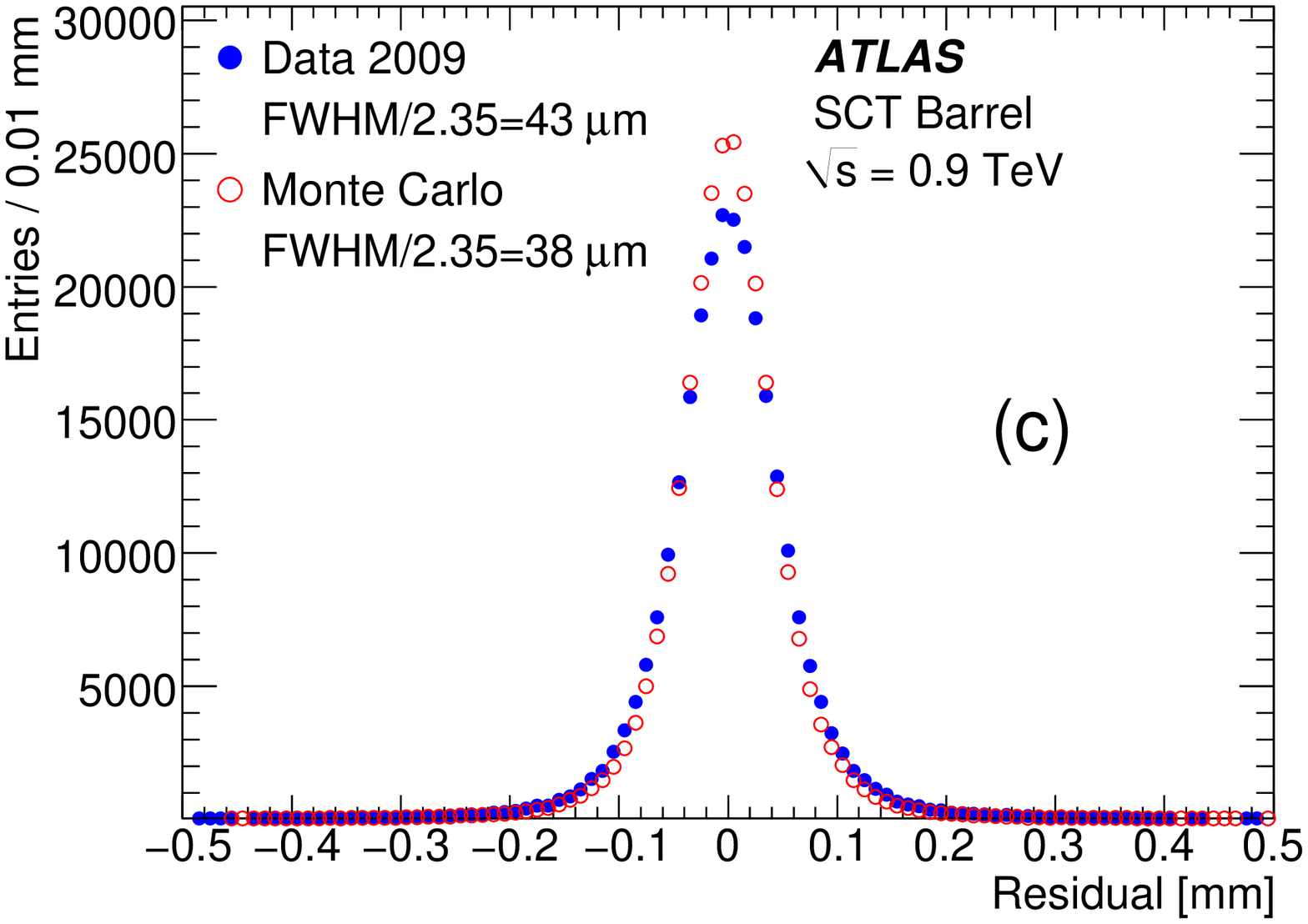}
    \includegraphics[width=0.49\textwidth]{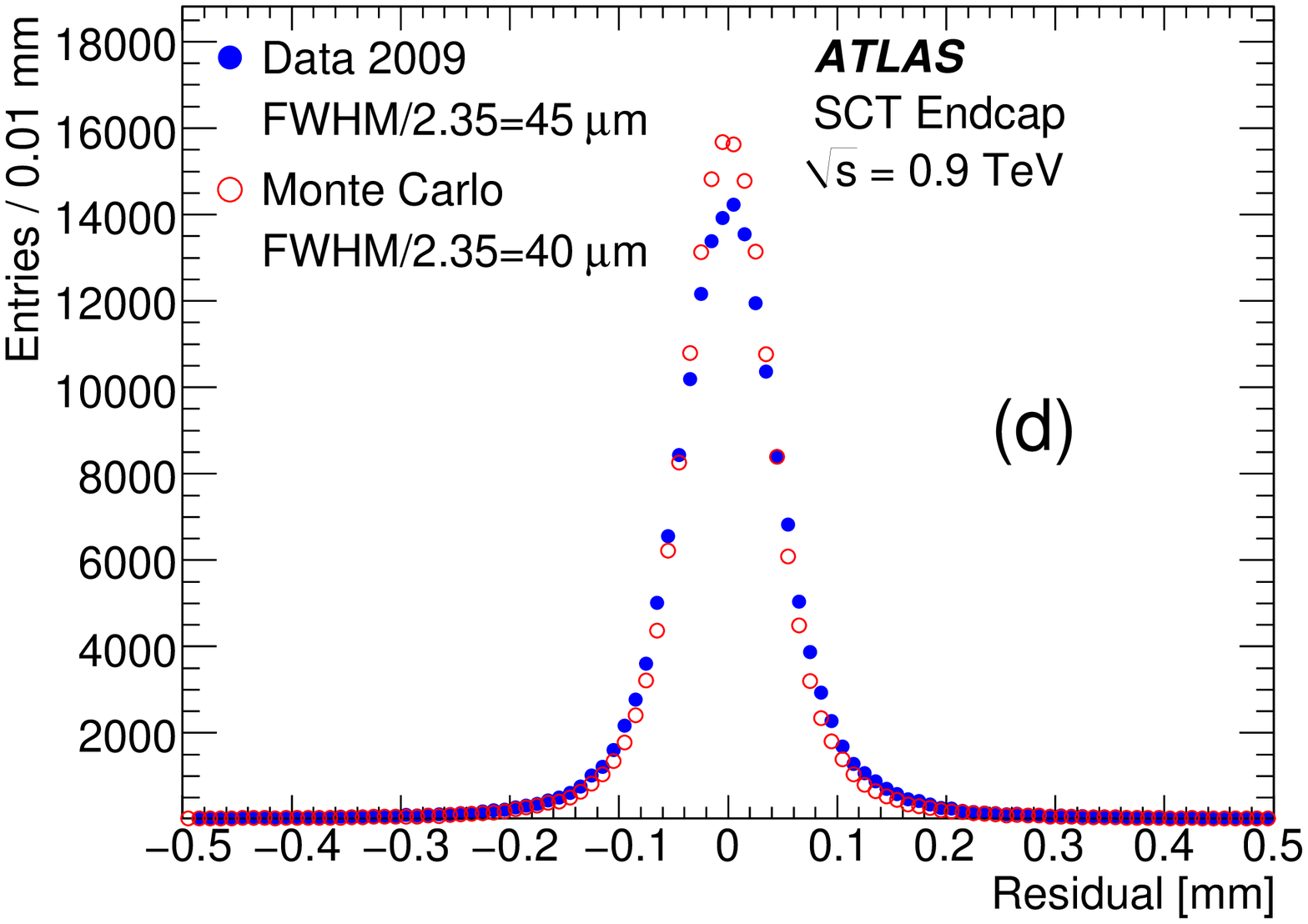}
   \vspace*{-0.7cm}
  \end{center}
  \caption{The distributions of the silicon detector unbiased residuals for
(a)  the pixel barrel,
(b)  the      pixel end-cap,
(c)  the      SCT barrel,
(d)  the      SCT end-cap.
    The data are in solid circles,
    the simulation, which has a perfect alignment, is shown with open ones.
  \label{fig:si-residuals}}
\end{figure}

Unbiased $x$ residuals from the
silicon detectors are shown in Fig.~\ref{fig:si-residuals}, where
$x$ refers to the more precise local coordinate on the detector.
Charged particles are selected to have  \pt$>2$ \GeV.
The equivalent Gaussian width is extracted from the full-width at half maximum.
The width of the
resulting distributions
in data are within about 15\% of those found in a  simulation
with no alignment errors, showing that the remaining impact on the
residual widths from 
imperfect alignment in data is at the level of approximately 10-15
\mum\ for the pixels and of 20~\mum\ for the SCT.


\subsection{ $\boldsymbol{K}^0_{\mathbf{S}}$ Studies}
\label{sec:k0}

\begin{figure}[htbp]
  \begin{center}
    \includegraphics[width=0.9\textwidth,angle=0]{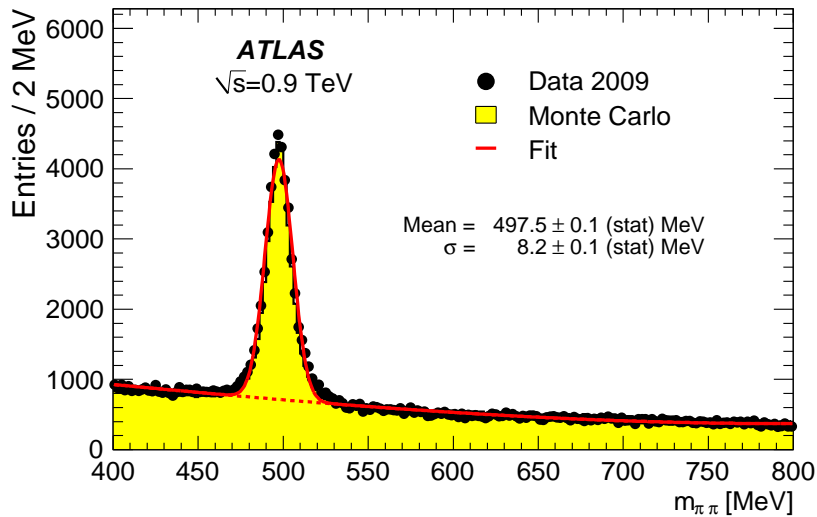}
   \vspace*{-0.7cm}
  \end{center}
  \caption{The  \kshort\ candidate mass distribution using impact parameter
    and lifetime selections. The simulated signal and background are
    separately normalized to the data.
  \label{fig:kShort}}
\end{figure}

The momentum scale  and resolution of the tracker, and energy loss
with in, were
all investigated by studying the \kshort\ to $\pi^+\pi^-$ decay.  The
reconstruction 
requires pairs of oppositely-charged particles compatible with coming
from a common vertex.  This vertex, in the transverse plane, must be
more than 0.2~mm from the primary vertex. The cosine of the angle
between the flight path relative to the primary vertex and the
momentum vector of the candidate, $\cos\theta_K$, is required to exceed 0.8. The
invariant mass distribution,
 calculated assuming that both charged particles  are pions
 is shown in Fig.~\ref{fig:kShort}.
 The simulated signal and background are
    separately normalized to the data, and the position and width of
    the \kshort\ mass peak are fitted using a Gaussian. The  peak in
    data is at $m_{\pi\pi} = 497.5\pm0.1$~\MeV, in agreement with the
    PDG average~\cite{Amsler20081}.

In order to test the momentum scale and resolution of the detector
the reconstructed pions in the simulation are adjusted by parameters
$\mu_{\rm tr}$, which scales the 1/\pt, and $\sigma_{\rm tr}$, a
Gaussian smearing on $\mu_{\rm tr}$. 
The values of these parameters
which best fit the observed \kshort\ mass and width in  the barrel
region  are $\mu_{\rm tr} = 1.0004 \pm 0.0002$ and $\sigma_{\rm tr}
= 0.0040 \pm 0.0015$. 
Thus the momentum scale for these barrel
charged particles  is known at better than the one per mille level, which is
as expected from the
accuracy of the solenoid magnet field-mapping performed before
installation of the inner detector~\cite{solref}.
This, and subsequent \kshort\ studies, use a tighter cut of 0.99 on
$\cos\theta_K$. 

 In the end-cap
regions there is evidence for a degraded resolution, especially at
low momentum. Charged particles  with $p_T$ below 500~\MeV\ require a
$\sigma_{\rm tr}$ of 0.024$\pm$0.004 and 0.022$\pm$0.004 in the
negative and positive end-caps, respectively, to match the data,
suggesting some material is missing in the description of the
end-caps. The momentum scale in the end-caps is compatible with
the nominal within errors of 1 to 2 per mille.

\begin{figure}[htbp]
  \begin{center}
    \includegraphics[width=0.7\textwidth,angle=0]{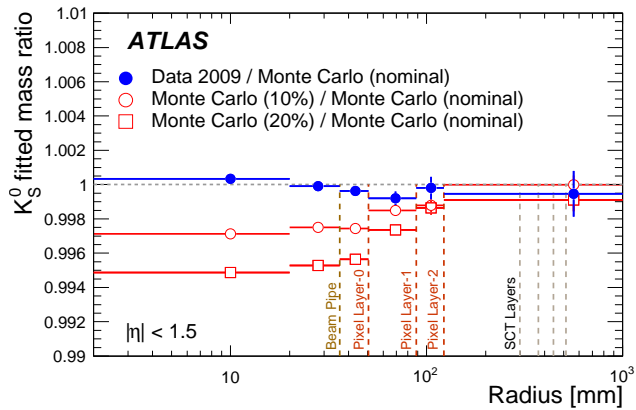}
   \vspace*{-0.5cm}
  \end{center}
  \caption{The fitted \kshort\ mass divided by the value found in nominal
MC simulation as a function of the reconstructed decay position.
The filled circles show the data, and the open symbols are for
simulation samples with approximately 
10\% and 20\% more silicon tracker material added.
The horizontal dotted line is to guide the eye.
  \label{fig:k0-v-r}}
\end{figure}

The \kshort\ peak was also used to investigate the amount of material
in the inner tracker as a function of radius. The mass  reconstructed in data,
divided by that found in
simulation,  is shown in
Fig.~\ref{fig:k0-v-r} as a function of decay radius.

Deviations of this ratio from unity would expose 
differences between the real detector and the model used for simulation.
The  results for  special simulation samples   with approximately 10\% and 20\%
fractional increase in the radiation length of the silicon systems included by
increasing the 
density of some of the support  structures are also shown in
Fig.~\ref{fig:k0-v-r}. These results suggest that discrepancies of
material between the data and the simulation  must be significantly smaller
than 10\% of the  material thickness in the inner silicon barrels.

\subsection{d$\boldsymbol{E}/$d$\boldsymbol{x}$ and $\boldsymbol{\phi}$(1020) Identification}

\begin{figure}[htp]
  \begin{center}
   \vspace*{0.5cm}
    \includegraphics[width=0.8\textwidth,angle=0]{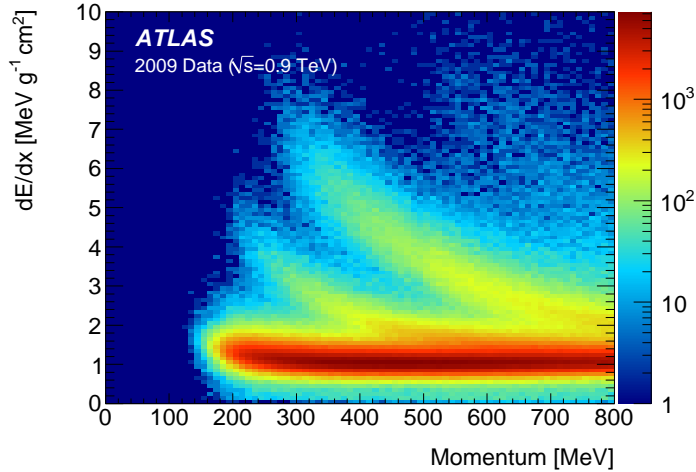}
   \vspace*{-0.5cm}
  \end{center}
  \caption{The \dedx\ measured in data as a function of momentum.
  \label{fig:dedx}}
\end{figure}

One  feature of the Pixel tracking system is a time-over-threshold
measurement
for the signal which was used to extract
the specific energy loss \dedx.
Tracks with  more than one Pixel hit were studied and the mean \dedx\ was
found for each after
the highest was removed to reduce the effect of Landau
fluctuations. Figure~\ref{fig:dedx} shows the  distribution
observed in the data. Bands corresponding to 
different particle species are clearly visible.

In the observation of $\phi \ra K^+K^-$ identification of the
$K^{\pm}$ reduces the combinatorial background.
The identification of kaon candidates through \dedx\ proceeds by finding the
probability density functions of
pions ($p_{\rm pion}$),  kaons ($p_{\rm kaon}$) and protons ($p_{\rm proton}$)
in the simulation as a function of momentum and \dedx. This is done via
fitting the observed value in simulation using a Gaussian
function whose parameters are momentum dependent. The simulation
models the data with an accuracy of about 10\%.

The tracks used in the reconstruction of  the $\phi$ meson
 must have more than
one hit in the Pixel system and an impact parameter
within 3$\sigma$ of the primary vertex. 
The track fit was re-run using the kaon mass hypothesis for the energy
loss. The simulation shows that after re-fitting the kaon momenta are
underestimated 
by up to 10~\MeV\ and a corresponding correction is applied.
This changes the reconstructed $\phi$ mass by approximately 0.3~\MeV.
All oppositely charged particle pairs 
where both momenta, reconstructed under the kaon hypothesis, are
below 800~\MeV\ are considered.

\begin{figure}[htp]
  \begin{center}
    \includegraphics[width=0.7\textwidth,angle=0]{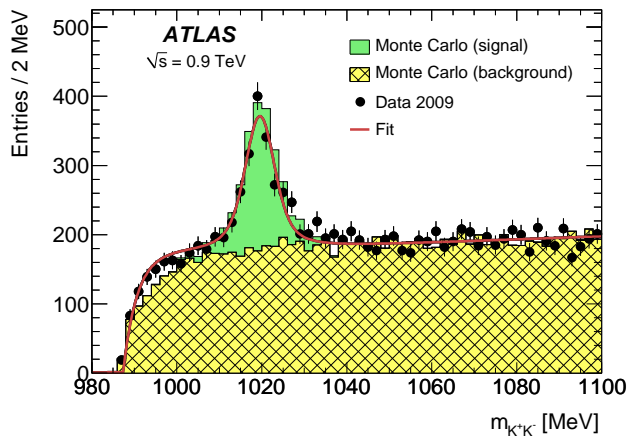}
   \vspace*{-0.5cm}
  \end{center}
  \caption{The  measured and simulated mass spectra of 
    $K^+K^-$ pairs. The $\phi$ peak is fitted with a Breit-Wigner
    with a fixed width convoluted with a Gaussian. Both kaons must be
    identified through the \dedx\ measurement.
  \label{fig:phi}}
\end{figure}

Figure~\ref{fig:phi} shows the resulting mass
 distribution for the $K^+K^-$ candidate pairs, selected using charged
 particles with
 200 $<$ \pT\ $<$~800~\MeV\ and a kaon \dedx\ tag.  The selection cuts
 were chosen to yield optimal signal significance
on simulated events; a measure which was 
 greatly improved using the \dedx\ information.

The background and signal levels in the simulation were scaled
independently to match the data.
The fit allowed the mass and experimental resolution to
vary, while keeping the natural width fixed to the
PDG~\cite{Amsler20081} average.
The mass was found to be
   1019.5$\pm$0.3~\MeV, in agreement with the expected value.
The fitted experimental resolution  in data is 2.5$\pm$0.5~\MeV\ and
matches the 2.4$\pm$0.3~\MeV\ found in Monte Carlo simulation.

\subsection{Secondary Vertex Tagging}


An important role of the tracking system is the identification
of heavy flavour hadrons. There are several tagging algorithms developed
in ATLAS. Some performance figures for two algorithms, the impact
parameter and the 
secondary vertex tagging algorithm, are presented in the following.


The  transverse impact parameter, $d_0$, is a key variable for
discriminating tracks originating from displaced vertices from those
originating from the primary vertex.  For studies of track impact
parameters  the $d_0$ was calculated with respect to a primary vertex
which was fitted excluding  that track in order to remove bias.

\begin{figure}[htp]
  \begin{center}
    \includegraphics[width=0.56\textwidth,angle=0]{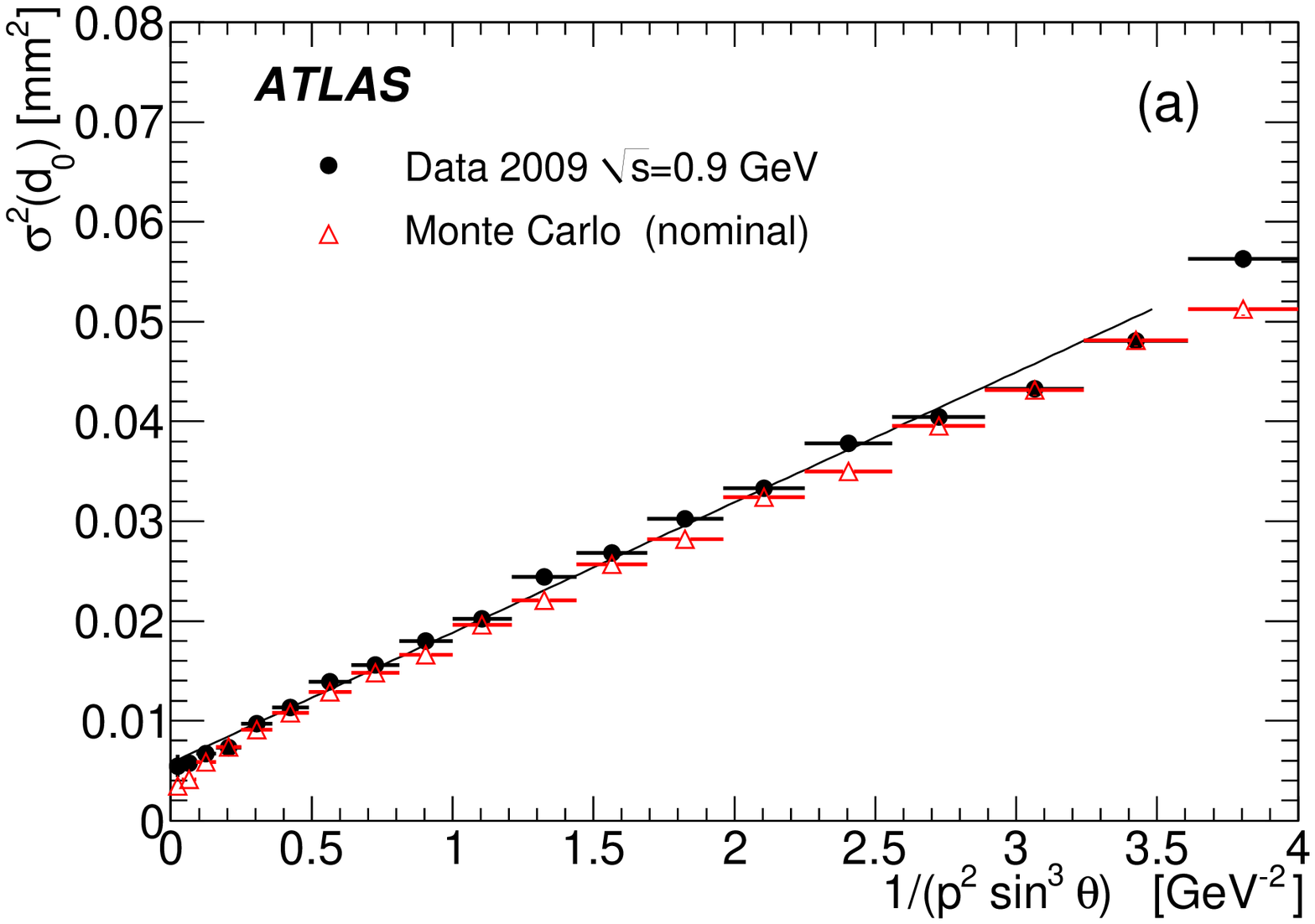}
    \raisebox{-0.11cm}{\includegraphics[width=0.43\textwidth]{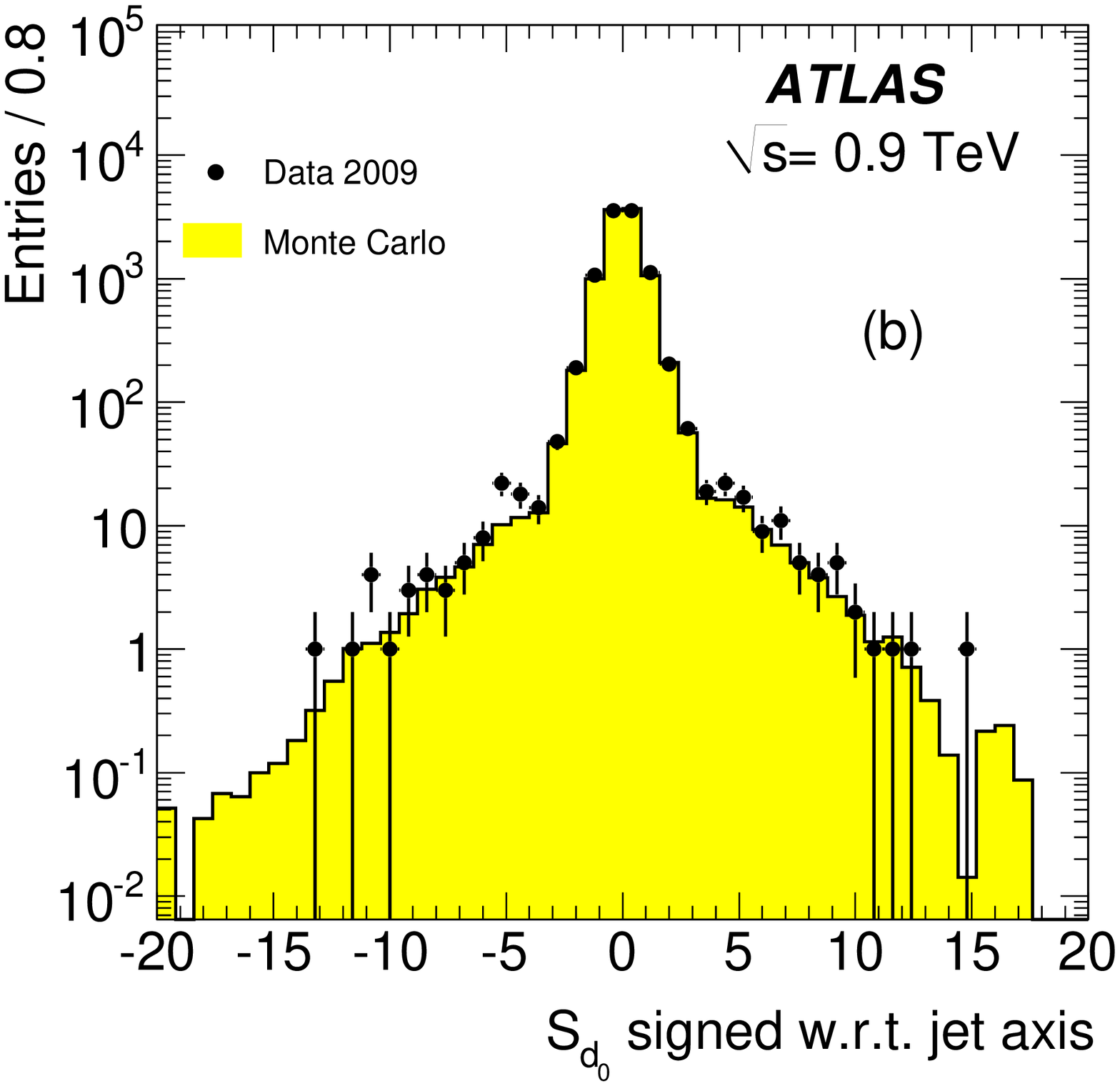}}
   \vspace*{-0.5cm}
  \end{center}
  \caption{(a) The variance of the $d_0$ distribution as a function of
    $1/(p^2 \sin^3\theta )$ of the tracks for 
   data (solid  points) compared to the nominal simulation
  (open points).  A straight line fit to the data points is also shown.
  (b) The lifetime-signed impact parameter  significance. 
 \vspace{-.2cm}
  \label{fig:ipsinth}}
\end{figure}

In order to study the effect of material on the $d_0$ resolution, 
Fig.~\ref{fig:ipsinth}(a) shows $\sigma^2(d_0)$ versus
$1/(p^2\sin^3\theta)$ for data and simulation using all 
selected charged particle tracks. 
The quantity $\sigma(d_0)$ is
determined by fitting the $d_0$ distribution in each bin of
$1/(p^2\sin^3\theta)$ with a Gaussian within $\pm 2 \sigma(d_0)$
about its mean.
The data lie approximately on a
straight line, 
as is expected if the scattering material is on a cylinder
and the match of the slope with the simulation implies a good
description of the material of the inner detector.
It should be noted  that the intercept on the $y$ axis has a
contribution from the 
primary vertex resolution.

The track selection for the $b$-tagging algorithms is designed to
select well-measured 
particles  and reject badly measured tracks, tracks from long-lived particles 
($\kzeros,\Lambda$ and other hyperon decays), and particles arising from
material interactions such as photon conversions or hadronic
interactions.

\begin{table}[htp]
\caption{Track selection criteria used for  the impact parameter and
secondary vertex tagging algorithms.
\label{tab:sv0_tracksel}}
\begin{center}
\begin{tabular}{llll}
\hline\hline
                         & Impact parameter & Standard vertex & Loose vertex  \\
\hline
\pt                      & $> 1$~\GeV           &  $> 0.5$~\GeV  & $> 0.5$~\GeV\ \\
$d_0$                    & $< 1$~mm         & $< 2$~mm     & $< 10$~mm \\
$z_0 \sin \theta$        & $<1.5$~mm        & $< 2$~mm     &  $< 50$~mm \\
\hline\hline
\end{tabular}
\end{center}
\end{table}

The track selection used by the impact parameter tagging algorithm is summarized in the
first column of Table~\ref{tab:sv0_tracksel}. Slightly different selections are used by
the secondary vertex algorithm (second column of Table~\ref{tab:sv0_tracksel}).

Calorimeter jets (see Section~\ref{jets}) are the reconstructed
objects the tagging algorithms are 
typically applied to.
Their direction is taken as estimator of the putative heavy flavour 
hadron direction. 
The impact parameter is then signed by whether the track perigee,
relative to the jet direction,
suggests a positive or negative flight distance.
The distribution of the lifetime-signed impact parameter significance
for tracks in 
jets is shown in Fig.~\ref{fig:ipsinth}(b).



Reconstructing explicitly the secondary decay vertices of heavy flavour
had\-rons adds substantial tagging information. 
There are expected to be few  b-quarks which can be tagged  in
this data set, so the algorithm was run with the standard as well as loose
settings, as described in the second and third columns of
Table~\ref{tab:sv0_tracksel}, respectively.
The loose setting selects vertices originating from \kshort\ as well
as from b-hadron decays whereas in the standard configuration
any pair of tracks consistent with a \kshort, $\Lambda$ or photon conversion
is explicitly removed.


%
Secondary vertices are reconstructed in an inclusive way starting from two-track
vertices which are merged into a common vertex. 
Tracks giving large $\chi^2$ contributions are then iteratively removed
until the reconstructed vertex fulfils certain quality criteria. 

\begin{figure}[htp]
  \begin{center}
    \includegraphics[width=0.6\textwidth,angle=0]{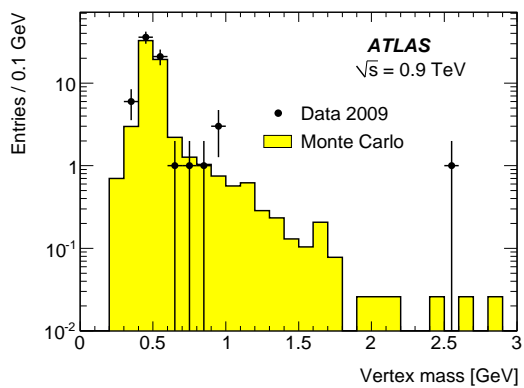}
   \vspace*{-0.5cm}
  \end{center}
  \caption{The vertex mass distribution for all secondary vertices
    with positive decay length selected in data.
The expectation from simulated events, normalized to the number of
jets in the data, is superimposed.
  \label{fig:vtx_mass}}
\end{figure}

The mass distribution of the resulting vertices for the loose configuration, assuming a pion mass
for each track, is shown in Fig.~\ref{fig:vtx_mass}.

Running the algorithm in the standard configuration 
results in the reconstruction of 9  secondary vertices
with positive decay length significance.  This is in good agreement
with the $8.9 \pm 0.5$(stat.) vertices expected from the same number
of jets,
10\,503,
in non-diffractive minimum-bias simulation.  The  
vertices reconstructed with the standard version of the tagging
algorithm are predominantly those with higher masses as the low-mass
region is dominated by \kshort\ mesons.

\begin{figure}[htp]
  \begin{center}
    \includegraphics[width=\textwidth,angle=0]{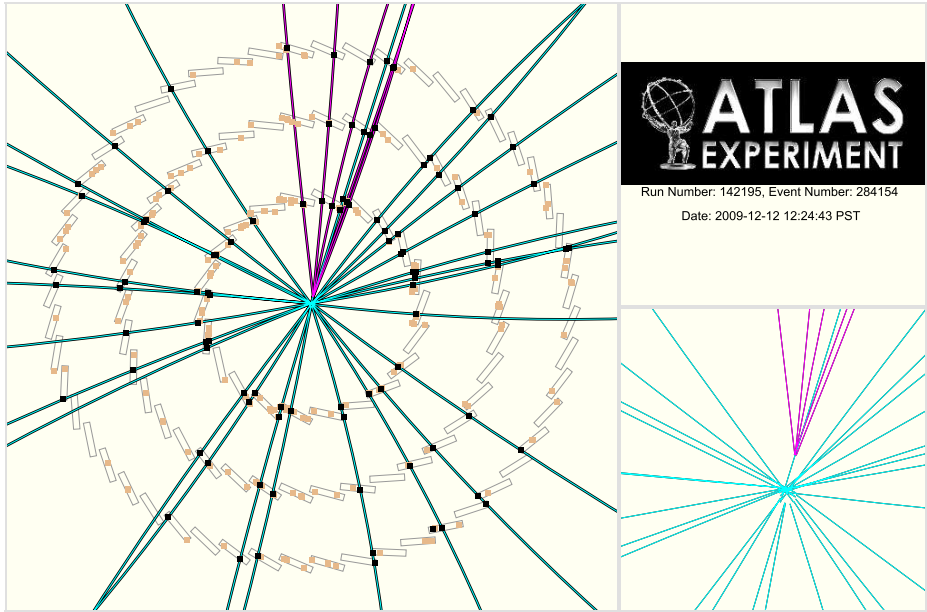}
  \end{center}
  \caption{An event containing a secondary vertex selected by the
    secondary vertex algorithm. The pixel detector can be seen on the
    left and an expansion of the vertex region on the right. 
    Unassociated hits, in a lighter colour,  are predominantly due to
    unreconstructed 
    particles such as those with  transverse momenta
    below 0.5~\GeV.
  \label{fig:b-cand}}
\end{figure}

An event display of the highest-mass candidate is shown
in Fig.~\ref{fig:b-cand}.
The secondary vertex consists of five tracks and
has a mass of 2.5~\GeV.
The vertex is
significantly displaced from the primary vertex, with a signed decay
length significance $L/\sigma(L) = 22$. From the vertex mass, 
momentum and $L$  a proper  lifetime of 3.1~ps is estimated.
The data was also tested by the impact-parameter based b-tagging
algorithm and this jet 
is  assigned a  probability below $10^{-4}$ for originating from a
light quark jet. 

\subsection{Particle Identification using Transition Radiation}
\label{sec:trttr}

\begin{figure}[htp]
  \begin{center}
    \includegraphics[width=0.8\textwidth,angle=0]{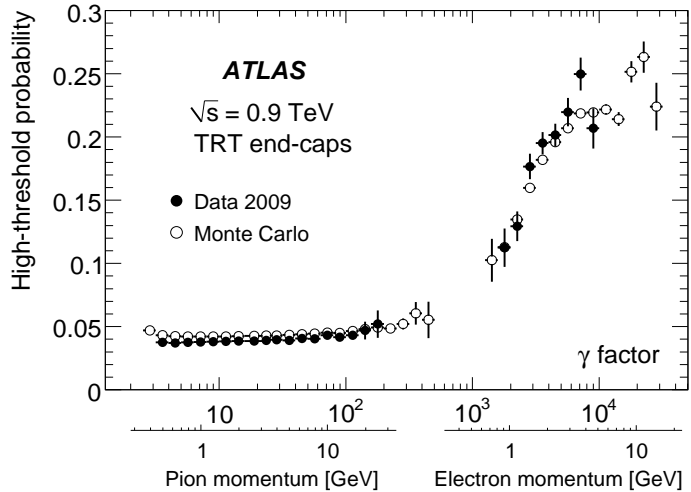}
   \vspace*{-1cm}
  \end{center}
  \caption{The fraction of high-threshold transition radiation hits on
    tracks as a 
    function of the relativistic $\gamma$ factor (see  text for details).
\label{fig:trt-tr}}
\end{figure}

The TRT  provides substantial discrimination
 between electrons and pions over the wide energy range between
$1$~and $200$~GeV by utilizing transition radiation in
foils and fibres. The readout discriminates at two thresholds, the
lower set to register minimum-ionising particles and the higher
intended for transition radiation (TR) photon interactions.
The fraction of high-threshold TR hits as a function
of the relativistic $\gamma$ factor is shown in Fig.~\ref{fig:trt-tr} for
particles in the forward region. This region is displayed
because there are more conversion candidates and they have higher
momenta than in the barrel.

The high-$\gamma$ part of the distribution is constructed using electrons from
photon conversions while the low-$\gamma$
component is made using   charged 
particle tracks with a hit in the B-layer and treating them as pions.
All tracks 
   are required to have at least 20 hits in the TRT.
The photon conversions are found similarly to  those in 
Section~\ref{sec:phot-conv} with at least one silicon hit, but the
transition radiation electron 
identification was  not applied to the  electron that was being
plotted. To ensure high purity 
   (about 98\%), the conversion candidates are also required to have a 
   vertex more than 40 mm away from the beam axis.
The pion sample excludes any photon conversion
candidate tracks. 

\subsection{Tracking Efficiency for Level-2  Trigger}

\begin{figure}[htp]
  \begin{center}
    \includegraphics[width=0.6\textwidth,angle=0]{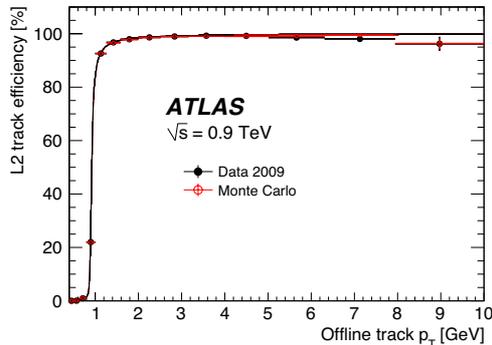}
   \vspace*{-1cm}
  \end{center}
  \caption{The efficiency for reconstruction  of a L2 
    track candidate
    as a function of the \pt\ of the matched offline track in data and
    Monte Carlo simulation. A fit of the threshold curve is superimposed.
\label{fig:l2track}}
\end{figure}

The L2 track trigger is one component of the HLT whose performance can
be  tested with current data.
 The trigger runs custom track reconstruction algorithms at L2,
 designed to produce fast and efficient tracking using all tracking subdetectors.
Tracking information forms an integral part of many
 ATLAS triggers including electron, muon and tau
 signatures~\cite{CSC}. These use 
 L1 information to specify a region of interest to examine. In
 the 2009 data there were few high-\pt\ objects, so the results here are
 taken from a  mode which searches for tracks across the
 entire tracking detector and is intended for B-physics and
 beam-position determination at L2.

Offline tracks with $|d_0| <$ 1.5 mm and $|z_0| <$ 200~mm are matched
to L2 tracks if they are within
$\Delta R = \sqrt{\Delta \eta^2 + \Delta \phi^2}  < 0.1$.
 The
efficiency is defined as the fraction of offline tracks which are 
matched and is shown in Fig.~\ref{fig:l2track} as a function of the track \pt.



\section{Electrons and Photons}
\label{e-gamma}

The electron and photon reconstruction and identification algorithms
used in ATLAS are designed to achieve  a large background
rejection and a high and uniform efficiency over the full acceptance
of the detector for transverse energies above~20~\GeV. Using these
algorithms on the 0.9~\TeV\ data, a significant number of
low-\pt\ electron  and photon candidates were reconstructed.
The measurements  provide a quantitative test
of both the algorithms themselves and the reliability of the
performance predictions in the transverse energy range from the
reconstruction threshold of~2.5~\GeV\ to about~10~\GeV.




The electromagnetic calorimeter (EM) consists of the
barrel  (EMB) and two end-caps (EMEC). The barrel covers the pseudorapidity range $|\eta| < 1.475$; the end-cap calorimeters cover $1.375 < |\eta | < 3.2$.
In the forward direction energy measurements for both
electromagnetic and hadronic showers are provided by the Forward Calorimeter (FCal) in the
range $3.1 < |\eta| < 4.9$. The hadronic calorimetry in the range
$|\eta| < 1.7$ is provided by the scintillator-tile calorimeter
(Tile). For $1.5 < |\eta| < 3.2$ hadronic showers are measured by the
hadronic end-caps (HEC), which use LAr with a copper absorber.

The e/$\gamma$ algorithms make use of the fine segmentation of the
EM calorimeter in both the lateral
and longitudinal directions of the showers~\cite{DetectorPaper}. At
high energy, most of the EM~shower energy is collected in the second
layer which has a lateral granularity of~$0.025\times 0.025$ in
$\eta\times\phi$~space. The first layer consists of finer-grained
strips in the $\eta$-direction (with a coarser granularity
in~$\phi$), which improves $\gamma$-$\pi^0$~discrimination. A third
layer measures the tails of very highly energetic EM showers and
helps in rejecting hadron showers. In the range~$|\eta|<1.8$ these
three layers are complemented by a presampler layer placed in front
with coarse granularity to correct for energy lost in the material
before the calorimeter.

The algorithms also make use of the precise track reconstruction
provided by the inner detector.  The TRT also provides substantial
discriminating power between electrons and pions over a wide energy
range (between 1~and~200~\GeV). The Pixel  B-layer provides
precision
vertexing and significant rejection
of photon conversions through the requirement of a track with a hit in
this layer.

\subsection{Electron and Photon Reconstruction}
\label{egamma-reconstruction}
The basic algorithms for electron and photon reconstruction are
described in detail in Ref. \cite{CSC}.
The first stage of the search for EM objects is to look for significant
deposits in the EM calorimeter cells inside a sliding window as it
is moved across the detector.
The size of the sliding window cluster depends on
the type of candidate (electron, unconverted or converted photon) and
the location (barrel, end-caps).  The cluster energy is calculated
from the amplitudes observed in the cells of the three longitudinal
layers of the EM~calorimeter and of the presampler (where
present). The calculation sums the weighted energies in these
compartments, then takes into account several corrections for shower
depth, lateral and longitudinal leakage, local modulation etc. The
weights and correction coefficients were parameterized from
beam-tests~\cite{DetectorPaper} and simulation. 

 Electrons are reconstructed from the clusters if there is a suitable
 match with a particle track of \pt$>0.5$~\GeV. The best track is the one  with
 an extrapolation closest in ($\eta,\phi$) to the cluster barycentre
 (the energy-weighted mean position)
 in the middle EM~calorimeter layer. Similarly, photons are
 reconstructed  from the clusters if there is no reconstructed track
 matched to the cluster (unconverted photon candidates) or if there is
 a reconstructed conversion vertex matched to the cluster (converted
 photon candidates). ``Single track conversions'' (identified via tracks
 lacking a hit in the B-layer) are also taken into account.
First, electron
candidates with a cluster $|\eta|<2.47$ and photons with cluster
$|\eta|<2.37$ are selected and investigated (the cluster~$\eta$ is
defined here as the barycentre of the cluster cells in the middle
layer of the EM~calorimeter). Electron and photon candidates in the EM calorimeter transition region $1.37< |\eta|< 1.52$ are not considered.
At this stage,
 879 electron and 1\,694 photon candidates are reconstructed in the
 data with $E_T$  above 2.5~\GeV.

\subsection{Electron and Photon Identification}
\label{Identification}

The isolated electron and photon identification
algorithms rely on selections based on variables which provide good
separation between  electrons/photons and fake signatures from
hadronic jets. These variables include information from the calorimeter
and, in the case of electrons, tracker and combined
calorimeter/tracker information. There are three classes of electrons
defined: loose, medium and tight, and two for
photons: loose and tight. The selection criteria  were optimized in bins of
$E_T$ and $\eta$, separately for electrons, unconverted and converted
photons.

The loose selection criteria are based on the shower shape and are common to
electrons and photons. For electrons, the medium requirements make use
of the track 
information while in the tight ones the particle track selections are
more stringent 
and use the particle identification capability of the TRT. For photons
the tight selection criteria make full use
of the EM calorimeter strip layer information, mainly to reject merged
photon pairs 
from high energy $\pi^0$'s. 

In the following  all reconstructed electron and photon candidates
with cluster $E_T>2.5$~\GeV\ at the sliding window level are considered.

\subsection{Electron Candidates}

\label{electroncandidatesection}

\begin{figure}[htp]
{\includegraphics[width=0.49\textwidth]{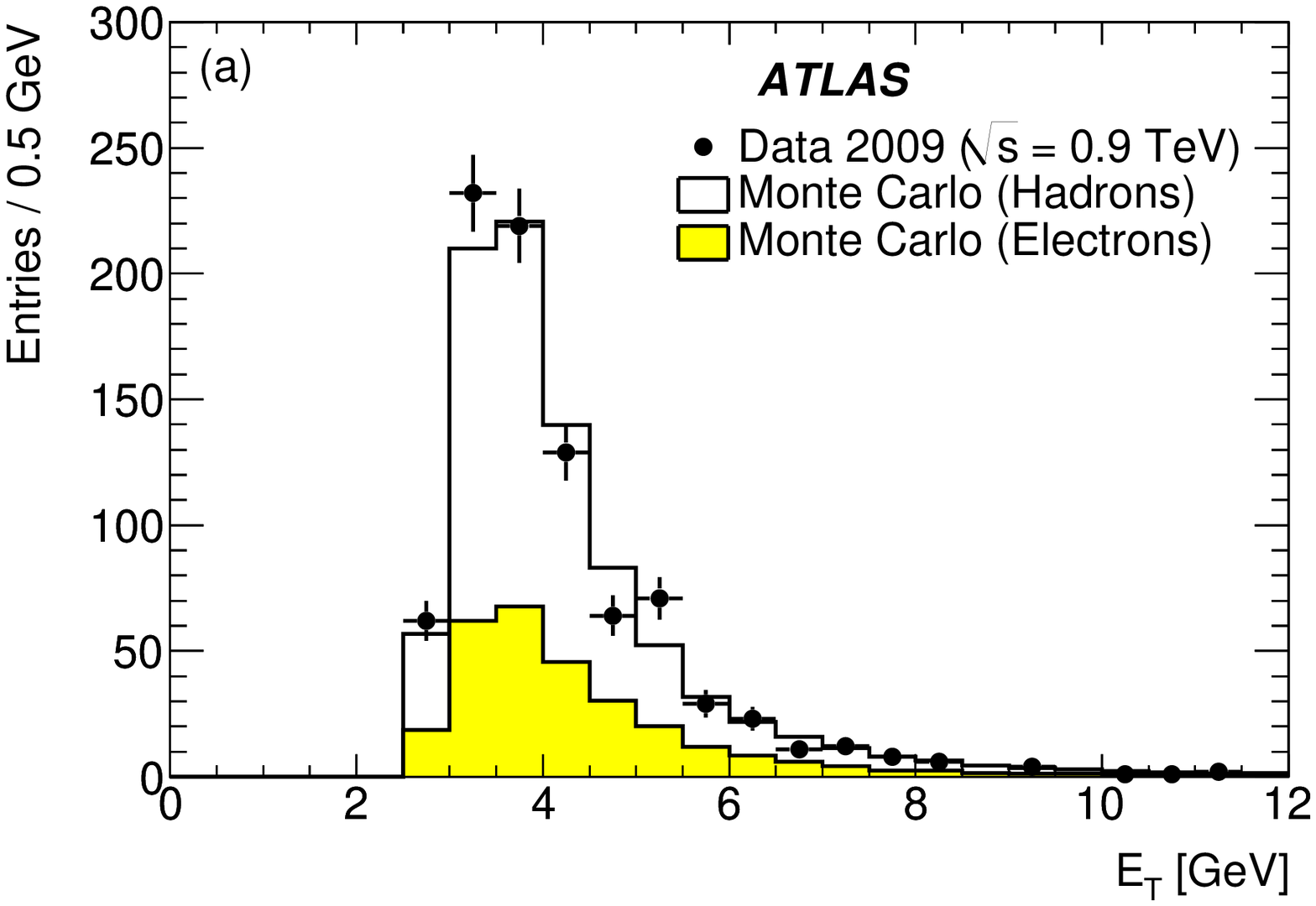}}
{\includegraphics[width=0.49\textwidth]{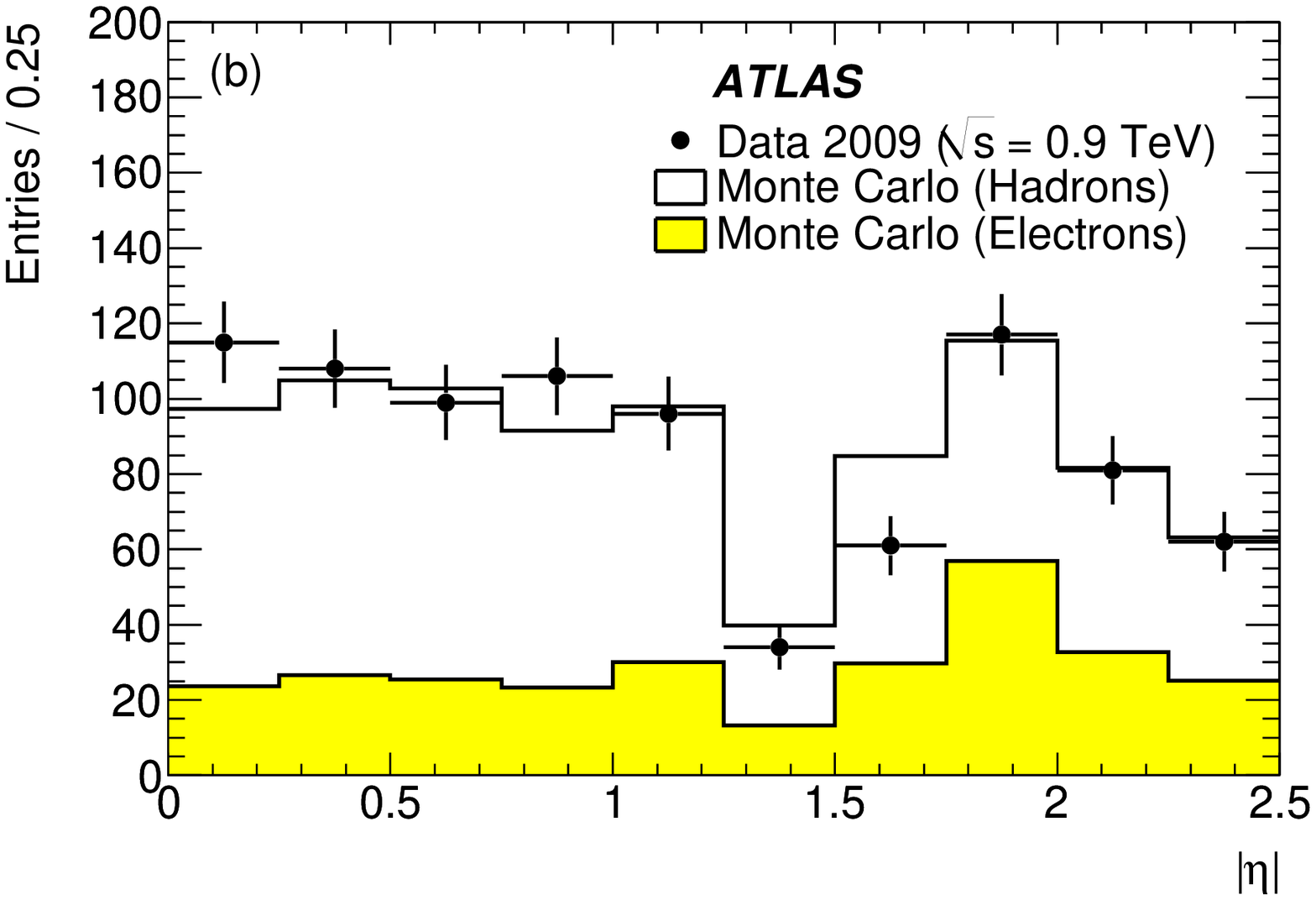}}
\caption{Distribution of cluster $E_T$~(a) and $|\eta|$~(b)
  for all selected electron candidates. The simulation is normalized to
  the number of data events.
}
\label{fig:ElectronKinematics}
\end{figure}

 Figure~\ref{fig:ElectronKinematics} displays, for all of the 879
 electron candidates from 384\,186 events,
  the transverse energy and pseudorapidity
 spectra. Table~\ref{ElectronCandidateNumbers} presents the percentage
 of these candidates which pass the successive selection criteria
 both for data and simulation. These criteria were not optimized for such low-energy electron candidates (see Section~\ref{Identification}).
 Both Fig.~\ref{fig:ElectronKinematics}
 and Table~\ref{ElectronCandidateNumbers} show similar behaviour in data
 and simulation. The remaining discrepancies in the first stages of
 background rejection may be related to the small differences observed
 in shower variables (see Section~\ref{sec:calovariables} below).  
In Fig.~\ref{fig:ElectronKinematics}(b) the drop in efficiency around $|\eta|=1.5$ corresponds to the barrel/end-cap transition.

\begin{table}[htp]
\caption{The fraction of electron and photon candidates passing the different
 selection criteria, compared to those predicted by Monte
 Carlo~(MC). Statistical error are quoted. 
}
\label{ElectronCandidateNumbers}
\begin{center}
\begin{tabular}{lcccc}
\hline\hline
        & \multicolumn{2}{c}{Electron candidates} &
   \multicolumn{2}{c}{Photon candidates} \\
\cline{2-5}
        & Data ($\%$) & MC ($\%$) & Data ($\%$) & MC ($\%$) \\
\hline
Loose   & 46.5$\pm$1.7  & 50.9$\pm$0.2  &  25.4$\pm$1.0 & 30.5$\pm$0.1 \\
Medium  & 10.6$\pm$1.0 & 13.1$\pm0.2$  & n.a.    & n.a.  \\
Tight & 2.3$\pm$0.5  & 2.4$\pm$0.1 &  4.1$\pm$0.5 &  6.6$\pm$0.1 \\
\hline\hline
\end{tabular}
\end{center}
\end{table}

 In these figures the Monte Carlo prediction is sub-divided into its two
 main components: hadrons and real electrons. The latter is largely
 dominated by electrons from photon conversions, but also includes a
 small fraction ($\sim3\%$) of electrons from other sources, such as
 Dalitz decays, and an even smaller one~(below~1\%) of electrons from
 $\mathrm{b},\mathrm{c} \to \mathrm{e}$~decays.
There are twenty electron candidates passing the tight selections in the
data. Approximately~15\% of such candidates in the Monte Carlo  are
from heavy flavour decays.

\subsection{Photon Candidates}
\label{photoncand}

\begin{figure}[htp]
{\label{Kinematics_photonEt}\includegraphics[width=0.49\textwidth]{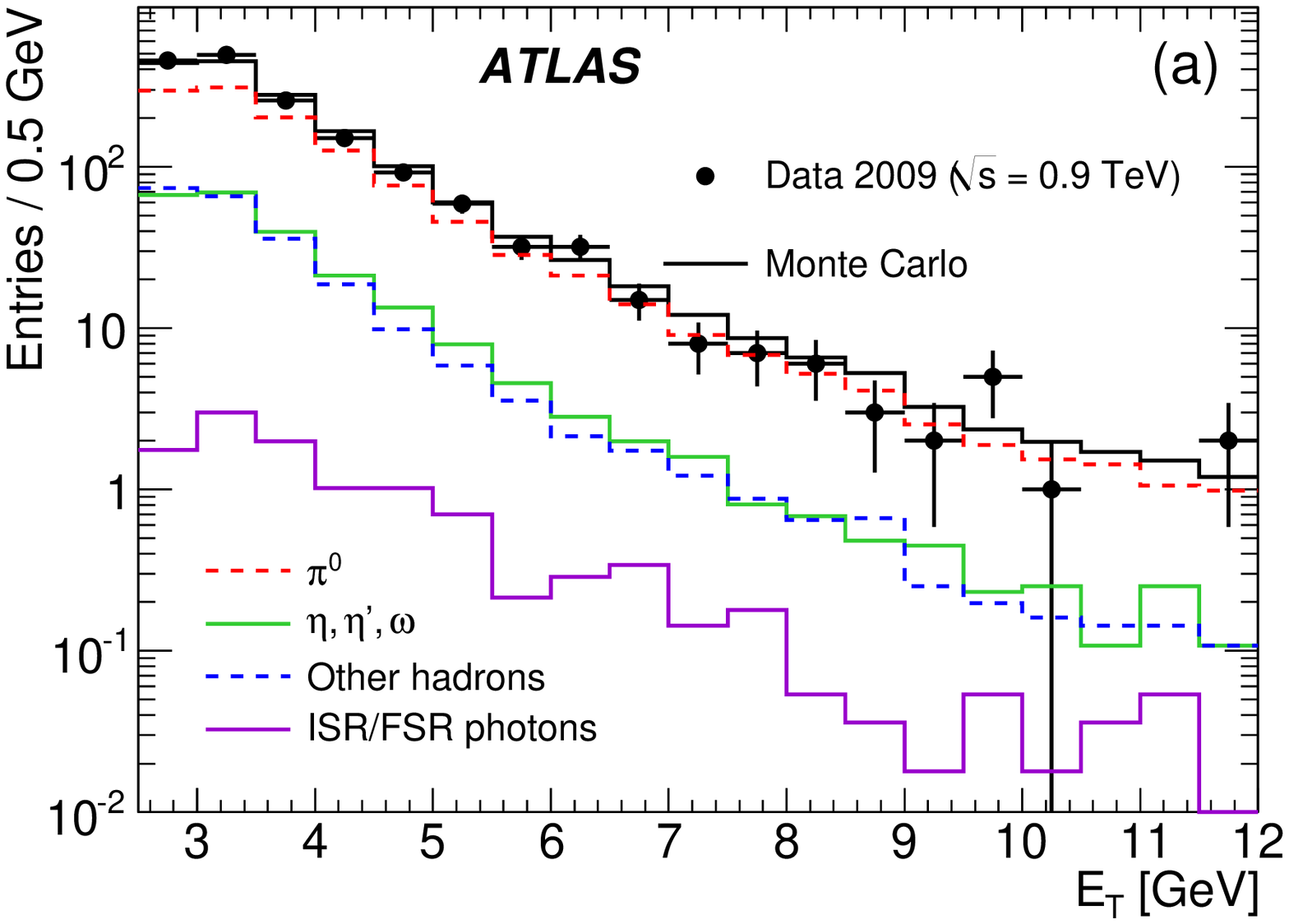}}
{\label{Kinematics_photonEta}\includegraphics[width=0.49\textwidth]{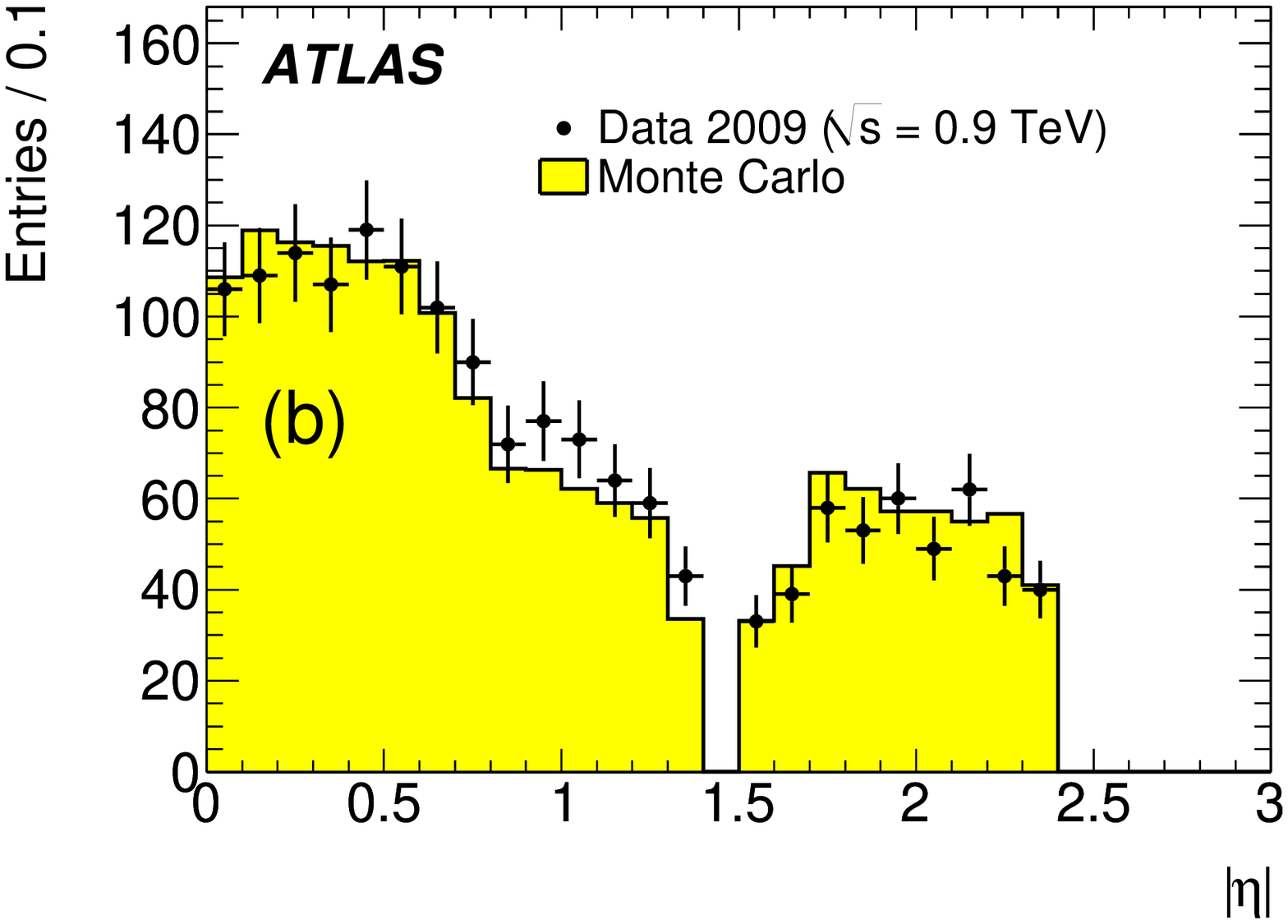}}
\caption{Cluster $E_T$~(a) and $|\eta|$~(b) for all selected photon
  candidates. The simulation is normalized to
  the number of data events.
}
\label{fig:PhotonKinematics}
\end{figure}

Transverse energy and pseudorapidity spectra for all 1\,694  photon
candidates are displayed in
Fig.~\ref{fig:PhotonKinematics}. Table~\ref{ElectronCandidateNumbers}
presents the percentage of photon candidates, as a function of the
selection level applied. 
Of the selected candidates,  14\% 
are reconstructed as converted photons and almost all of
these,~$\sim$~98\%, are also selected as electron
candidates.

The Monte Carlo prediction is sub-divided in this case into four
components of decreasing importance: approximately~71\% of the
candidates correspond to photons from $\pi^0$~decay,
whereas~$\sim$~14\% are from $\eta$, $\eta^{'}$ or $\omega$ decays into
 photons;~$\sim$~14\% are from other hadrons with complex decay
processes and particles interacting in the tracker material. At these energies,
only a very small fraction, $\sim$~1\%, of all photon candidates are
expected to be primary products of the hard scattering.

\subsection{First-level Electron and Photon Trigger Performance}

 The L1 $e/\gamma$ selection algorithm
searches for narrow, high-$E_T$ electromagnetic showers and does not
 separate electrons from photons.  The
primitives for this algorithm are towers which sum the transverse
energies of all electromagnetic calorimeter cells in
$\Delta\eta\times\Delta\phi = 0.1 \times 0.1$. The trigger
examines adjacent pairs of towers 
and tests their total energy against several trigger thresholds.
Isolation requirements  were not yet employed. The lowest threshold $e/\gamma$
trigger for the 2009 data-taking period required a transverse energy of at least
$4$~\GeV.

Clusters consistent with originating
 from an electron or photon are selected by requiring at least 30\% of the
 cluster energy to be deposited in the second layer of the
 electromagnetic calorimeter, where the maximum of an electromagnetic
 shower is expected. The selected $e/\gamma$ candidates are matched to
 L1 clusters in $\eta$ and $\phi$ by requiring $\Delta
 R < 0.15$. The
 efficiency is then calculated from the fraction of
 reconstructed clusters which have a matching L1 cluster.

\begin{figure}[htp]
\begin{center}
\includegraphics[width=0.6\textwidth]{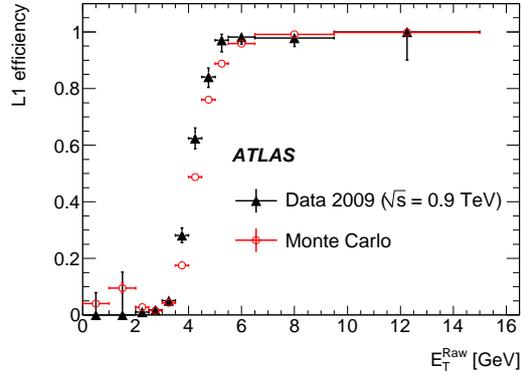}
\caption{Efficiency for the lowest threshold L1 electromagnetic
  trigger, a nominal 4~\GeV, as a
function of the uncalibrated offline cluster transverse energy.
The turn-on is shown for
data (solid triangles) and non-diffractive minimum-bias simulation
(open circles). 
}
\label{fig:EM-LVL1}
\end{center}
\end{figure}

The resulting L1 trigger efficiency for the lowest threshold
component
is shown in Fig.~\ref{fig:EM-LVL1}. 
The sharpness of the efficiency turn-on curve around threshold agrees
with the Monte Carlo expectation. 
Low energy reconstructed  clusters occasionally fire the trigger, especially
when the coarser granularity used at L1 merges two
separate offline clusters.

\subsection{Electron and Photon Identification Variables}

\subsubsection{Calorimeter Variables}
\label{sec:calovariables}

\begin{figure}[htp]
\begin{center}
{\includegraphics[width=0.49\textwidth]{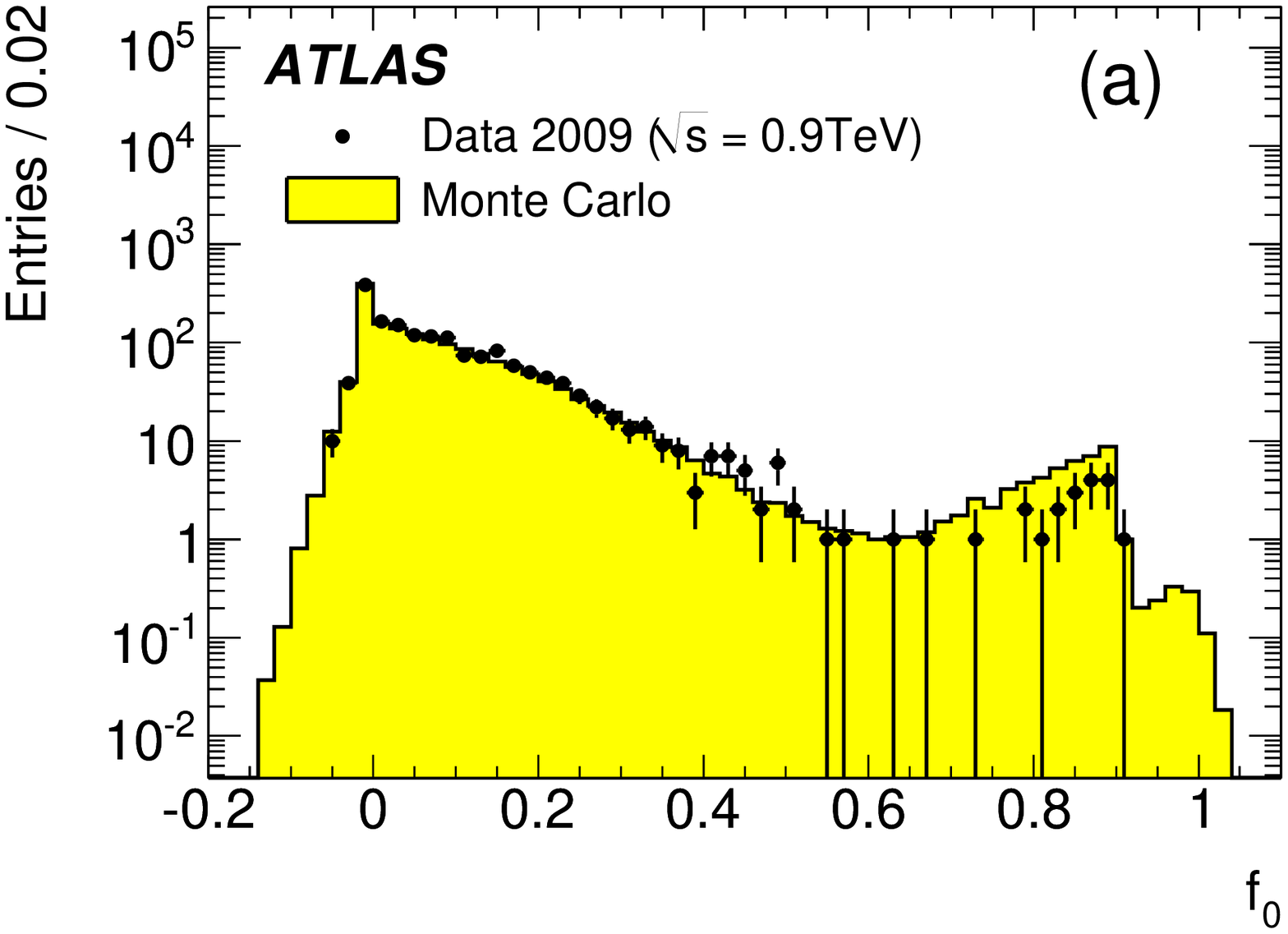}}
{\includegraphics[width=0.49\textwidth]{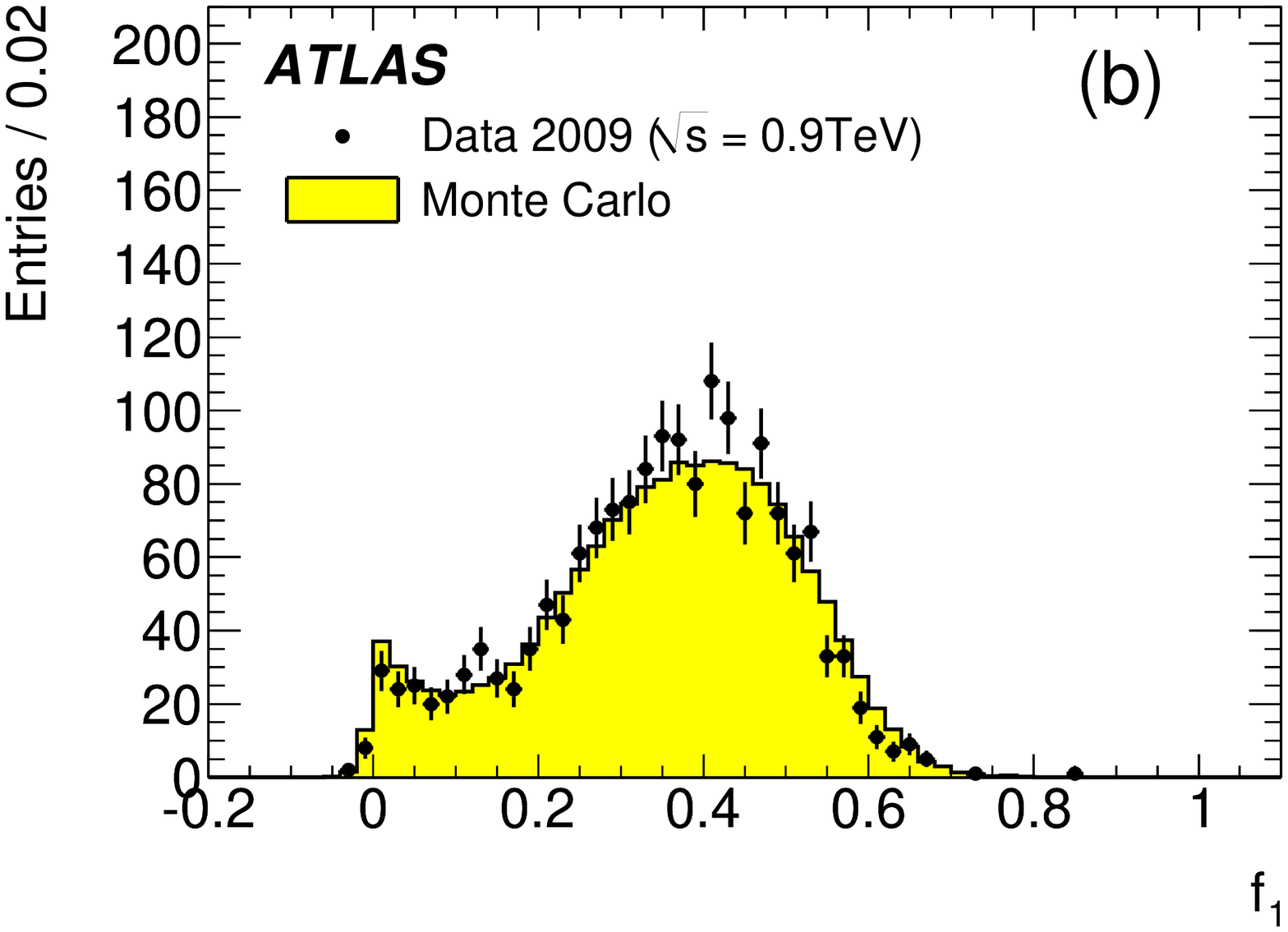}}
{\includegraphics[width=0.49\textwidth]{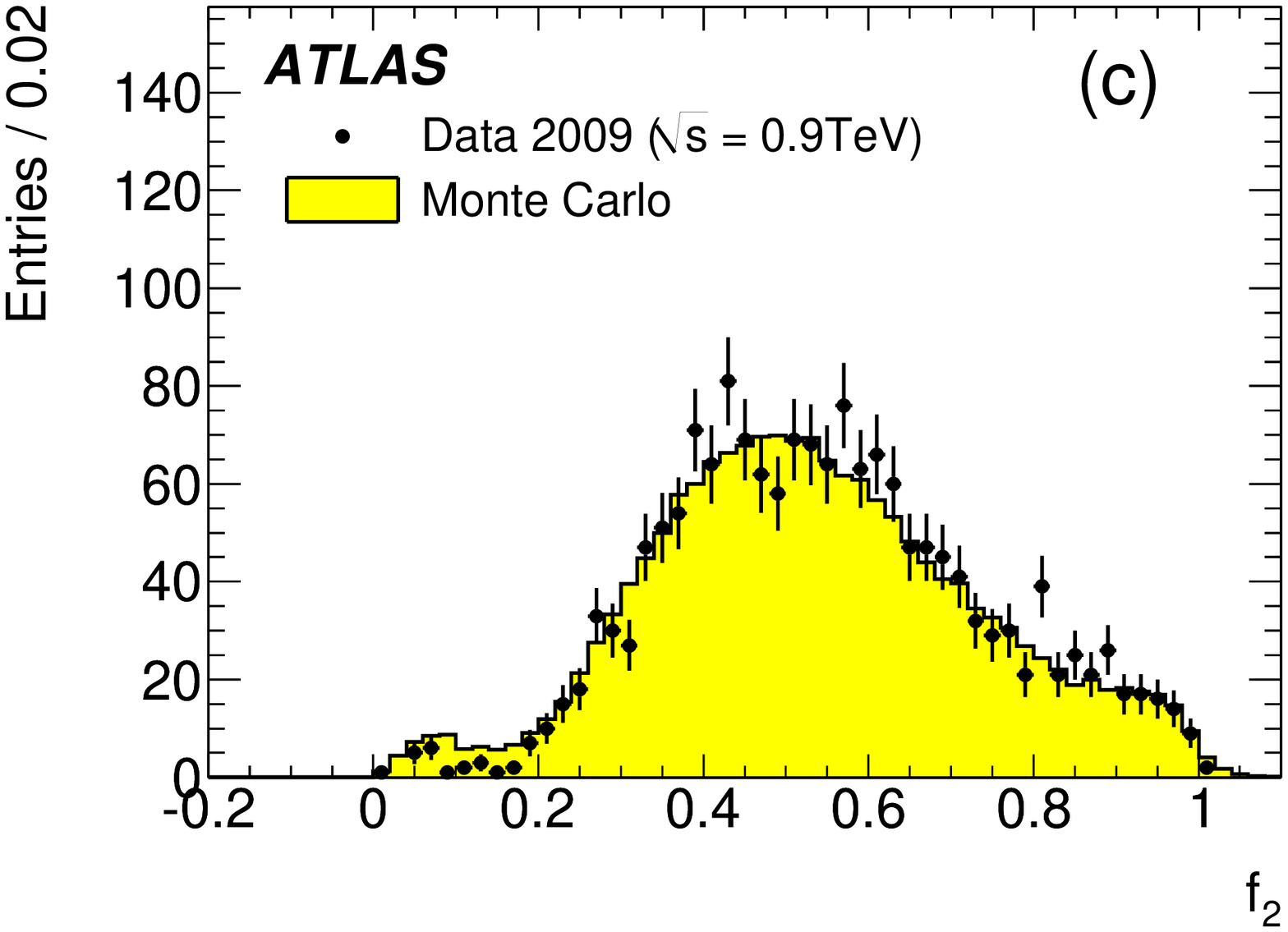}}
{\includegraphics[width=0.49\textwidth]{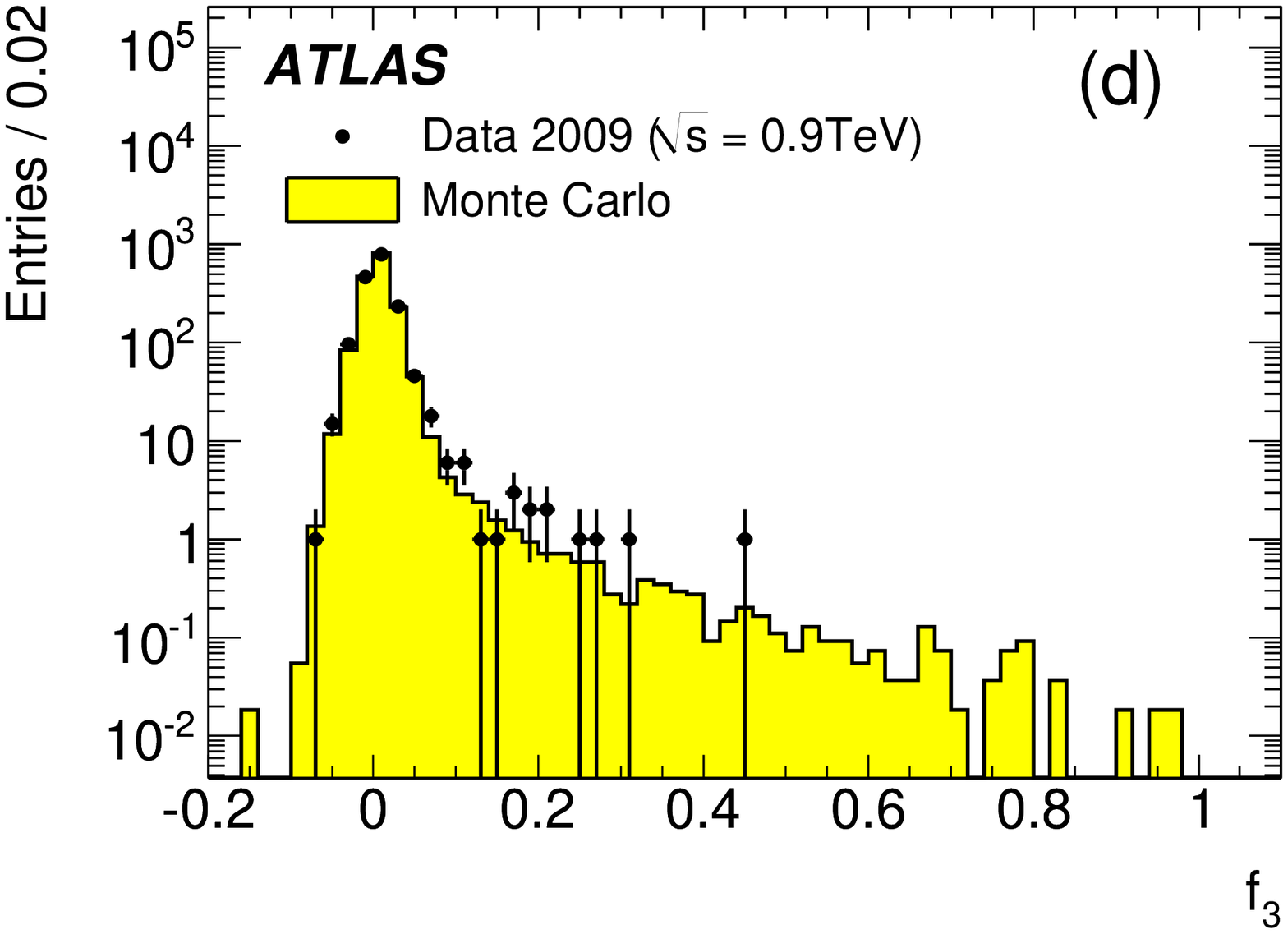}}
\caption{Fraction of energy deposited by photon candidates with $E_T >$ 2.5~\GeV\
  in each
  layer of the electromagnetic calorimeter for data and
  simulation. These fractions are labelled as (a) $f_0$~for the presampler
  layer, (b) $f_1$~for the strip layer, (c) $f_2$~for the middle
  layer and $(d)~f_3$~for the back layer.
Fractions can be negative due to   noise  fluctuations.
The simulation is normalized to the number of data events.
}
\label{fig:fractionSampling}
\end{center}
\end{figure}

In this section, various calorimeter-based quantities are displayed for the
photon candidates. These are preferred to the similar electron
distributions because of the higher purity.

 Figure~\ref{fig:fractionSampling} illustrates the
longitudinal development of the shower in the successive layers of the
EM~calorimeter, based on the measured layer energies before
corrections are applied. For the observed photon candidates,
which in simulation are predominantly  from
$\pi^0$~decays, the energy is deposited in earlier  calorimeter layers
than typical for high energy photons.
In the presampler part, the simulation points are higher than the
data for fractions above 0.6. This is at least in part because the presampler
simulation does not 
describe the recombination of electron-positron pairs
by highly ionizing hadrons or nuclear fragments that lower
the LAr response, 
an effect which is included in the
accordion calorimeter simulation.
This feature also
explains the observed disagreement in the first
bins for the fractions in the other layers, since the various
fractions are correlated.

\begin{figure}[htp]
\begin{center}
\includegraphics[width=0.49\textwidth]{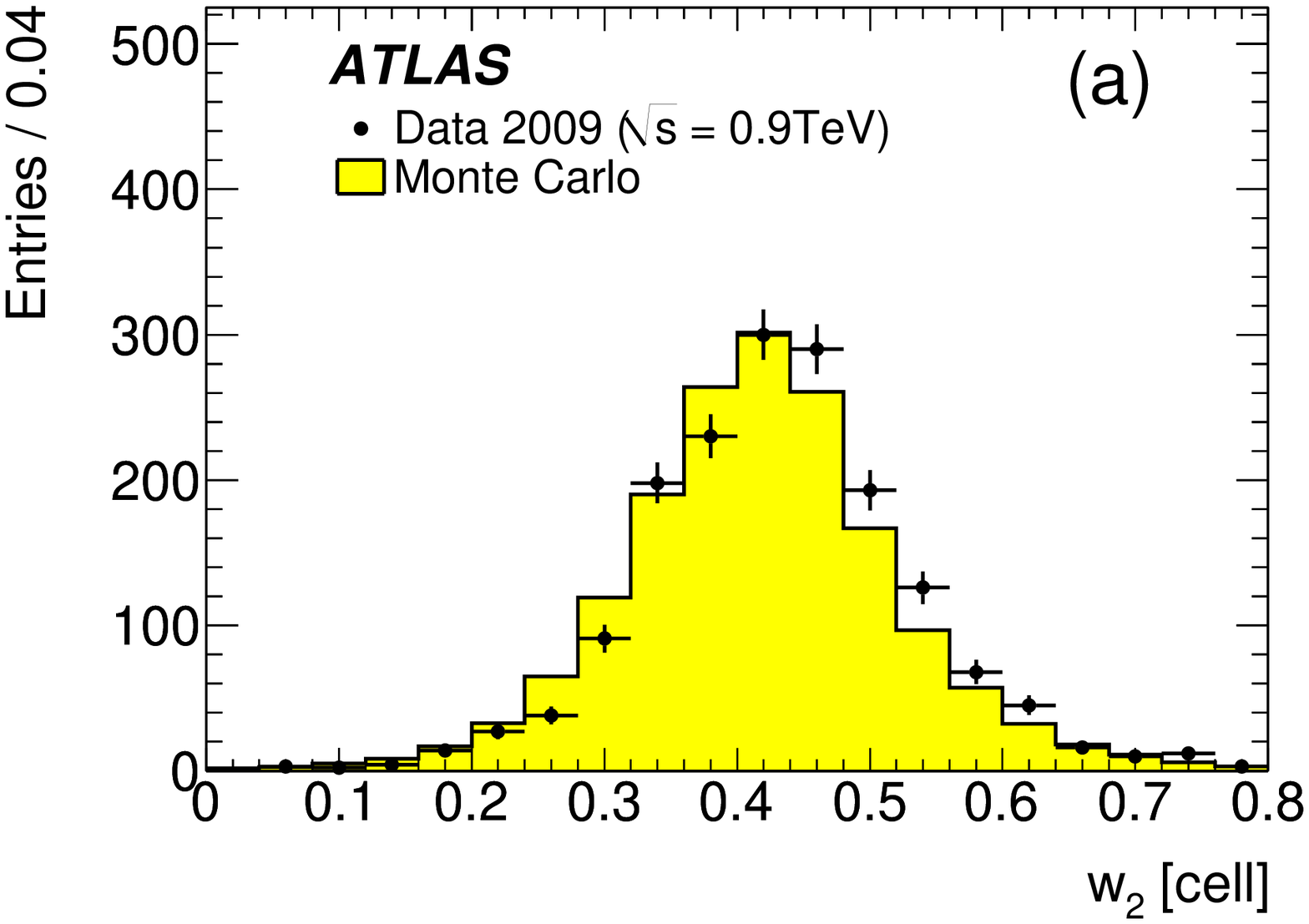}
\includegraphics[width=0.49\textwidth]{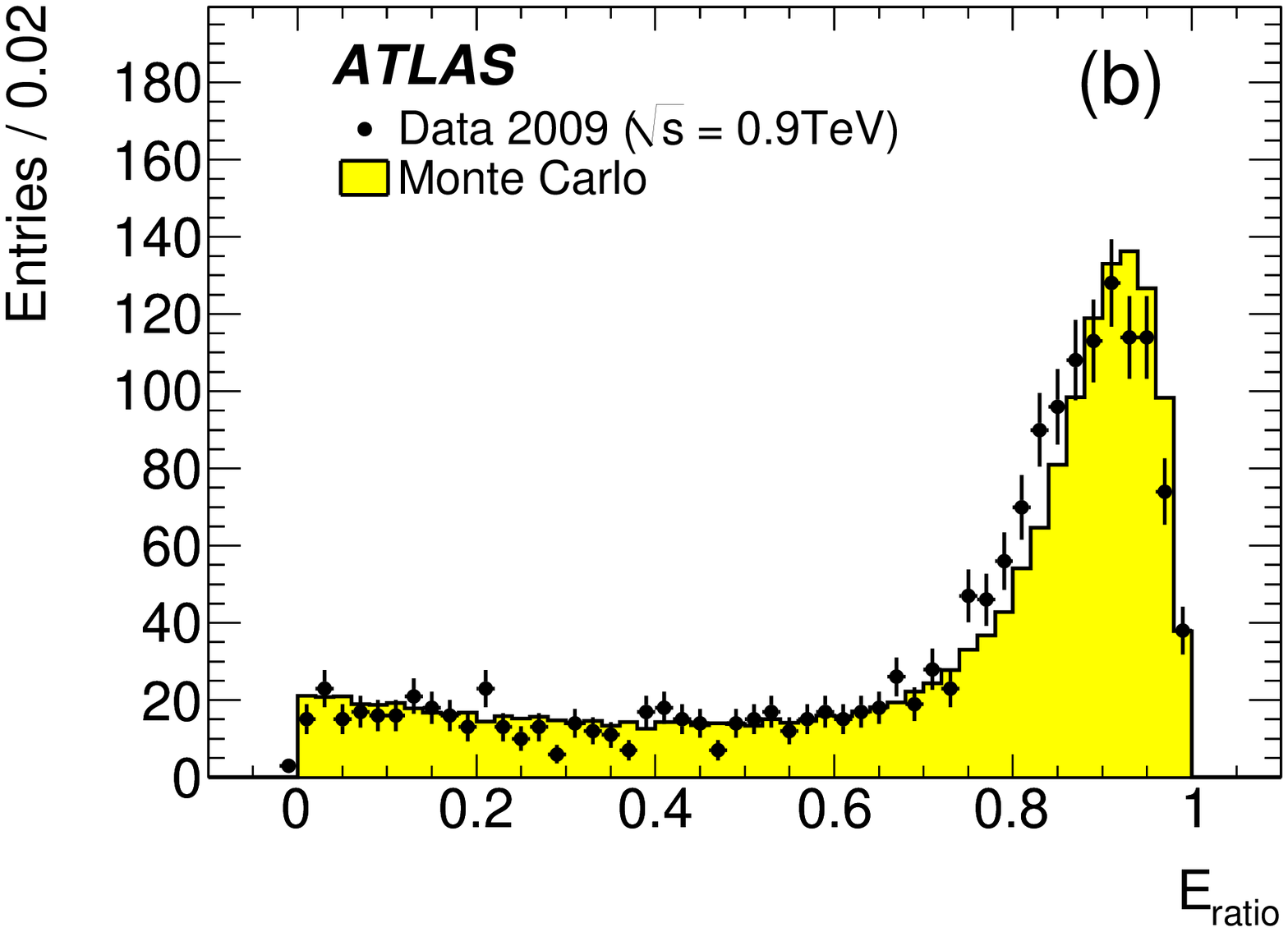}
\includegraphics[width=0.49\textwidth]{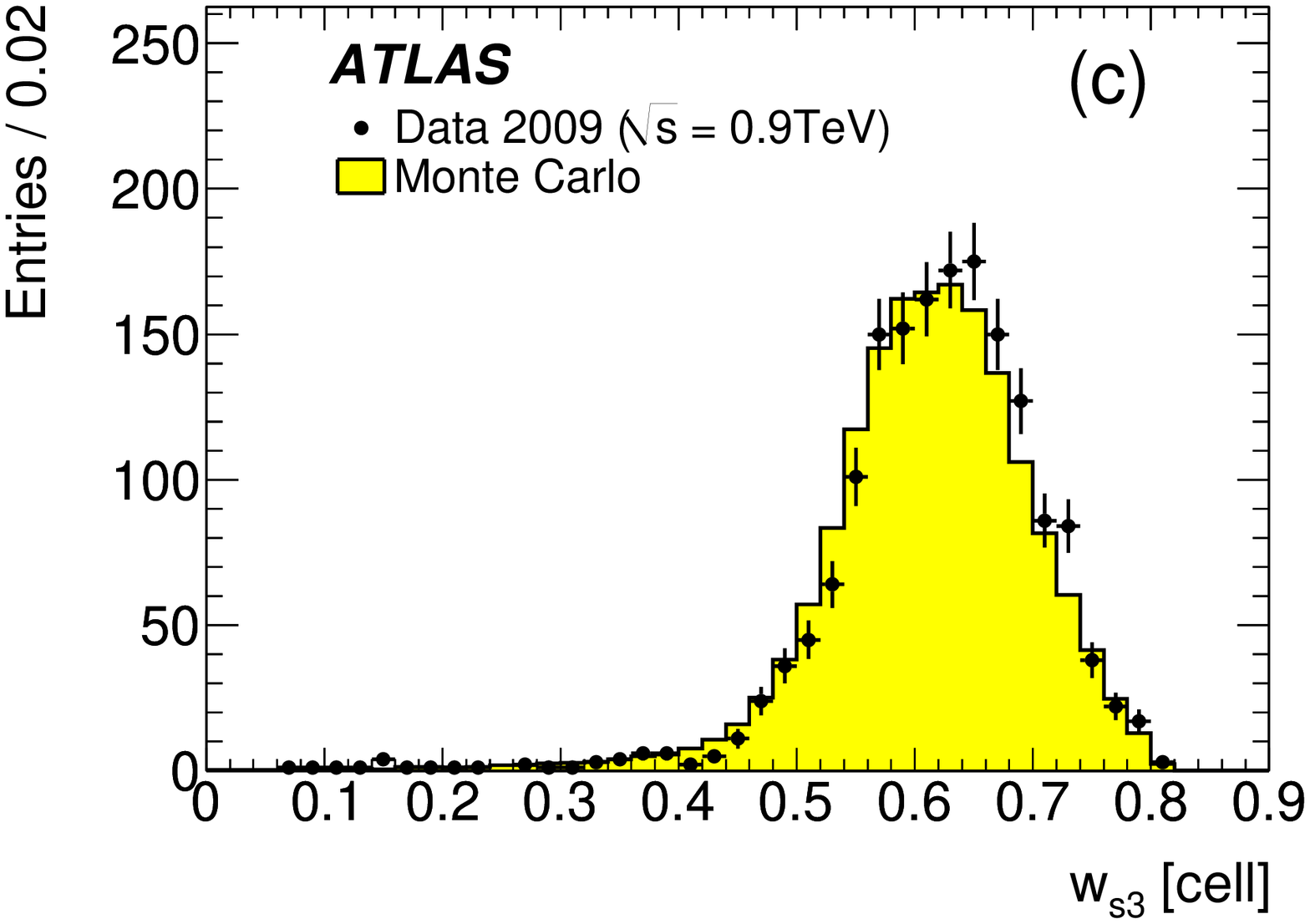}
\caption{\label{fig:ShowerShapes} Distributions of calorimeter variables compared between data
  and simulation for all photon candidates with \pt\ above 2.5~\GeV.
 Shown are
the  shower width in the middle layer of the EM~calorimeter, $w_2$~(a),
  and the variables $E_{\rm ratio}$~(b) and $w_{s3}$~(c), which
  characterize the shower shape in the first (strips) EM~layer.
The simulation is normalized to  the number of data events.
}
\end{center}
\end{figure}

Several variables are used to quantify the lateral development of the
shower. From these, the distribution of $w_2$, the shower width measured in the second
layer of the EM calorimeter,  is shown in~Fig.~\ref{fig:ShowerShapes}(a).
The shower width $w_2$ is slightly larger in the data. Preliminary
studies show that including the cross-talk between neighbouring middle
layer cells ($\sim0.5\%$)~\cite{CTBpapers_2} in the simulation 
explains part of the observed difference.

The distribution of two
 variables used for  $\pi^0/\gamma$ separation in the tight photon
 selection, $E_{\rm ratio}$ and $w_{s3}$,
 are shown in Figs.~\ref{fig:ShowerShapes}(b) and \ref{fig:ShowerShapes}(c).
$E_{\rm ratio}$ is the difference of  the highest and second highest
 strip energies, divided by 
their sum. $w_{s3}$ is the shower width measured in three strips around
the maximum energy strip. For this variable the data show a
slightly wider profile than the simulation, although in this case
 the simulation
already includes the measured cross-talk. In general, all the
shower shape variables show  good agreement between data and
simulation.

\subsubsection{Tracking and Track-Cluster Matching Variables}
\label{sec:matching}

\begin{figure}[htp]
\begin{center}
\includegraphics[width=0.49\textwidth]{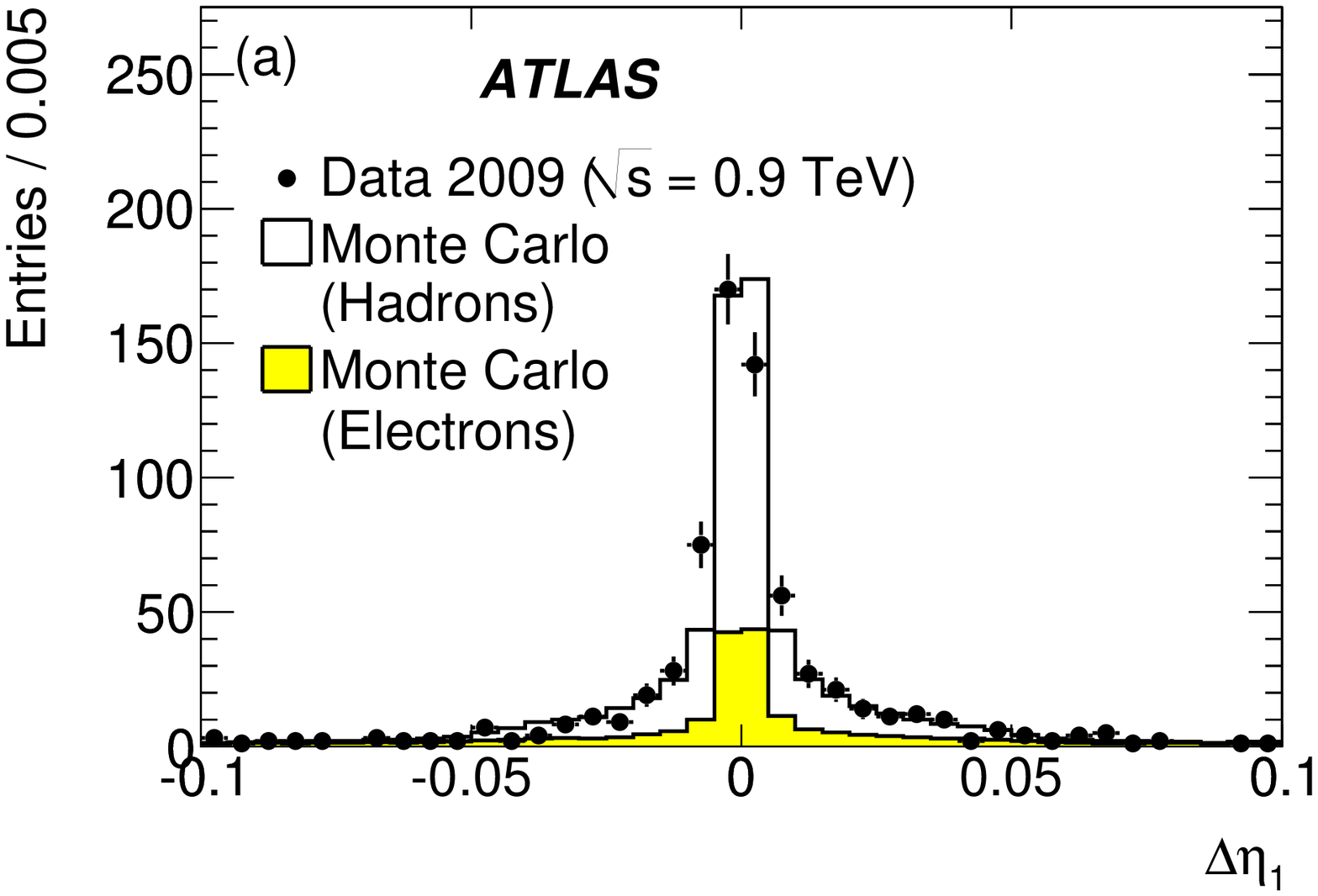}
\includegraphics[width=0.49\textwidth]{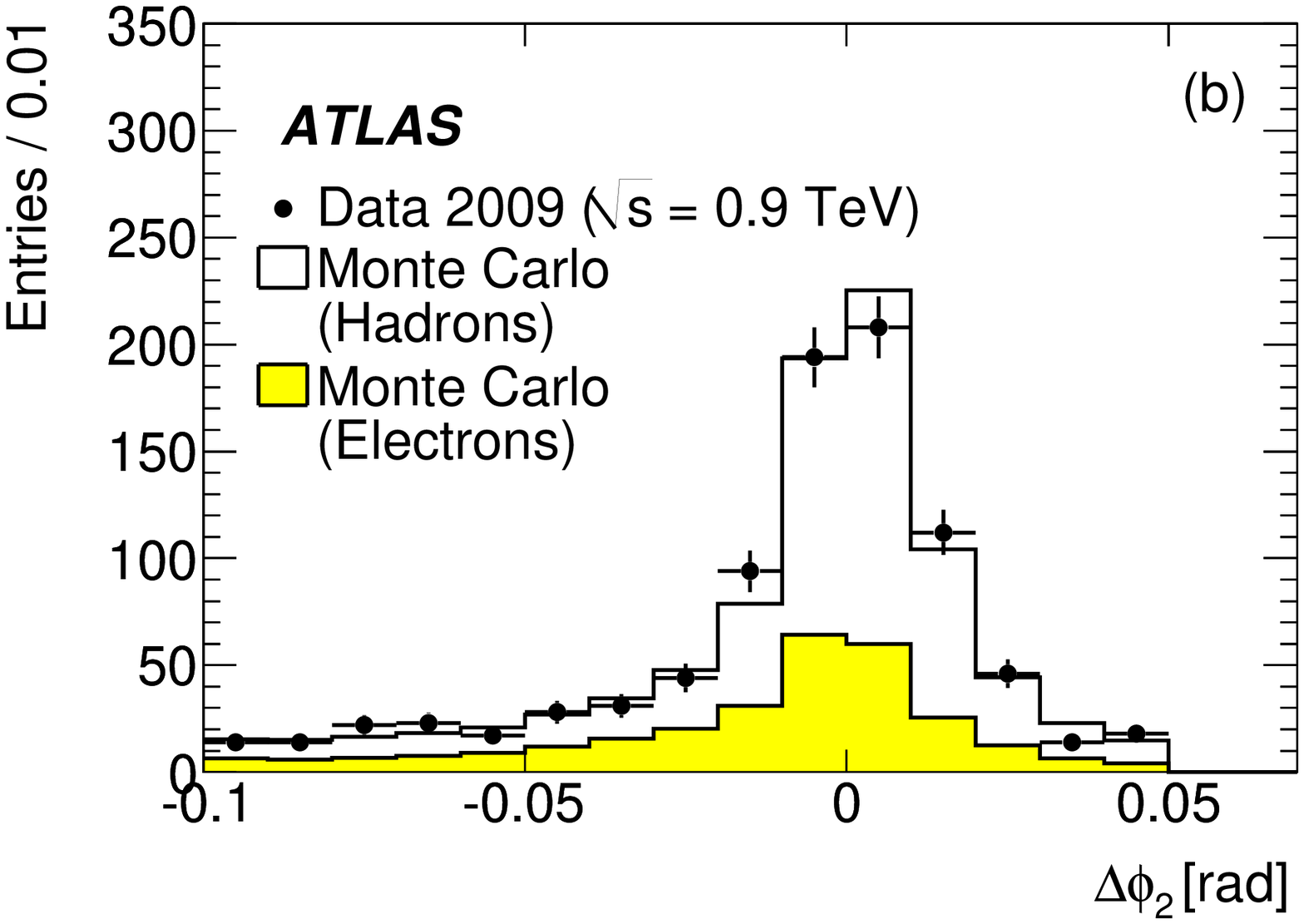}
\caption{\label{fig:TrackMatching}Distributions of track-calorimeter
  matching variables for all electron candidates compared 
  between data and simulation.
  (a) shows the difference in  $\eta$ in the first calorimeter layer (see
  Section~\ref{sec:matching})
 and (b) shows the match in charge-signed-$\phi$ in the second.
The simulation is normalized to the number of data events.
}
\end{center}
\end{figure}

Electron and converted photon  identification rely
heavily on tracking performance. Figure~\ref{fig:TrackMatching}
illustrates two of the track-calorimeter matching variables used in
the identification of electron candidates  in data
and simulation. For simulation, hadrons and real electrons are shown
separately. Fig.~\ref{fig:TrackMatching}(a) shows the difference 
in~$\eta$,~$\Delta\eta_1$, between the track extrapolated to the strip
layer of the EM~calorimeter and the barycentre of the cell energies in
this layer. Figure~\ref{fig:TrackMatching}(b) shows the difference in
azimuth,~$\Delta\phi_2$, between the track extrapolated to the middle
layer of the EM~calorimeter and the barycentre of the cell energies in
this layer. This variable is signed by the charge of the particle to account
for the position of any radiated photons with respect to the track
curvature, and an asymmetric cut is applied in the selection. 
 The asymmetric tails at
large negative values of~$\Delta\phi_2$ are more pronounced for the
electrons than for the hadrons.

\begin{figure}[htp]
\begin{center}
\includegraphics[width=0.49\textwidth]{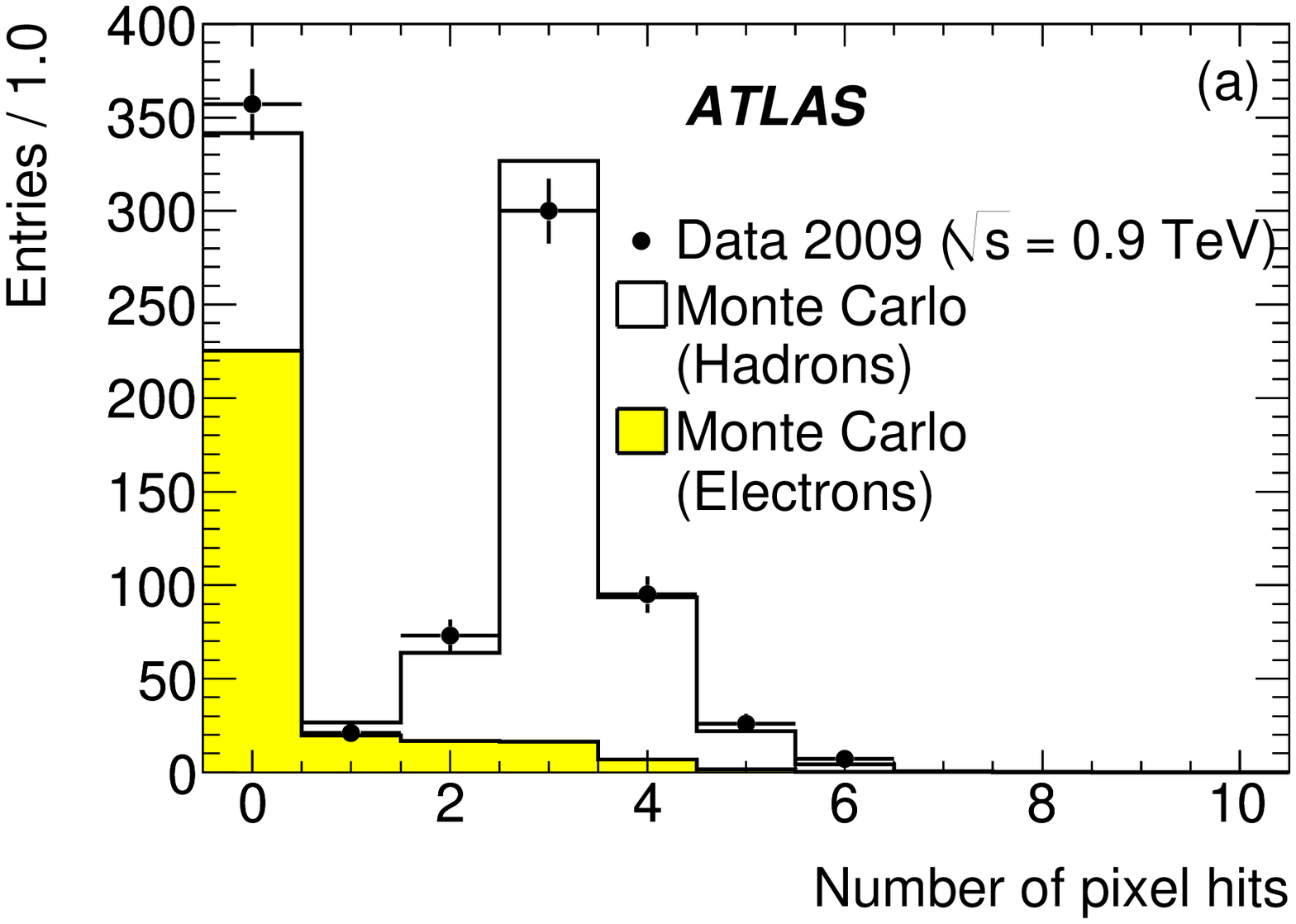}
\includegraphics[width=0.49\textwidth]{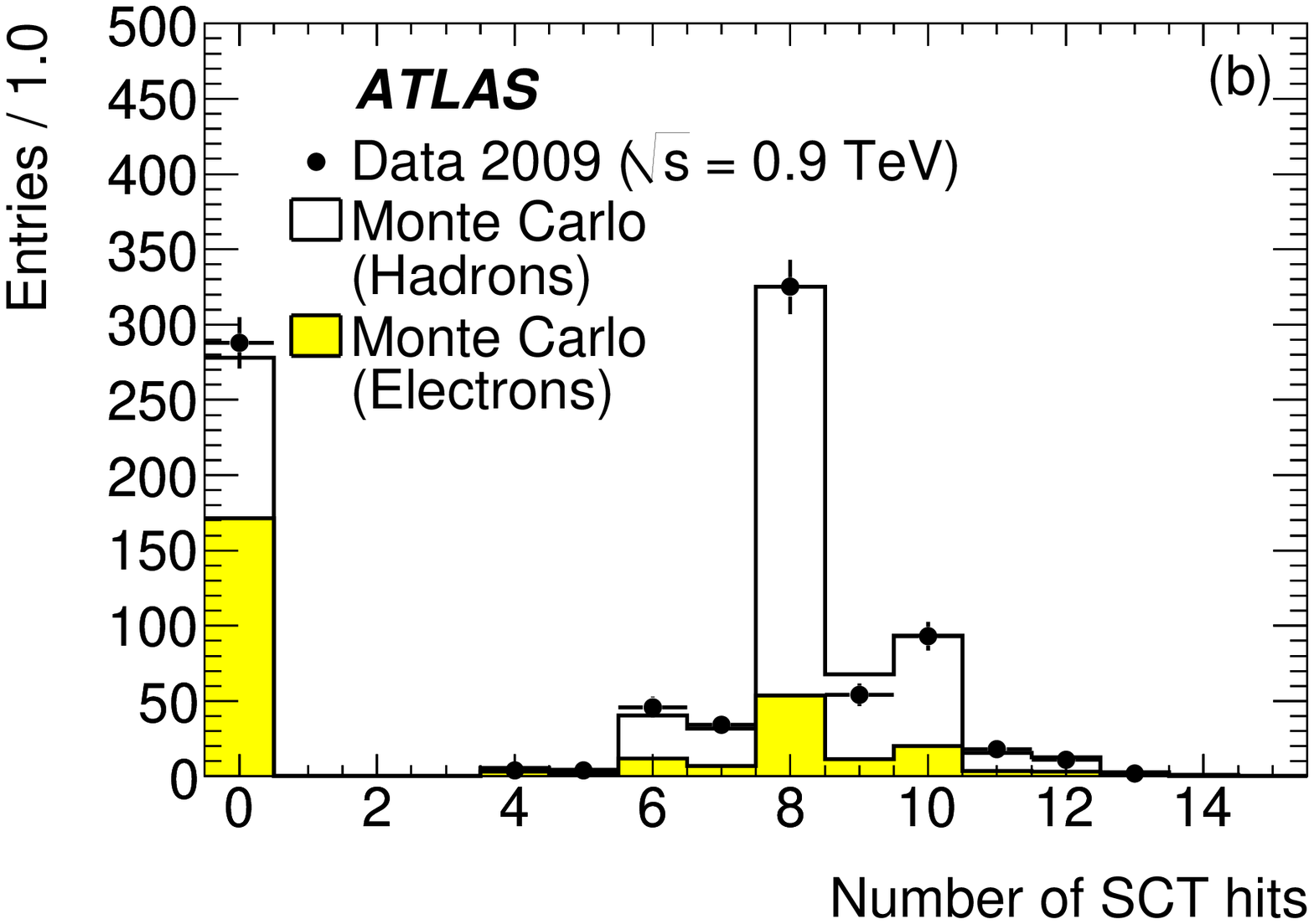}
\includegraphics[width=0.49\textwidth]{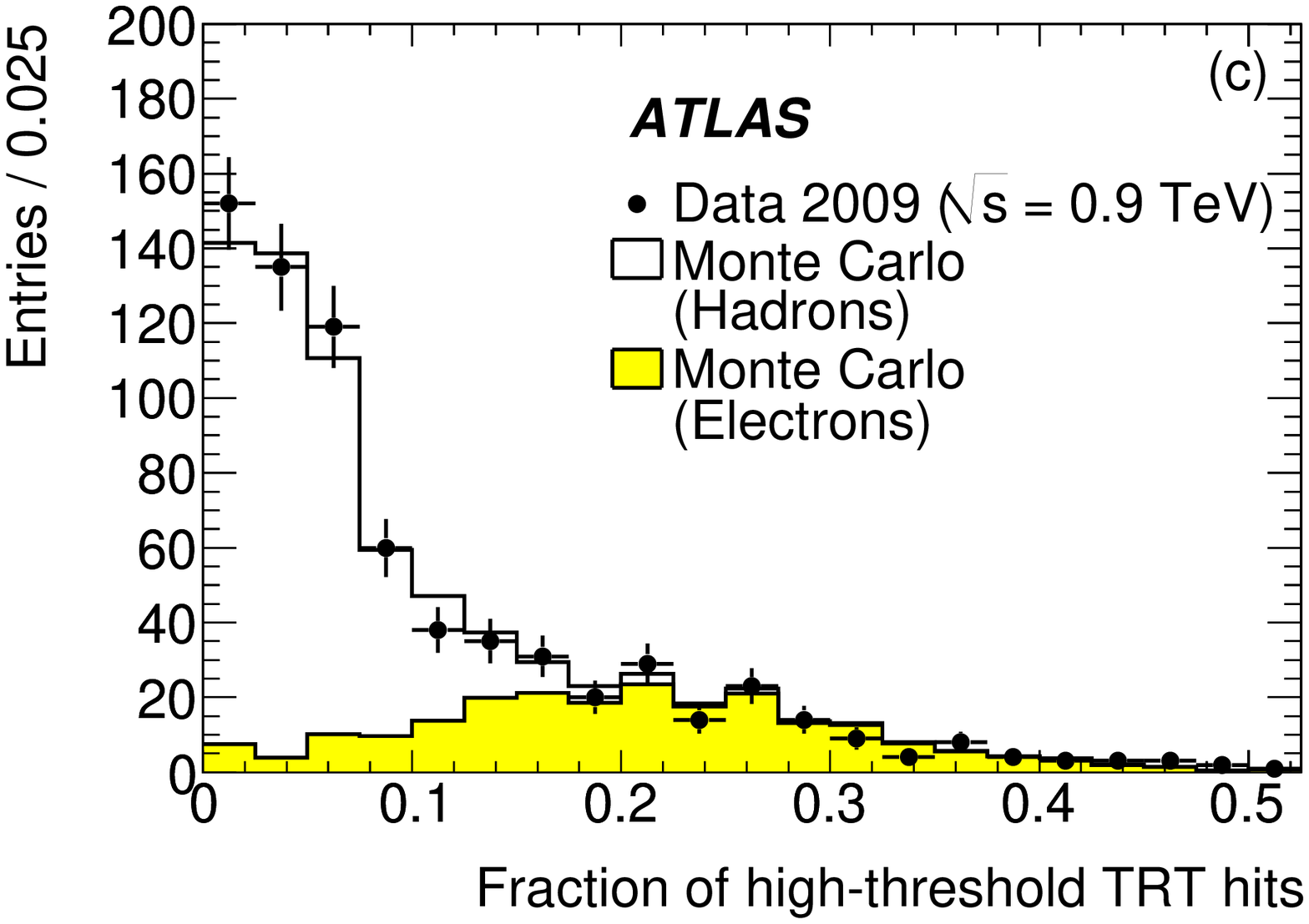}
\includegraphics[width=0.49\textwidth]{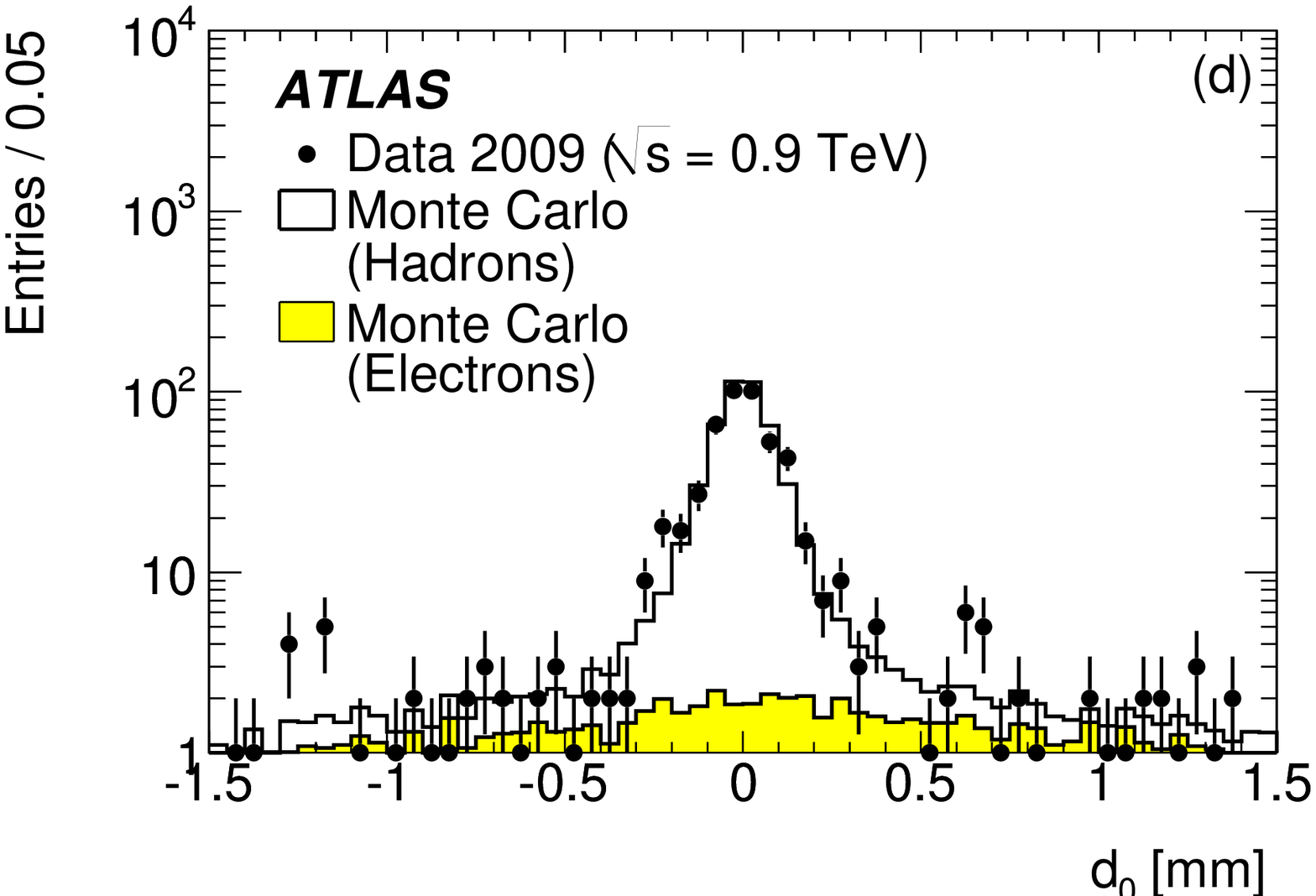}
\caption{\label{fig:TrackingHits} Distributions of tracking variables for all electron
  candidates compared between data and simulation. The
  number of Pixel~(a) and SCT~(b) hits on the electron tracks are shown, the
  fraction of high-threshold TRT~hits for candidates with~$|\eta|<2.0$
  and with a total number of TRT~hits larger than ten~(c), and the
  transverse impact parameter,~$d_0$, with respect to the
  reconstructed primary vertex~(d).
The simulation is normalized to  the number of data events.
}
\end{center}
\end{figure}

Figure~\ref{fig:TrackingHits} shows a comparison of four of the tracking variables
 between data and simulation for all electron candidates.
Figures~\ref{fig:TrackingHits}(a) and~\ref{fig:TrackingHits}(b) show the numbers of
hits on the  tracks in the Pixel and SCT~detectors,
respectively. 
The fraction of high-threshold
TRT~hits belonging to the track for electron candidates
with~$|\eta|<2.0$ and with a total number of TRT~hits larger than
ten is shown in Fig.~\ref{fig:TrackingHits}(c).
At these low energies  the transition radiation yield of
electrons is not optimal and yet a very clear difference can be seen
between the distributions expected for hadrons  and for electrons from
conversions.
 Finally, Fig.~\ref{fig:TrackingHits}(d) shows the distribution of the
 transverse impact parameter,~$d_0$, of the electron track with
 respect to the reconstructed primary vertex position in the
 transverse plane; whereas the hadrons in the simulation display a
 distribution peaked around zero with a resolution
 of~$\sim100$~\hbox{$\mu$}m, the electrons from conversions have
 large impact parameters.
The
agreement between data and simulation is good,
despite the complications expected at these low
energies due to material effects and track reconstruction
inefficiencies.

\paragraph{}

\begin{figure}[htp]
\begin{center}

\includegraphics[width=0.49\textwidth]{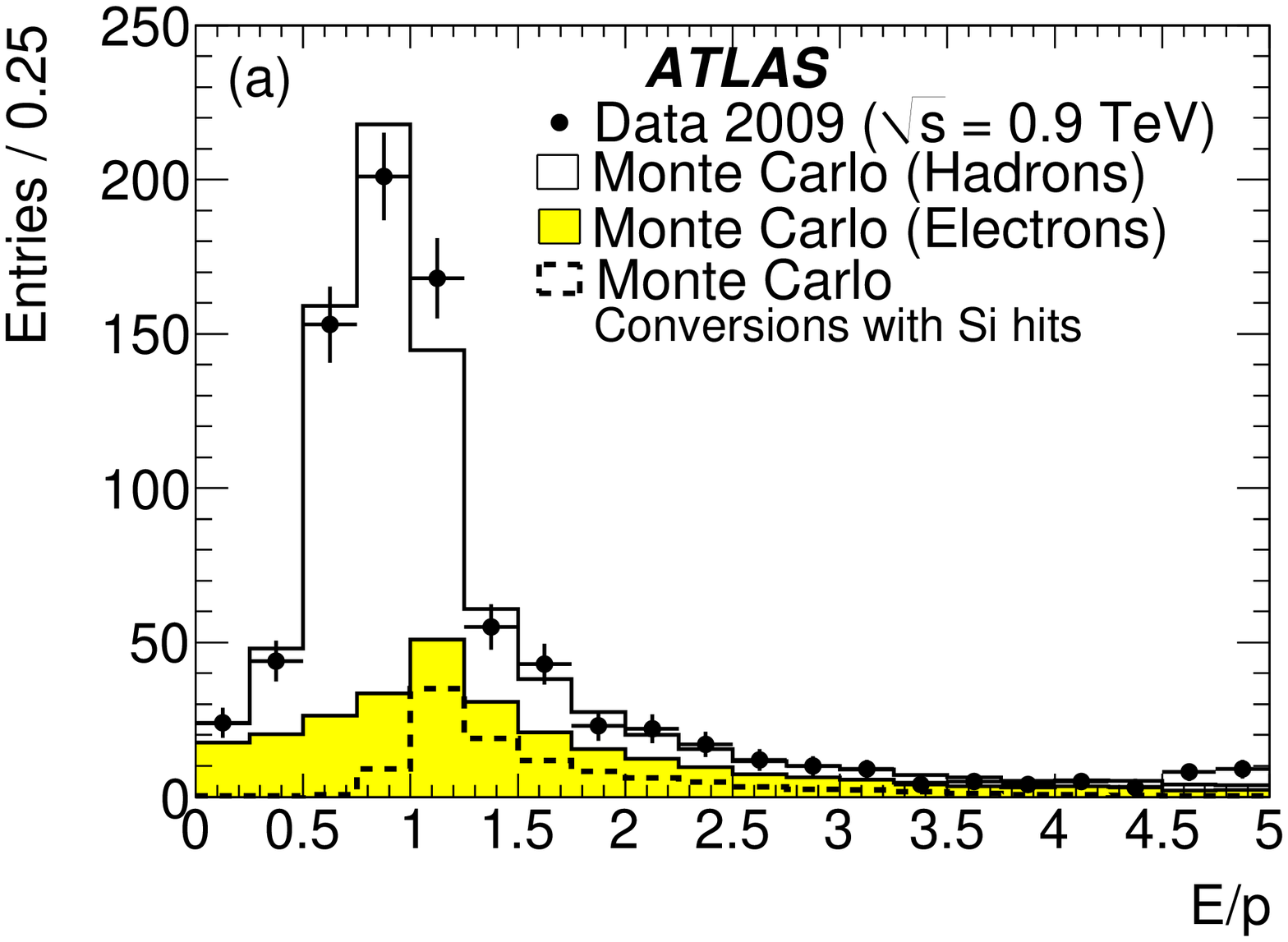}
\includegraphics[width=0.49\textwidth]{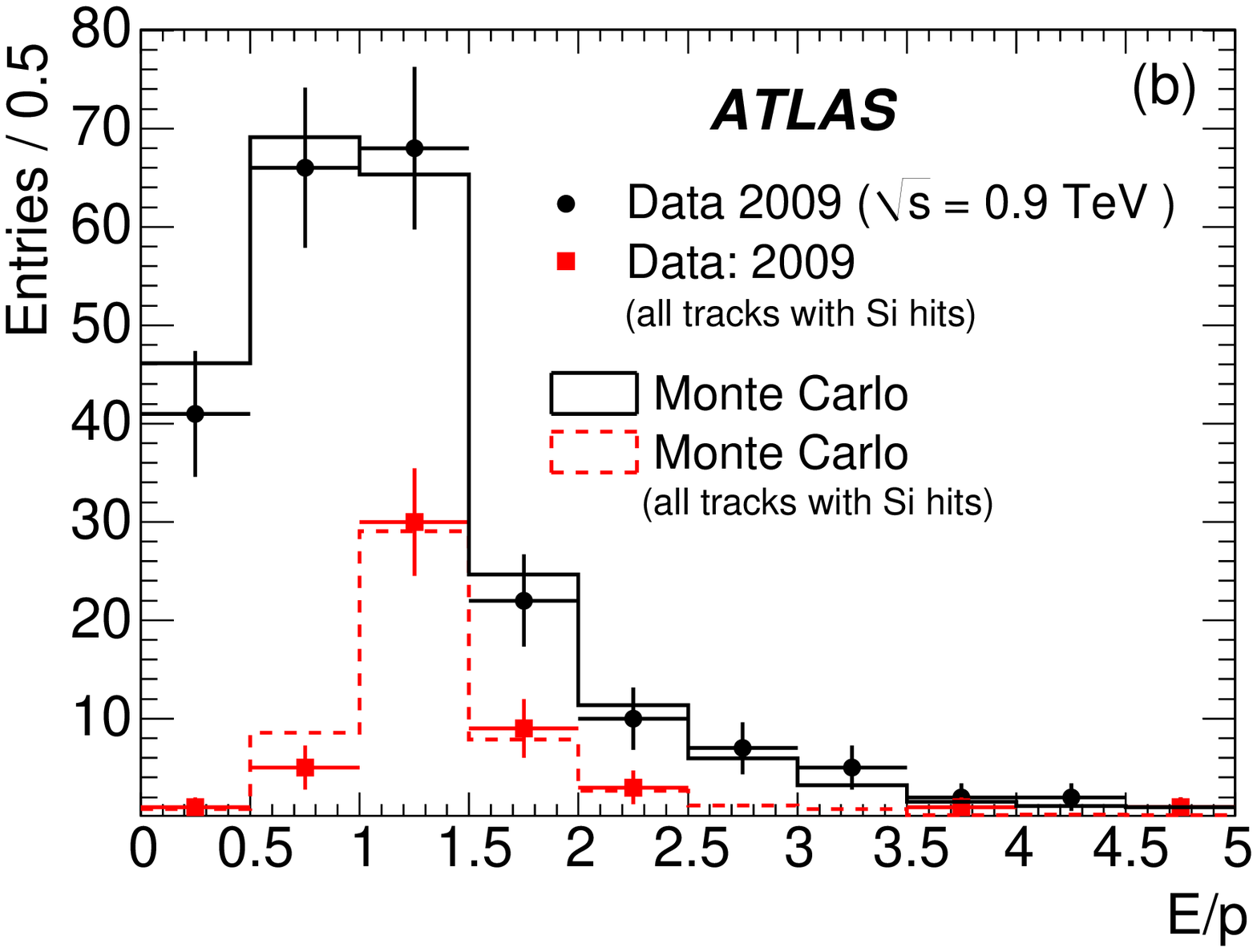}
\caption{\label{fig:EoverP} Ratio, $E/p$, between cluster energy and
  particle track momentum (a) for
  electron candidates and
(b) for electrons from converted photons.
In each case candidates with \pt\ above 2.5~\GeV\ in the calorimeter are shown.
Sub-figure (a) is dominated by real electrons.
The simulation is normalized to  the number of data events.
}
\end{center}
\end{figure}
Figure~\ref{fig:EoverP}(a) shows the distribution of the ratio $E/p$ of cluster
energy in the calorimeter to track momentum for all electron
candidates and for data and simulation. 
Electrons from conversions have a broad $E/p$ distribution as their
shortened tracks have a large momentum error.
The hadron component 
peaks at values near unity: this behaviour, due to the selection bias
for these hadrons, is also observed in the simulation. 
In a similar
fashion, Fig.~\ref{fig:EoverP}(b) shows the $E/p$
ratio of the reconstructed converted photon candidates,
 where the converted photon momentum is
estimated from the combination of the particle momenta for double-track
conversions and from the particle momentum measurement available for
single-track conversions. Approximately $20\%$ of the converted photon candidates
are reconstructed as single-track conversions in this kinematic
regime.
Both the electron dominated and the hadron dominated distributions show
good agrement with the simulation.

\subsubsection{Use of the TRT for Electron Identification}
As already discussed in Section~\ref{electroncandidatesection},
 the electron candidate data sample is
expected to consist predominantly of two components: charged hadrons
misreconstructed as electrons  and
electrons from 
photon conversions. These two
components can be separated  by using
the measured fraction of high-threshold TRT hits on the electron tracks
(see~Fig.~\ref{fig:TrackingHits}(c)). To perform such a measurement, the electron
candidates are required to lie within the TRT acceptance,
i.e.~$|\eta|<2.0$, and to have a reconstructed track with a total of
at least ten TRT hits.



The distribution of the fraction of high threshold hits has been fitted
in 20 bins between 0  and 0.5 to extract the number of hadrons and
electrons observed in the data.
This relies upon the modelling of the response of the TRT to electrons
and pions in the simulation. The sample of electron candidates
considered  here is  predicted to contain 494$\pm$26 electron candidates
which are actually hadronic fakes and 226$\pm$21 genuine electrons.

\begin{figure}[htp]
{\includegraphics[width=0.49\textwidth]{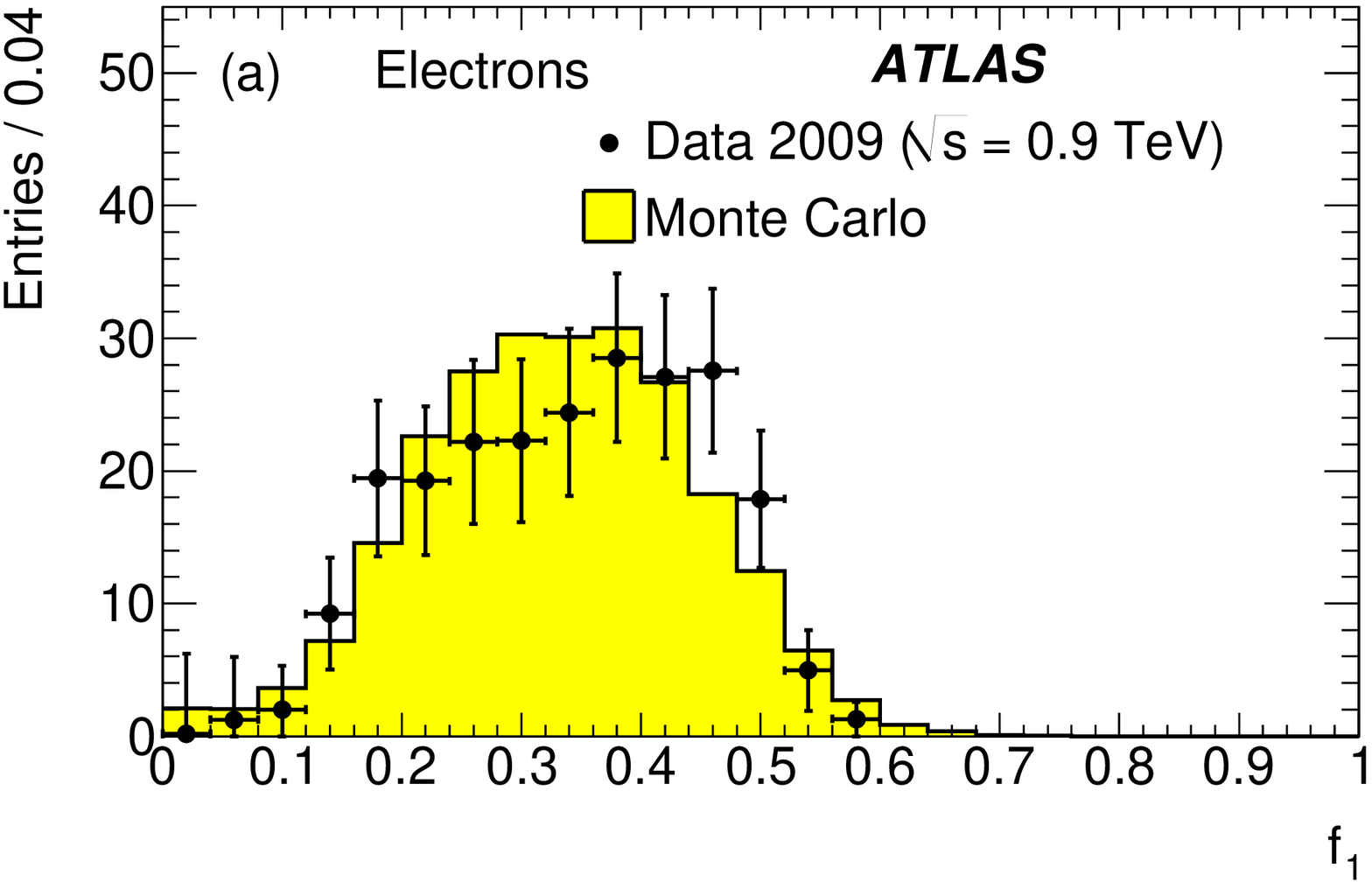}}
{\includegraphics[width=0.49\textwidth]{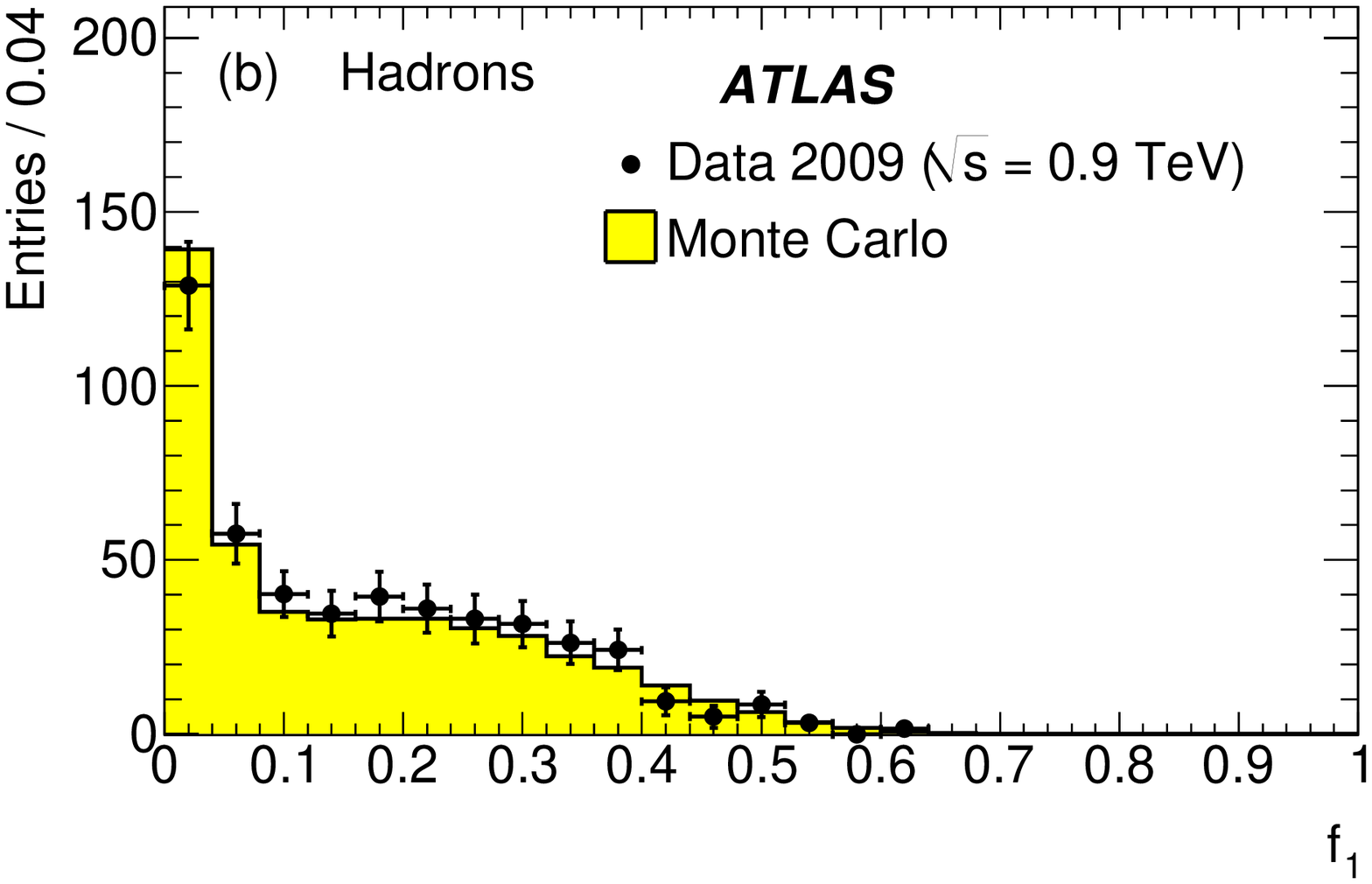}}
\caption{Distribution of the energy fraction in the strip layer of the
  EM~calorimeter as extracted from data compared to the truth from
  simulation. The results are shown for both components of the
  electron candidates: electrons from conversions~(a) and hadrons~(b).
The simulation is normalized to  the number of data events.
}
\vspace{-0.2cm}
\label{fig:ElectronBackground}
\end{figure}

\begin{figure}[htp]
{\includegraphics[width=0.49\textwidth]{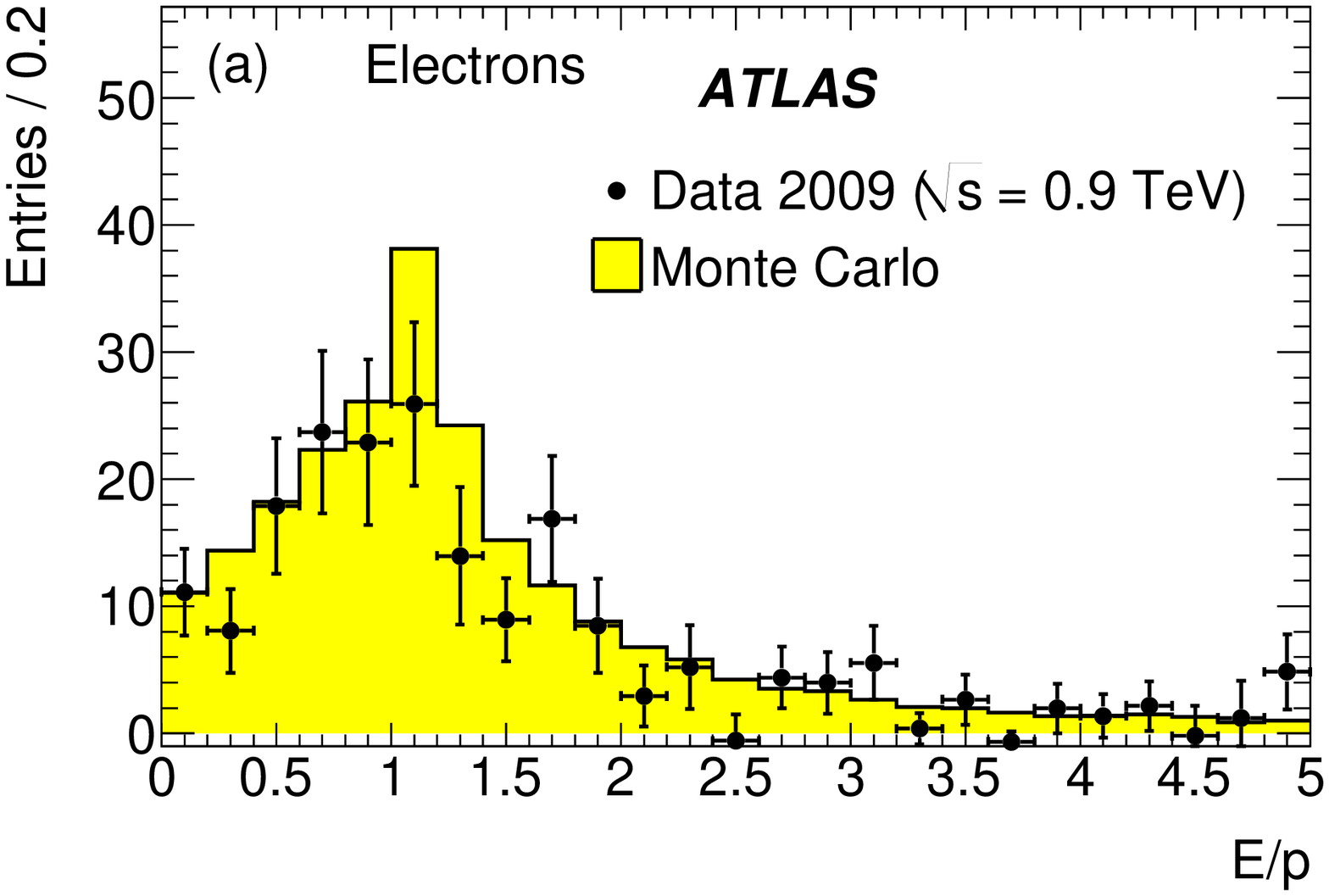}}
{\includegraphics[width=0.49\textwidth]{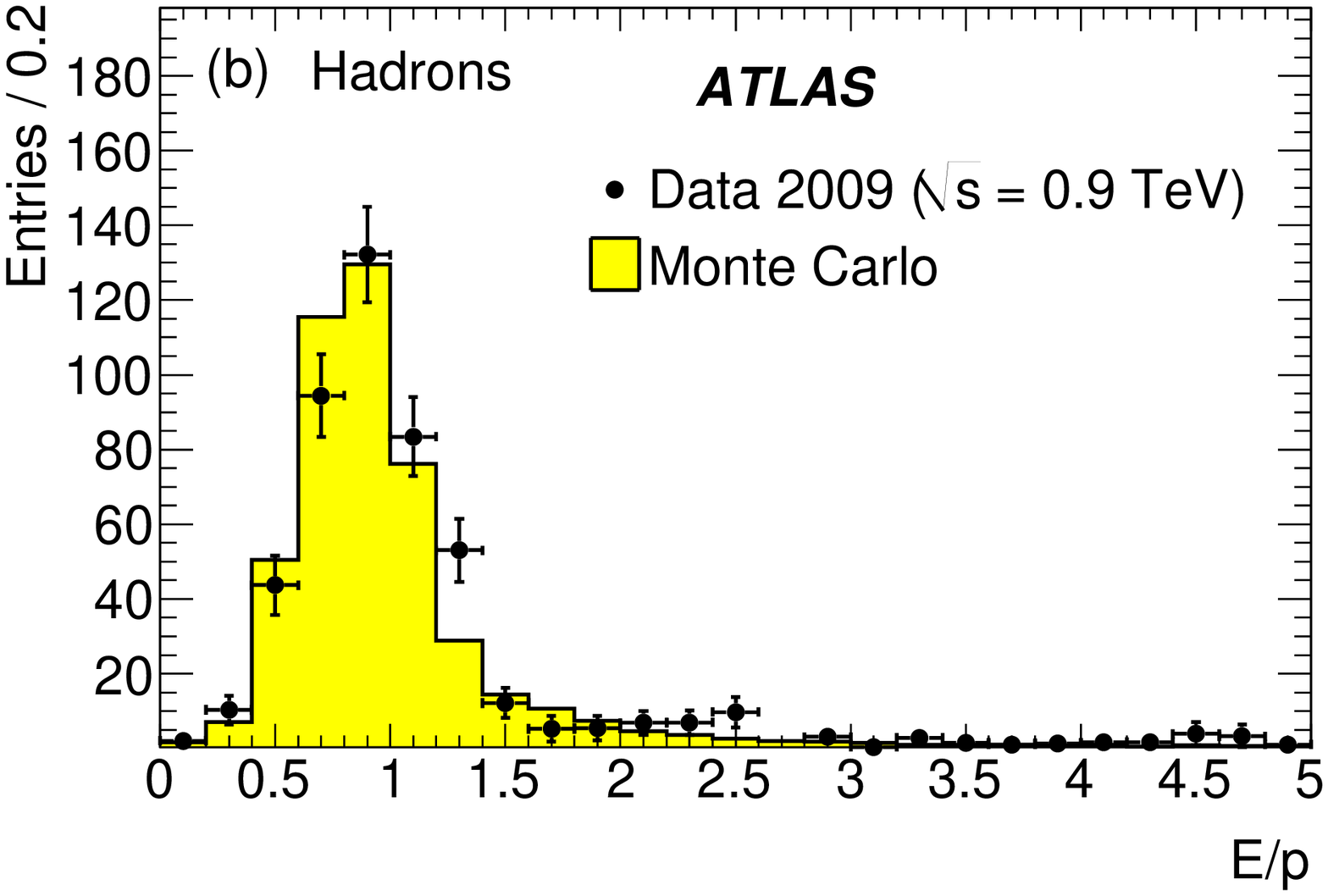}}
\caption{Distribution of the $E/p$ as extracted from data compared to
  the truth from simulation. The results are shown for both components
  of the electron candidates: electrons from conversions~(a) and
  hadrons~(b).
The simulation is normalized to  the number of data events.
\vspace{-0.2cm}
}
\label{fig:ElectronBackground2}
\end{figure}

Two examples of comparisons between the shapes of variables extracted
for each of the two components, using the method described above (on
each bin individually), and the shapes predicted for each component
 are shown in~Figs.~\ref{fig:ElectronBackground}
and~\ref{fig:ElectronBackground2}, respectively, for two of the most
sensitive variables: the fraction of the cluster energy measured in
the strip layer and the ratio~$E/p$. 
The $E/p$ distribution for electrons in both data and simulation shows
a peak close to unity and a tail at large values from bremsstrahlung
losses in the tracker material. 
The error estimates in these de-convolved plots come from toy Monte
Carlo trials and their size reflects the power of the TRT detector for
electron identification.  
 The rates of electrons and hadrons and the relevant distributions
 agree with the Monte Carlo simulation for each species illustrating
 the quality of the simulation modelling.

\subsection{Photon Conversions}
\label{sec:phot-conv}

An accurate and high-granularity map of the inner detector material is necessary for a
precise reconstruction of high-energy photons and electrons.
The location of the
conversion vertex can be used as a tool to map the position and
amount of material of the inner detector. In the following, photon
conversions are selected using only information from the inner
detector, enabling the use of very low momentum particle track pairs.
In addition, conversions give a source of electrons from which the TRT 
   detector response can be determined.

The conversion reconstruction algorithm is described in detail
elsewhere~\cite{CSC}.
In the following, the basic steps
of the algorithm are recalled together with an updated list of selection
criteria.
The algorithm begins by selecting single particle tracks with transverse momentum
\pt$>500$~\MeV.
These tracks must have a  probability
of being  an electron of more than $10\%$, calculated using the particle
identification capability of the TRT, see Section~\ref{sec:trttr}.

 Conversion candidates are then created by pairing oppositely charged
 particle tracks.
 The tracks are further required
to be close in space and to have a small opening angle.
The
selected particle track pairs are then fitted to a common vertex with the constraint
that they be parallel at the vertex. The final set of conversion candidates
is selected based on the quality of the vertex fit which must
have  $\chi^2$ smaller than $50$.

\begin{figure}[h]
\vspace{0.5cm}
\includegraphics[width=0.49\textwidth]{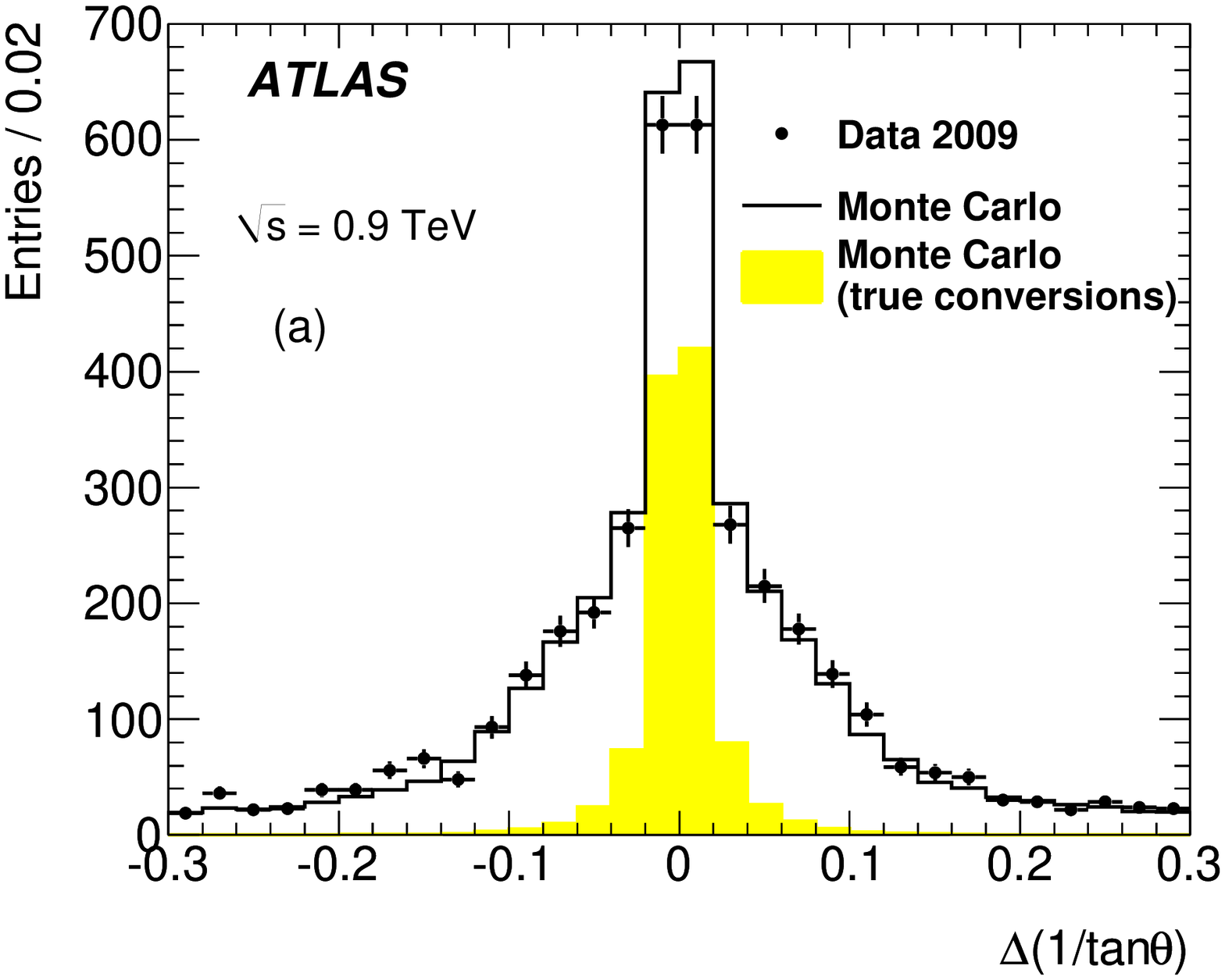}
\includegraphics[width=0.49\textwidth]{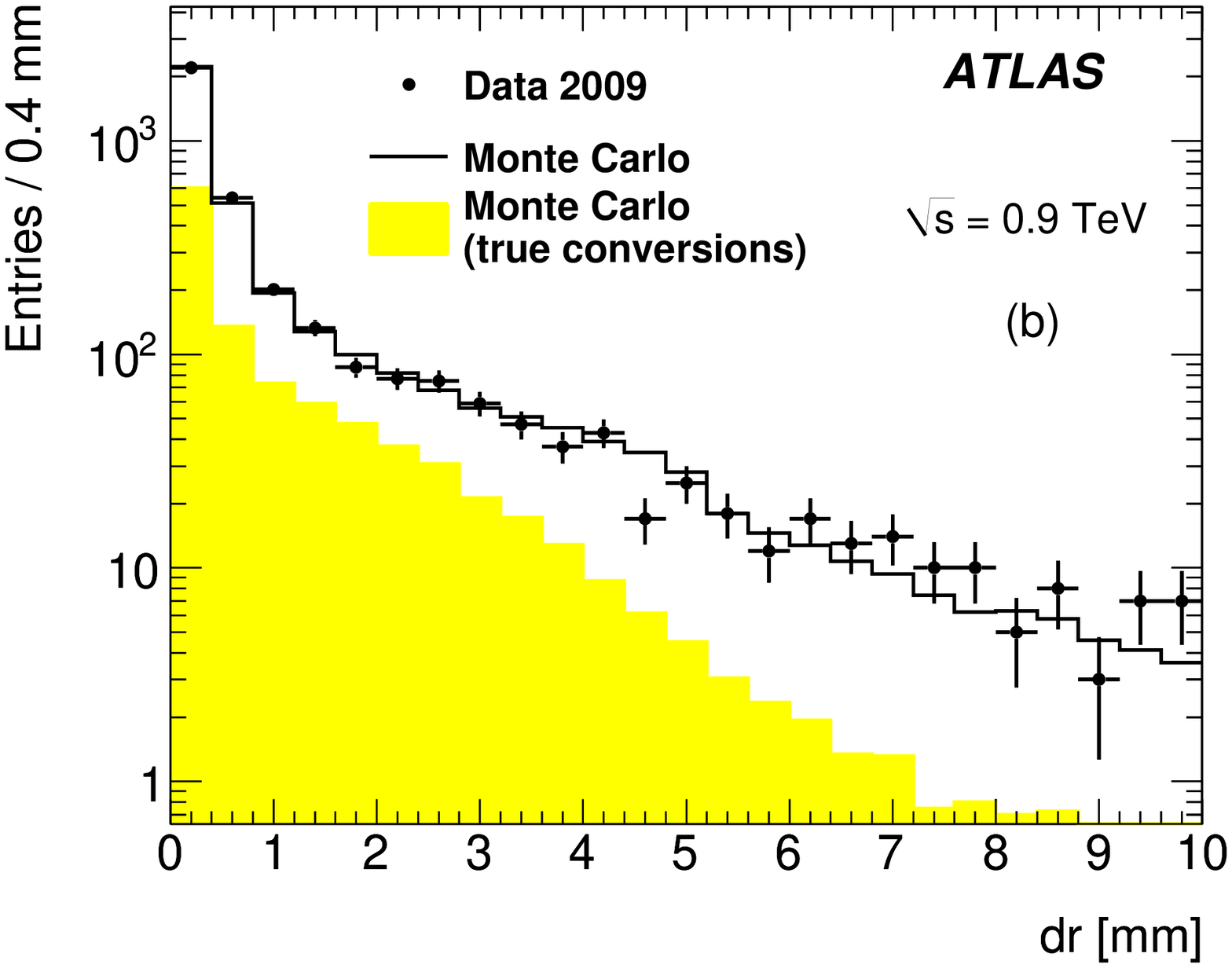}
\caption{Comparison between converted photon candidates, for which both
  tracks have silicon hits, in data and non-diffractive minimum-bias
  Monte Carlo
simulation. (a) Opening angle in the $rz$ plane between the two tracks
($\Delta(1/\mathrm{tan}\theta)$); (b) 3D distance of closest approach 
between the two tracks, dr.
The distributions are normalized to the
same number of conversion candidates in data and Monte Carlo
simulation.
}
\label{fig:dataMC}
\end{figure}

The tracks used for the reconstruction of conversions may be stand-alone TRT
tracks, or they may include silicon hits. 
In the data, 3\,662 vertices, 6.7\% of the total, have two tracks with
silicon information, to be 
compared with 10.4\% in the simulation. This class of vertices 
have much less background
 than the total and the following results are drawn from them.
Some properties of the candidates in  data and Monte Carlo simulation
are shown in
Fig.~\ref{fig:dataMC}.
Given the complexity of the reconstruction of converted
photons and the impact of bremsstrahlung of the electrons in the tracker
material, the consistency between the data and the simulation for the selection
variables is good.


\label{sec:estim-conv}
To measure the inner detector material the selection requirements are
tightened to $>$90\% TR electron probability and vertex $\chi^2<5$.

\begin{figure}[t]
\includegraphics[width=0.495\textwidth]{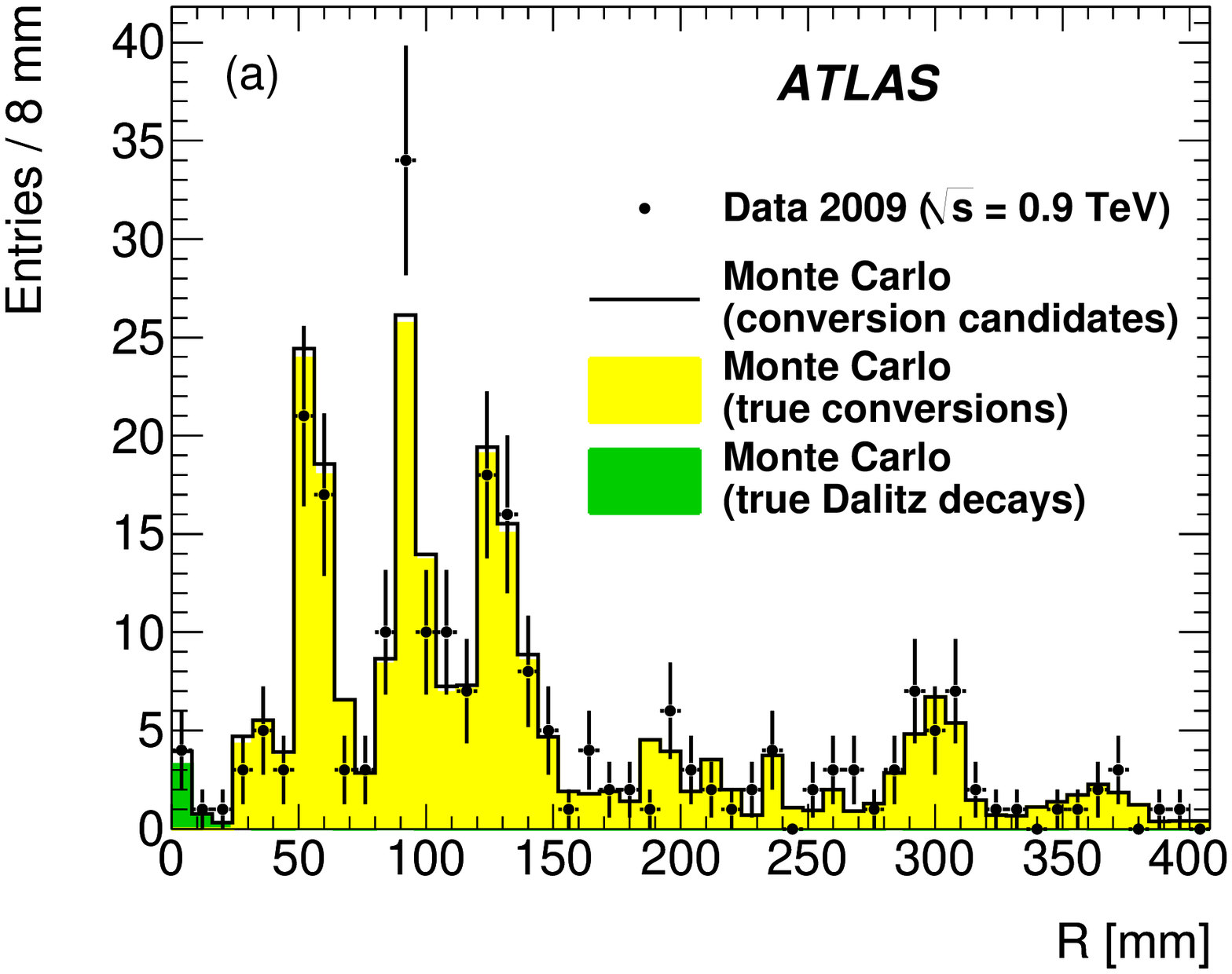}
\includegraphics[width=0.495\textwidth]{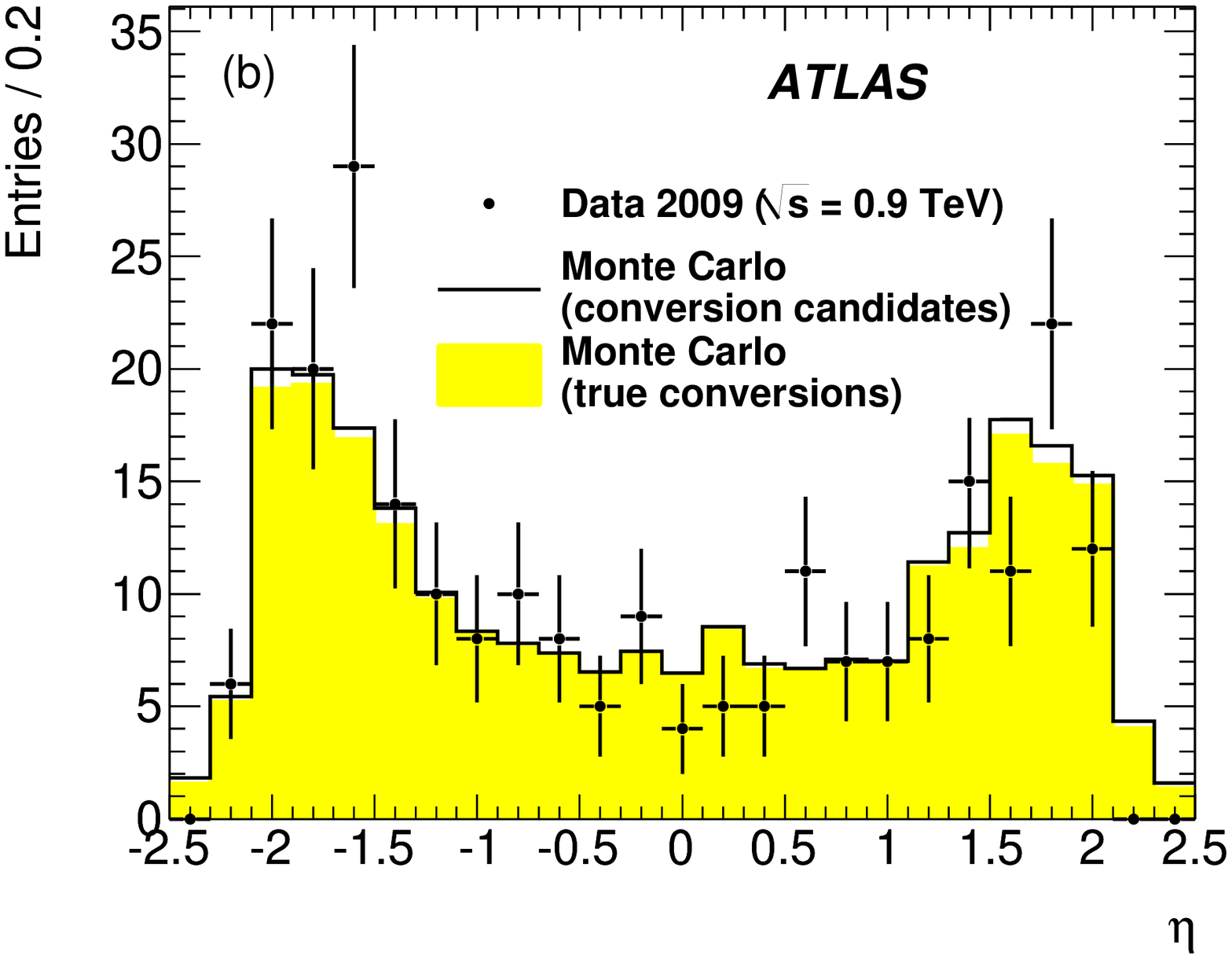}
\caption{Distribution of conversion candidate radius, 
(a), and $\eta$, (b). The points show
the distribution for  data; the open histograms,
the total from the Monte Carlo simulation and the filled
component  shows the expected contribution of true photon conversions.
The contribution from the Dalitz decays of neutral
mesons is shown in sub-figure (a). The Monte Carlo
simulation is normalized to number of conversion candidates in the
data, although in subsequent analysis normalization is to the number
in the beam pipe.
}
\label{fig:material}
\end{figure}

Figure~\ref{fig:material} shows the location in radius and $\eta$ of
conversion vertices. The simulation was
normalized to the same number of conversions as in the data and 
the agreement in shape is  satisfactory.

The amount of material, in multiples of the radiation length $X_0$, that
 the photons
 traverse can be calculated from the fraction of photon
conversions seen in it given  the reconstruction efficiency.
The combinatorial background and the error in
 determining the conversion radius must be accounted for.
To remove the dependence on the absolute flux of photons and the
overall reconstruction efficiency, the rate is normalized to that seen
in a well-understood reference material volume, which is chosen to be
the beam pipe. 
As shown in  Fig.~\ref{fig:material}, there were only 9 conversions
 in this reference volume and
so the absolute material determination has errors of at least 30\%.
The agreement between data and Monte Carlo is presented in Table~\ref{tab:material}.\\

\begin{table}[t]
\caption{$N_\mathrm{reco}$ is the number of reconstructed conversions
  in each  layer, and ${X}$/${X_0}_\mathrm{data}$ and
  ${X}$/${X_0}_\mathrm{MC}$ represent the amount of material in the
  different volumes estimated from data and Monte Carlo, normalized
  by the number of reconstructed converted photons in the beam pipe, whose
  material is assumed to be correct.
  The   normalization introduces an additional
statistical uncertainty of $30\%$ on ${X}$/${X_0}_\mathrm{data}$.\vspace{0.2cm} \newline
}
\begin{center}
\begin{tabular}{lccc}
\hline\hline
& $N_\mathrm{reco}$ & $\frac{X}{X_0}_\mathrm{data}$ & $\frac{X}{X_0}_\mathrm{MC}$ \\
\hline
Beam pipe     &  $9$ & $0.00655$ & $0.00655$  \\
Pixel B-layer & $46$ & $0.030 \pm 0.004$ & $0.032$ \\
Pixel layer 1 & $65$ & $0.035 \pm 0.004$ & $0.027$ \\
Pixel layer 2 & $55$ & $0.025 \pm 0.003$ & $0.023$ \\
SCT layer 1   & $25$ & $0.020 \pm 0.004$ & $0.016$ \\
\hline\hline
\end{tabular}
\end{center}
\label{tab:material}
\end{table}

\subsection{Reconstruction of $\boldsymbol\pizero$ and $\boldsymbol{\eta}$ Mesons}
\label{sec:pi0-eta}

For the analysis presented in this section, cells from the four layers are combined to form a cluster of
size $\Delta\eta \times \Delta\phi$ = 0.075 $\times$ 0.125, which
corresponds to an area of 3 $\times$ 5 cells in the middle layer of the EM calorimeter.
The EM
cell clusters are reconstructed with a seed cell threshold $|E_{\rm cell}
| = 4\sigma$ (where $\sigma$ corresponds to the
electronic noise in the cell) and with a cluster transverse
energy $E_T > 300$~\MeV~\cite{CSC}.
These clusters are used as photon candidates for $\pi^0$ and $\eta$
reconstruction.

The standard parameterization of energy response discussed in
Section~\ref{egamma-reconstruction} was performed for photons with
$E_T > 5$~\GeV. For the present study a dedicated parameterization was
extracted from the minimum-bias simulation sample using low-energy
photons coming only from {\pizero}s. 


\subsubsection{Extraction of $\boldsymbol{\pizero\rightarrow \gamma\gamma}$ Signal}
\label{sec:cuts}

In order to extract the \pizero\  signal from the combinatorial
background, well measured photons were selected inside an acceptance of
$|\eta | <$ 2.37, excluding a transition region 1.37 $< |\eta|<$  1.52.
The fraction of energy in the first layer,
   $E_1$/$(E_1 + E_2 + E_3)$, was required to be larger than 0.1
 and the clusters were required to have a
transverse energy, $E_{\mathrm T}$, above  400~\MeV.



All  pairs of photons  with $\pt^{pair} >$~900~\MeV\ are
selected. There  are about $8 \times 10^5$ of these  in the data.

\subsubsection{$\boldsymbol{\pizero}$ Mass Fit}
\label{sec:fit}

\begin{figure}[htp]
\begin{center}
\includegraphics[width=0.49\textwidth]{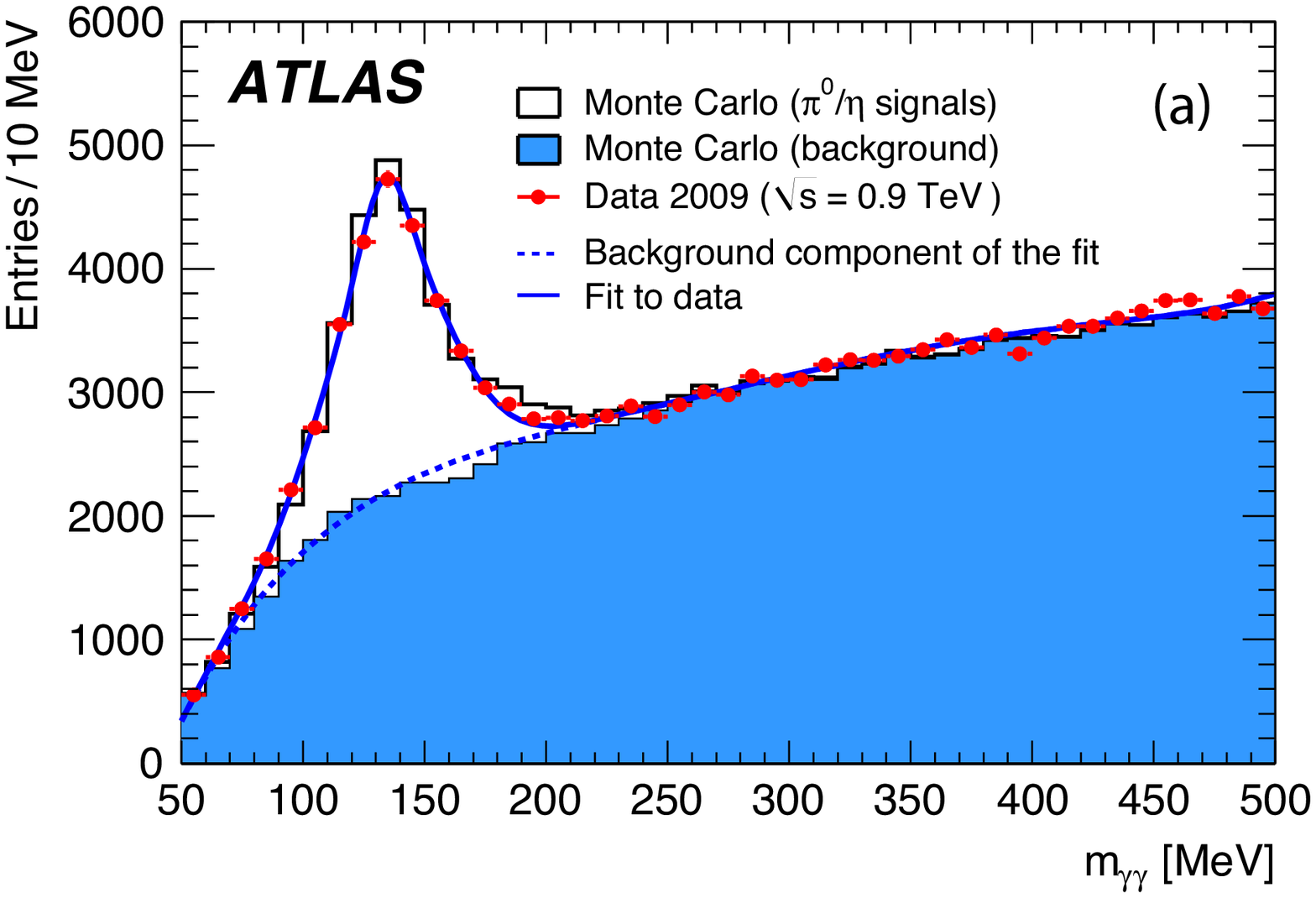}
\includegraphics[width=0.49\textwidth]{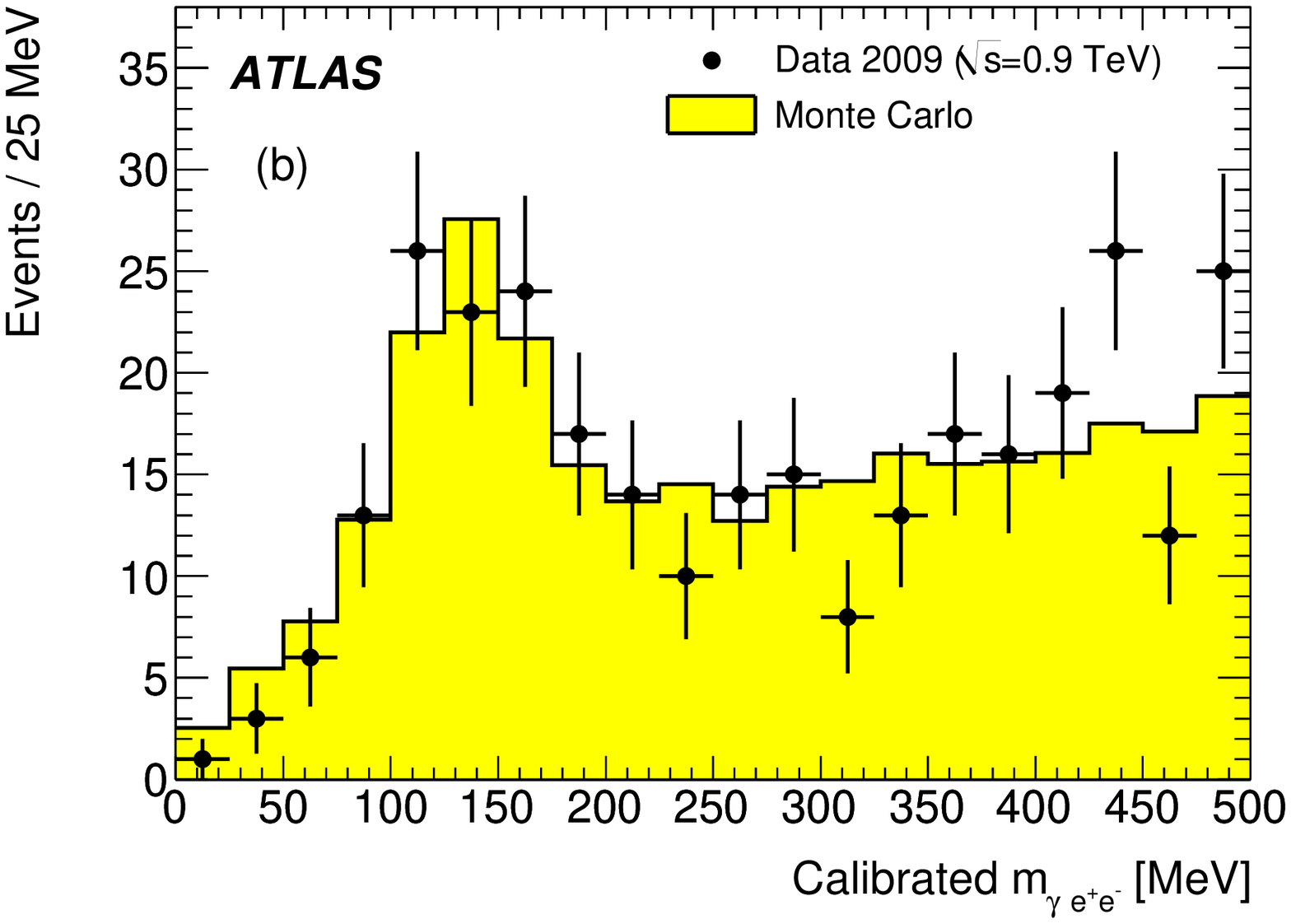}
\caption{(a) Diphoton invariant mass distribution for the
  \pizero\ selection  for data and Monte Carlo.
 The Monte Carlo is normalized to
  the same number of entries as the data.
(b) Invariant mass distribution from one converted and one unconverted
photon.
The data are represented by points and the Monte Carlo simulations are
shown as histograms.
}
\label{fig:pi_fig_05}
\end{center}
\end{figure}

The invariant mass distribution of the photon pairs is shown in
Fig.~\ref{fig:pi_fig_05} for both data and Monte Carlo. The diphoton
mass distribution is fitted using a maximum-likelihood fit. The
signal is described by the sum of a Gaussian and a ``Crystal-Ball
function" ~\cite{ref:crystal-ball}, which are required to have the same mean. The
combinatorial background is described with a 4$^{\rm th}$ order
Chebyshev polynomial. The parameters of the signal and the
background normalization are varied in the fit to the data,
 while the parameters of the
polynomial were extracted from the Monte Carlo.

The fitted \pizero\ mass is 134.0$\pm$0.8~\MeV\ for the data
and 132.9$\pm$0.2~\MeV\ for the Monte Carlo where the errors
are statistical only. The mass resolution in
the data is 24.0~\MeV, to be compared with 25.2~\MeV\ in
the simulation, and the number of \pizero's
is (1.34 $\pm$ 0.02) $\times$ 10$^4$.
This fit is
sensitive to the modelling of the background shape near the
\pizero\ mass.
Varying the background
shape under the peak leads to a differences of up to 1\% for the fitted
\pizero\ mass, up to 10\% for the fitted \pizero\ mass resolution and
up to 20\% for the fitted total number of signal events.

The 1\%
agreement of energy scale between data and Monte Carlo is well within
the 2 - 3\% uncertainty on the energy scale transported from test-beam
data analysis. 
 The 1.5\% discrepancy of the mass found in Monte
Carlo  with respect to the PDG nominal \pizero\  mass is
consistent with the  accuracy (as evaluated with simulation) of the  cluster
calibration procedure used for the low-energy photons,
and the 1\% uncertainty arising from the background modelling.


The converted photons reconstructed in Section~\ref{sec:estim-conv} can also
be used to search for the \pizero. This is done here using one
photon reconstructed in the calorimeter, with a track veto applied and
one conversion candidate. The conversion candidates are required to have four
silicon hits on both tracks and must be in the same hemisphere as the
calorimeter cluster. 
Figure~\ref{fig:pi_fig_05}(b)  shows the $\gamma e^+ e^-$ mass spectrum; the
\pizero\ peak is clearly visible. 

The uniformity of the EM calorimeter response  was studied in ten $\eta$ bins,
where both photons  are in the same bin.
The diphoton mass distribution
in each $\eta$ bin is fitted
separately with the background shape constrained from simulation.
The reconstructed \pizero\ mass is constant within 3\% for both data
and Monte Carlo for all $\eta$ bins, and the ratio of data to Monte Carlo
is consistent within the 2\%  statistical uncertainties.


\subsubsection{Extraction of the $\boldsymbol{\eta \rightarrow \gamma \gamma}$ Signal}

The number of $\eta \rightarrow \gamma \gamma$ events is expected to be one order of magnitude
smaller than \pizero\ $\rightarrow \gamma \gamma$ in the minimum-bias event
sample. Therefore, the combinatorial background contribution in the
$\eta$ mass region needs to be significantly reduced. This can be
achieved by adding the following criteria to the \pizero\  analysis:

\begin{itemize}
\item Tighter kinematic selections: E$_T^{cluster} > 800$~\MeV, p$_T^{\rm pair} >2200$~\MeV.
\item A track veto:
no track, extrapolated into the calorimeter, should be within $ -0.1 < (\phi_{\rm clus} - \phi_{\rm extr} ) < 0.05$ and
$|\eta_{\rm clus}-\eta_{\rm extr}| <$  0.05 of the cluster being
considered.

\end{itemize}

\begin{figure}[t]
\begin{center}
\includegraphics[width=0.7\textwidth]{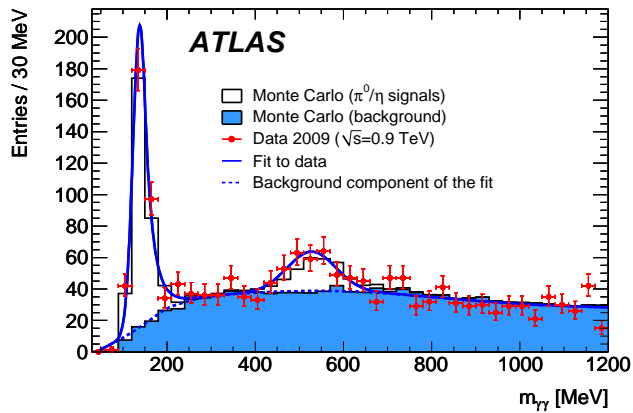}

\vspace{-.2cm} 
\caption{Diphoton invariant mass spectrum with tighter selection criteria to extract the $\eta$ peak with the fit
superimposed to the data. The Monte Carlo simulation sample is normalized to the number of entries in
the distribution for data.
}
\label{fig:pi_fig_09}
\end{center}
\end{figure}

The diphoton invariant mass spectrum of this sample is
shown in Fig.~\ref{fig:pi_fig_09} for both data and Monte Carlo. In
addition to the \pizero\ peak, the $\eta\rightarrow \gamma\gamma$
signal can be observed on top of the combinatorial background. The
mass spectrum was fitted using the sum of a Gaussian and a
Crystal-Ball function with the same mean for the \pizero\ peak, a
Gaussian for the $\eta$ peak and a 4$^{\rm th}$ order Chebyshev polynomial for the
background. The \pizero\ and background shape parameters are taken
from the Monte Carlo simulation while their normalizations are free in
the fit, as are the parameters of the Gaussian describing the $\eta$
peak.

As can be seen from Fig.~\ref{fig:pi_fig_09}, the number of $\eta$
candidates per photon pair  agrees between  the data and the Monte Carlo
simulation. The $\eta$ mass extracted from the data, 527 $\pm$ 11~(stat)~\MeV, agrees with the  mass obtained using
the same fitting function on the Monte Carlo simulation,
544 $\pm$ 3~(stat)~\MeV, within the statistical and energy scale
uncertainties. 

\newcommand{\antikt}{anti-k$_{T}$}
\newcommand{\ptjet}{\ensuremath{p_{\rm T}^{\rm jet}}}
\newcommand{\Etjet}{\ensuremath{E_{\rm T}^{\rm jet}}}

\section{Jets}

\label{jets}
Many of the final states which will be studied in high energy
collisions contain jets of hadrons produced by strong interactions.
The ATLAS analysis chain applies the same jet algorithm
 to the $0.9$~\TeV\ and 2.36~\TeV\ collision data and to the
Monte Carlo simulation. 
The following
comparison between data and Monte Carlo simulations should not be
taken as a precise analysis of the underlying physics in the
simulation, but rather as an assessment of the general behaviour of
the detector and software chain (reconstruction and simulation).

Results are presented using  clusters of calorimeter cells  calibrated
to correctly measure the energy deposited by electrons and photons in
the calorimeter. This is known as the
electromagnetic scale. There was, at this stage, no allowance for
energy loss in inert material.
From this starting point  jets were reconstructed using
 the \antikt~algorithm \cite{Cacciari:2008gp},
 which is safe against  infrared and collinear divergences. The
parameter R, which controls the size of jets in the $\eta-\phi$ plane,
was set to R=0.6.

\subsection{Jets from Calorimeter Clusters}
\label{jetscalo}
\begin{figure}[htp]
{\includegraphics[width=0.49\textwidth]{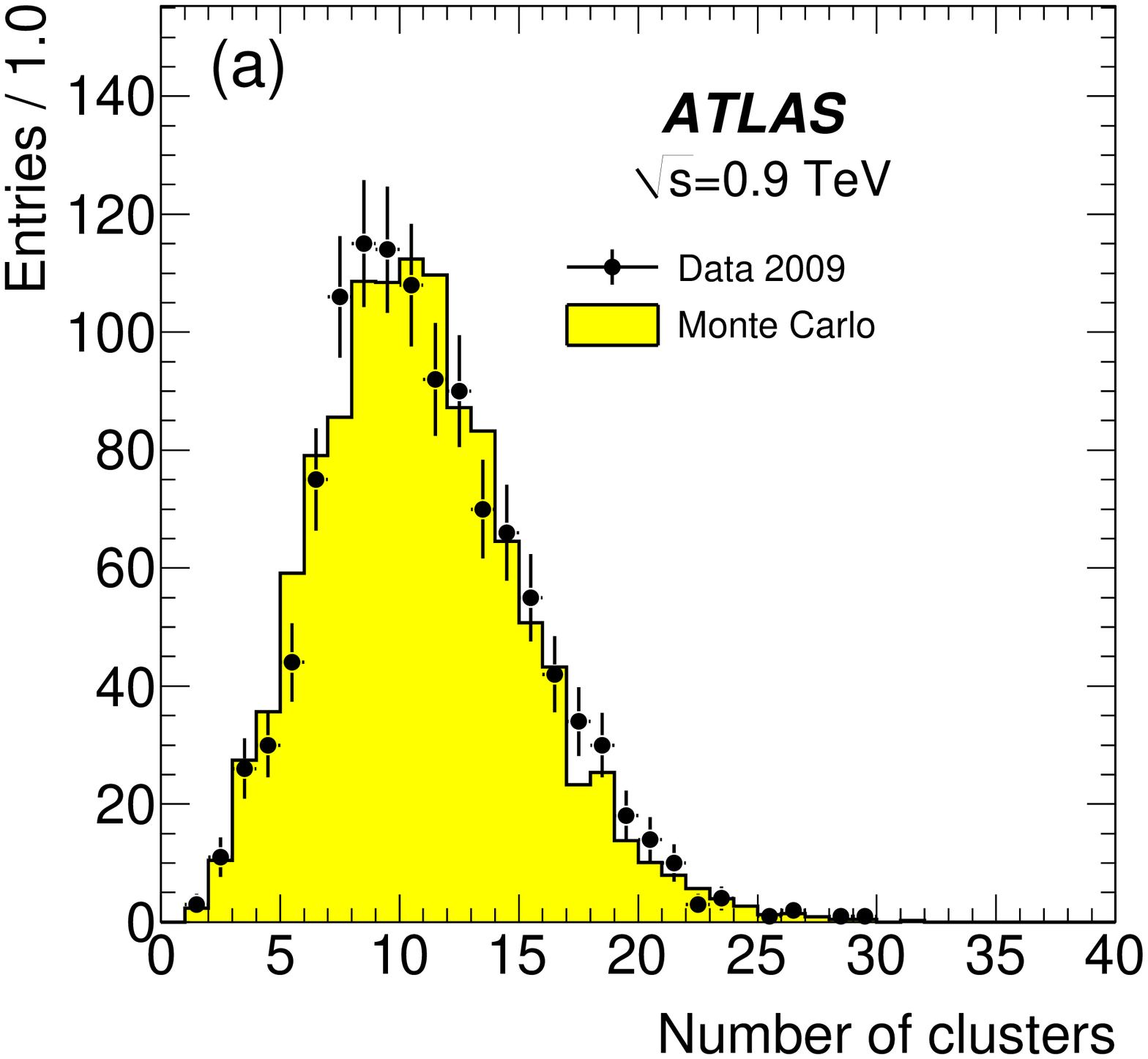}}
{\includegraphics[width=0.49\textwidth]{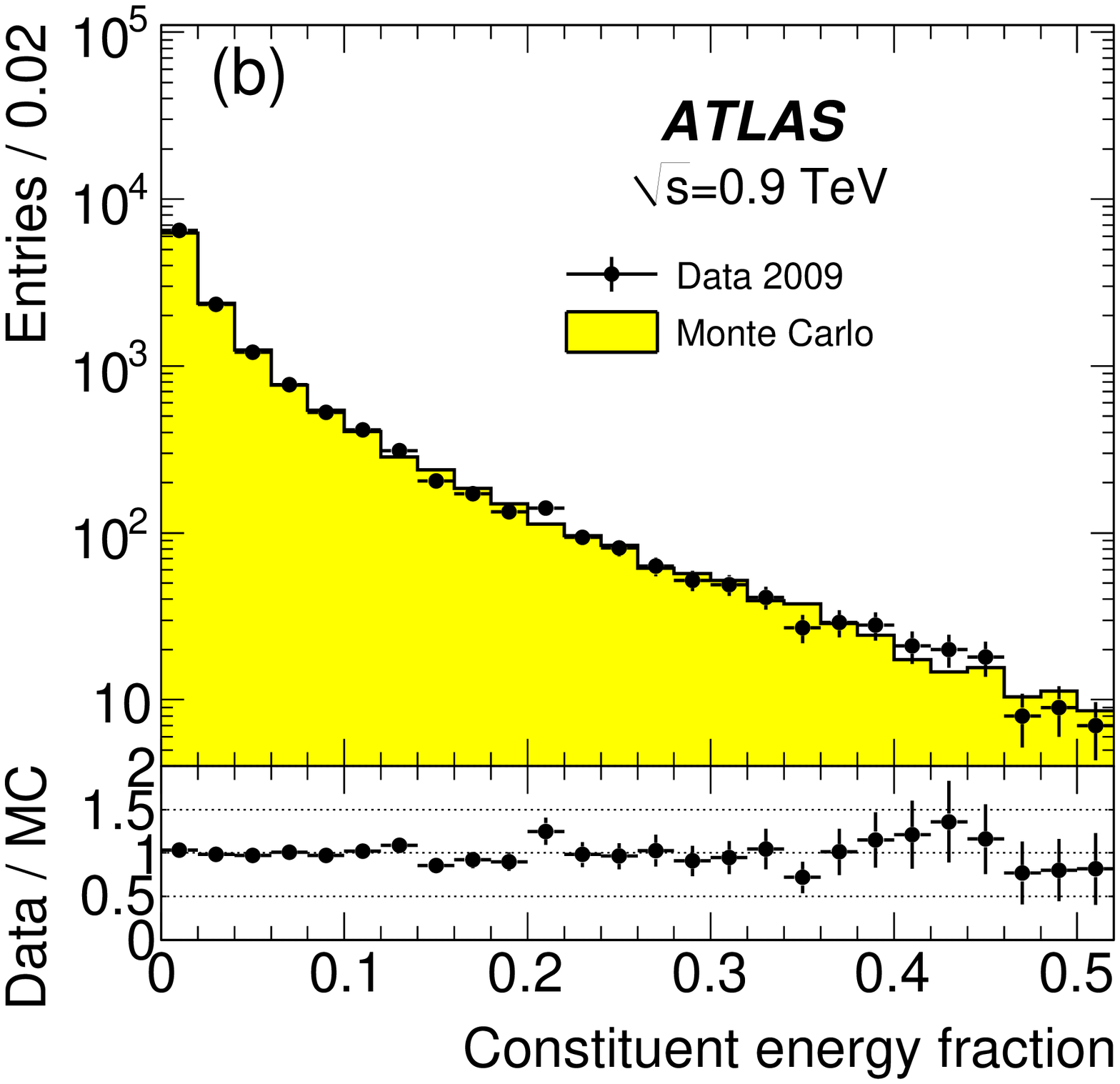}}
\vspace{-.2cm} 
\caption{Distributions of (a) the number of clusters per jet and (b) the
  fraction of energy per cluster for 
jets    reconstructed with topological   clusters using the 
\antikt~algorithm with R=0.6.
}
\label{fig:Jetconst_topo}
\end{figure}

The inputs to the jet algorithm are topological clusters \cite{CSC}
which 
attempt to reconstruct the three-dimensional shower topology of each
particle. 
These clusters were built starting from seed cells
with energies $|E_i | > 4 \sigma_{\mathrm noise}$, where
$\sigma_{\mathrm noise}$ is the  electronic noise measured by
iteratively gathering neighbouring cells with $|E_j | > 2
\sigma_{\mathrm noise}$ and, in a final step, adding all direct
neighbours of these accumulated secondary cells. 
The noise in the EM calorimeter, for example, was in the range of
10-40~\MeV\ per cell, depending upon the compartment and pseudorapidity.
Approximately 0.1\% of all  cells were  classified as noisy  and were removed. 
Clusters  built from the remaining cells 
 were then used to create jets, which have to satisfy:
\ptjet$>$7~\GeV~and $|\eta|<$2.6 where \ptjet\ is the transverse jet
momentum at the electromagnetic  scale.
In order to remove  cosmic muons and some
residual effects from cells in the calorimeter that exhibit large
noise fluctuations the jets are required to pass quality criteria.
Furthermore, if
 the jet energy corrections compensating for excluded calorimeter
regions  exceed 20\%,  the corresponding candidates are not considered. 
Figure~\ref{fig:Jetconst_topo} presents  example  distributions
of the internal structure of the jets, namely the number of
topological clusters, and the fraction of the jet energy
carried by each of them.

\begin{figure}[htp]
{\includegraphics[width=0.49\textwidth]{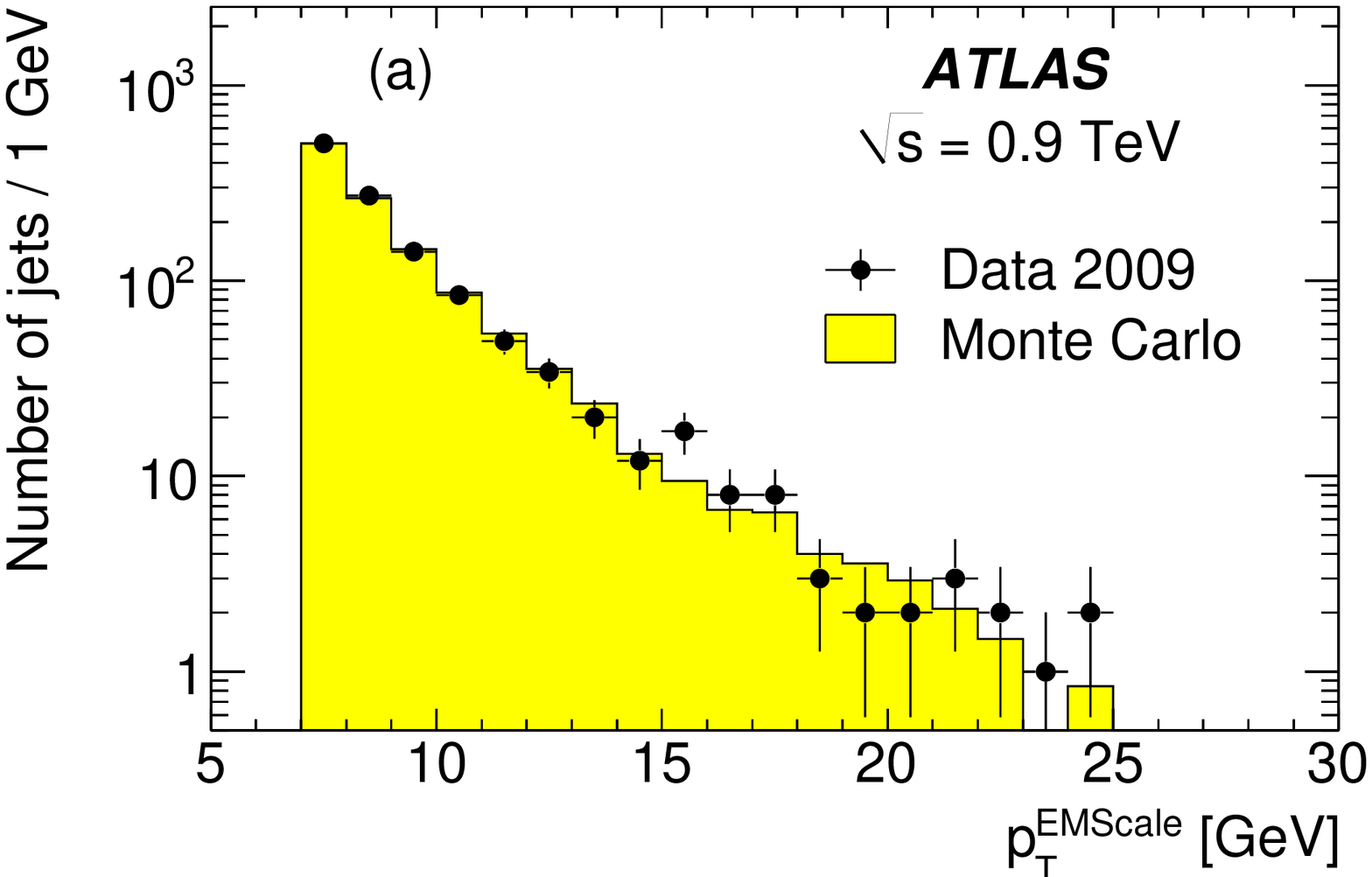}}
{\includegraphics[width=0.49\textwidth]{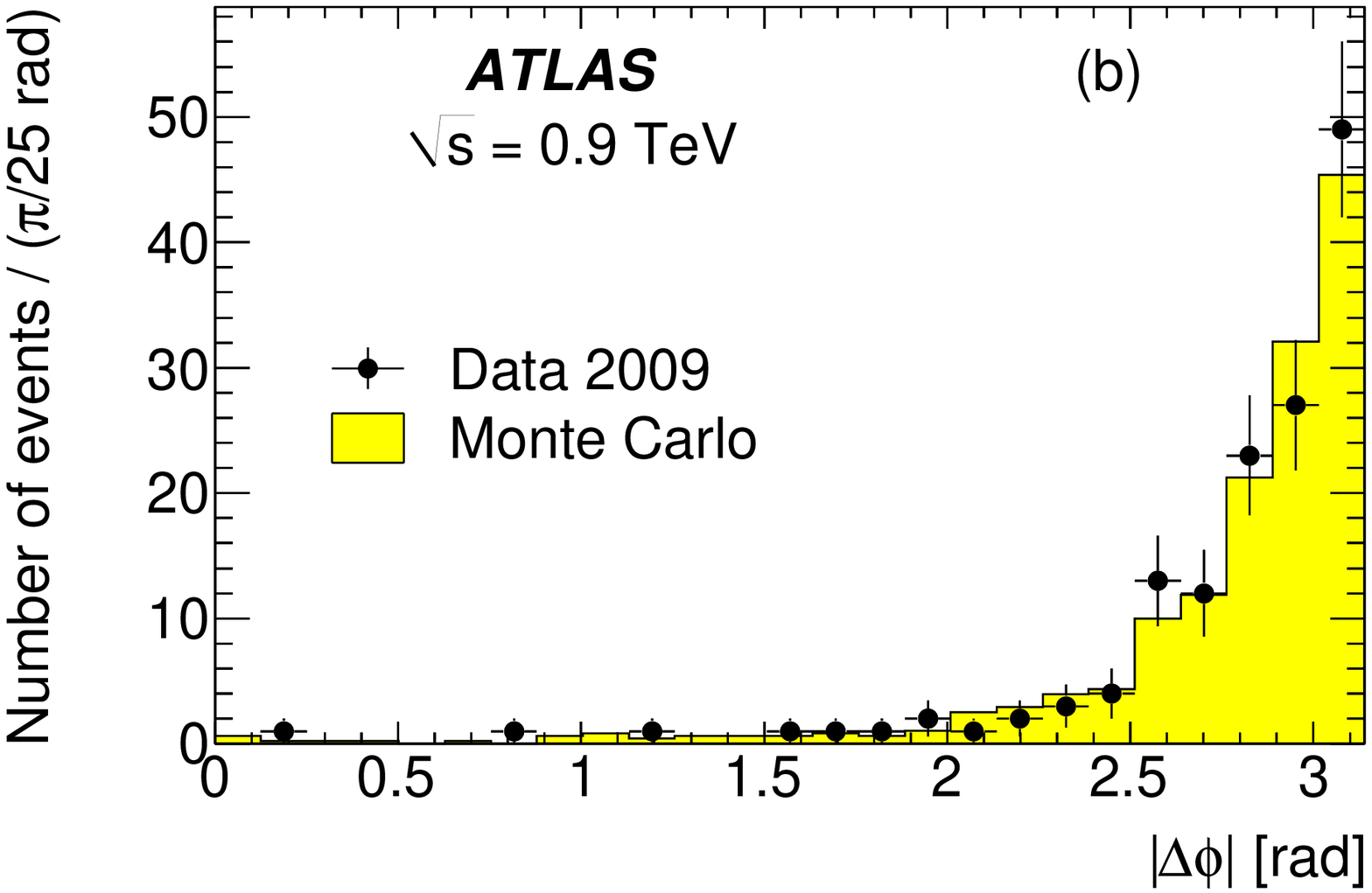}}
\vspace{-.2cm} 
\caption{Distributions of (a) \pt\  for all jets and (b) $\Delta \phi$
  for events with two or more jets. These are shown 
for jets    reconstructed with topological   clusters using the 
\antikt~algorithm with R=0.6.
}
\label{fig:Jets_pt}
\end{figure}

Figure~\ref{fig:Jets_pt} shows the jet \pt\ in data and Monte Carlo
simulation,  normalized to the  number of jets in data. 
Figure~\ref{fig:Jets_pt}(b) shows the difference of the
azimuthal angle of the two leading jets ($\Delta\phi$) in events
 with at least two reconstructed jets in data.
The
distribution peaks at
$\pi$, corresponding to a topology where the two reconstructed jets are
back-to-back in the transverse plane.
In all distributions, the agreement between data and simulations is
good, demonstrating that
the description of the material and detector response in the
 simulation provides an adequate model of basic jet quantities.

\subsection{Performance of the First Level Jet Trigger}

The efficiency of the L1 jet
trigger has been studied. Jets were reconstructed as before, but with a
lower energy threshold, \Etjet\ $> 4$~GeV.
For the efficiency determination the jet signatures identified by the
first level calorimeter trigger were matched to those of reconstructed jets
by requiring $\Delta R < 0.6$. L1 jet
signatures are based on sums of trigger towers in the electromagnetic
and hadronic calorimeter both calibrated to the electromagnetic scale. Trigger
towers are formed by analogue summation on the detector and mostly
have a 
size of $0.1 \times 0.1$
in $\eta \times  \phi$, being larger in parts
of the end-caps and in the FCal. For the lowest threshold jet trigger
 the
trigger towers are summed over $0.4 \times 0.4$ in $\eta \times  \phi$
and all higher
threshold jet triggers use $0.8 \times 0.8$ in $\eta \times  \phi$.
The special treatment of the lowest threshold jet trigger  is
motivated by its higher susceptibility to noise.

\begin{figure}[htp]
\begin{center}
\includegraphics[width=0.49\textwidth]{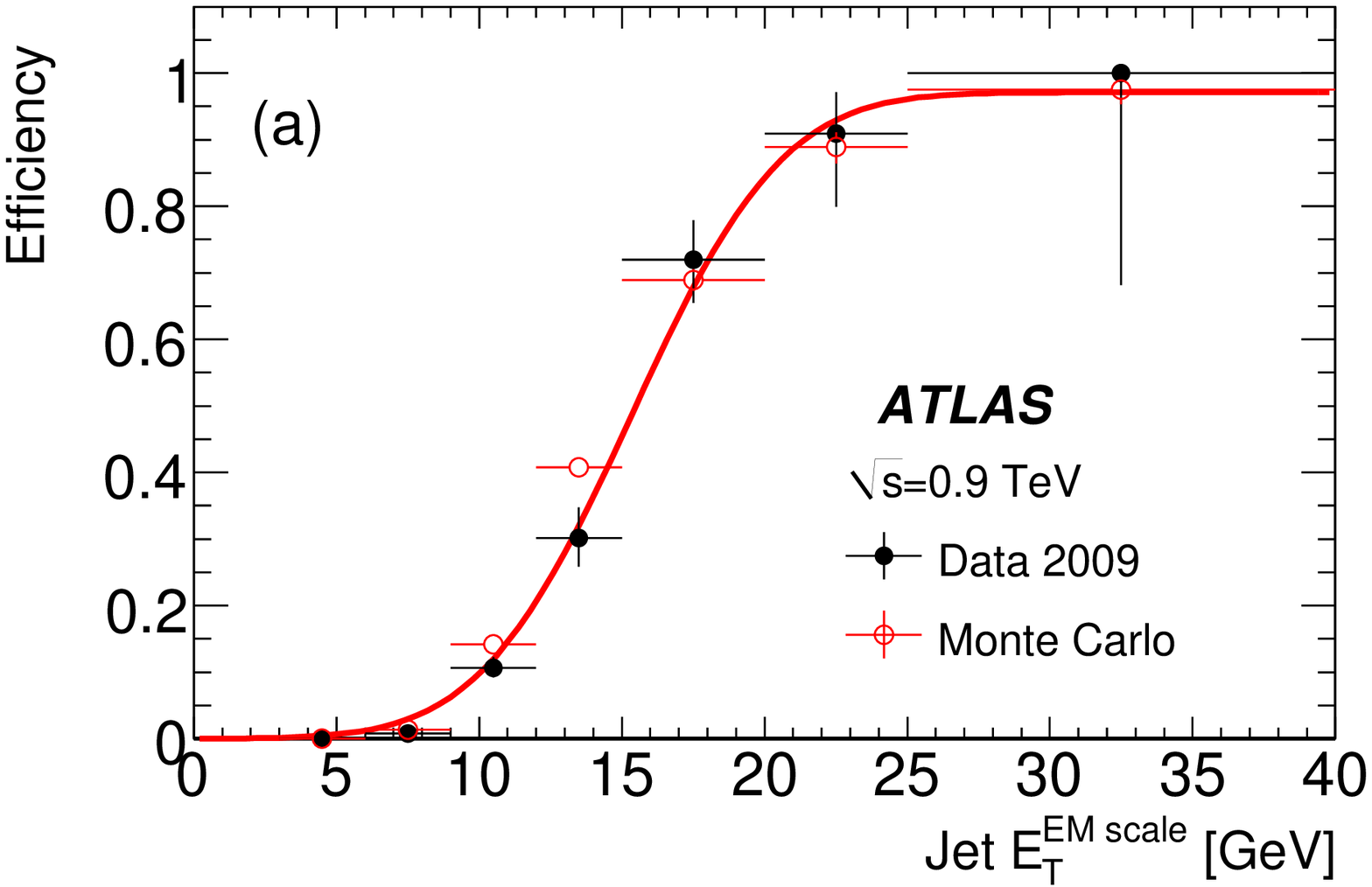}
\includegraphics[width=0.49\textwidth]{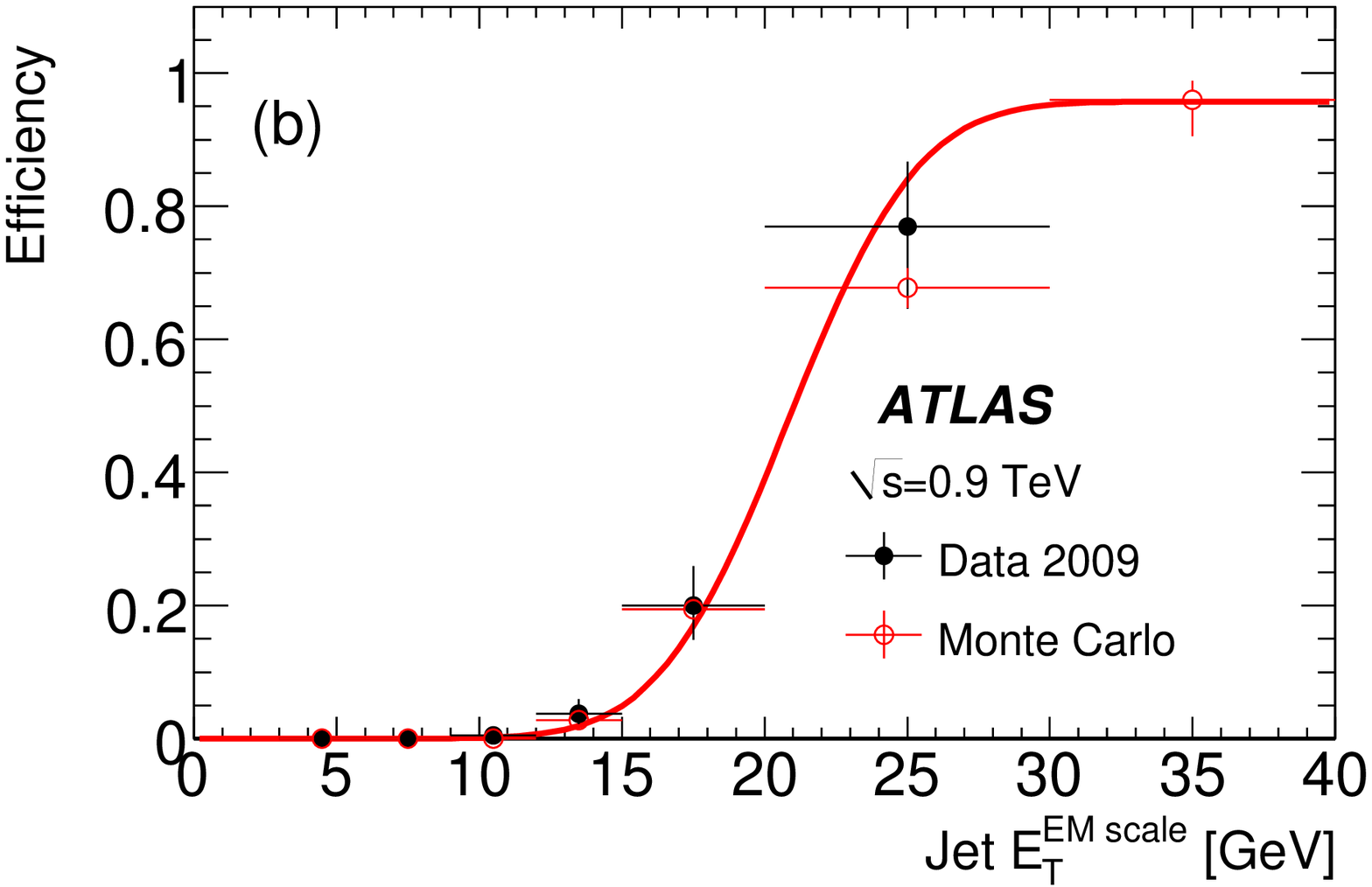}
\caption{L1 jet trigger efficiency for the triggers with
  50\% efficiency at around 15~\GeV\ and 20~\GeV\  for data (solid)
and  simulation (open) together with a fit to the Monte Carlo 
using a standard threshold
function (see text). 
}
\label{l1-eff}
\end{center}
\end{figure}

The trigger efficiency is calculated as the
fraction of reconstructed jets passing the quality requirements described
which have a matched trigger jet.
The results for the two jet triggers with 
50\% efficiency at  around 15 \GeV\ and 20 \GeV\ are
shown for data and Monte Carlo simulation as a function of the
reconstructed jet transverse energy, \Etjet\ in Fig.~\ref{l1-eff}, where
\Etjet\ is given at the electromagnetic scale. 
The curves are
fits to the simulation using a standard trigger turn-on
parameterization with an error function. The data are in agreement with the fit,
which shows that the thresholds are as they are expected to be.
This is a  step to understanding the
initial jet selection performance of the first-level calorimeter
trigger.

\subsection{Tau Studies Using Jets}

\begin{figure}[htp]
{\includegraphics[width=0.49\textwidth]{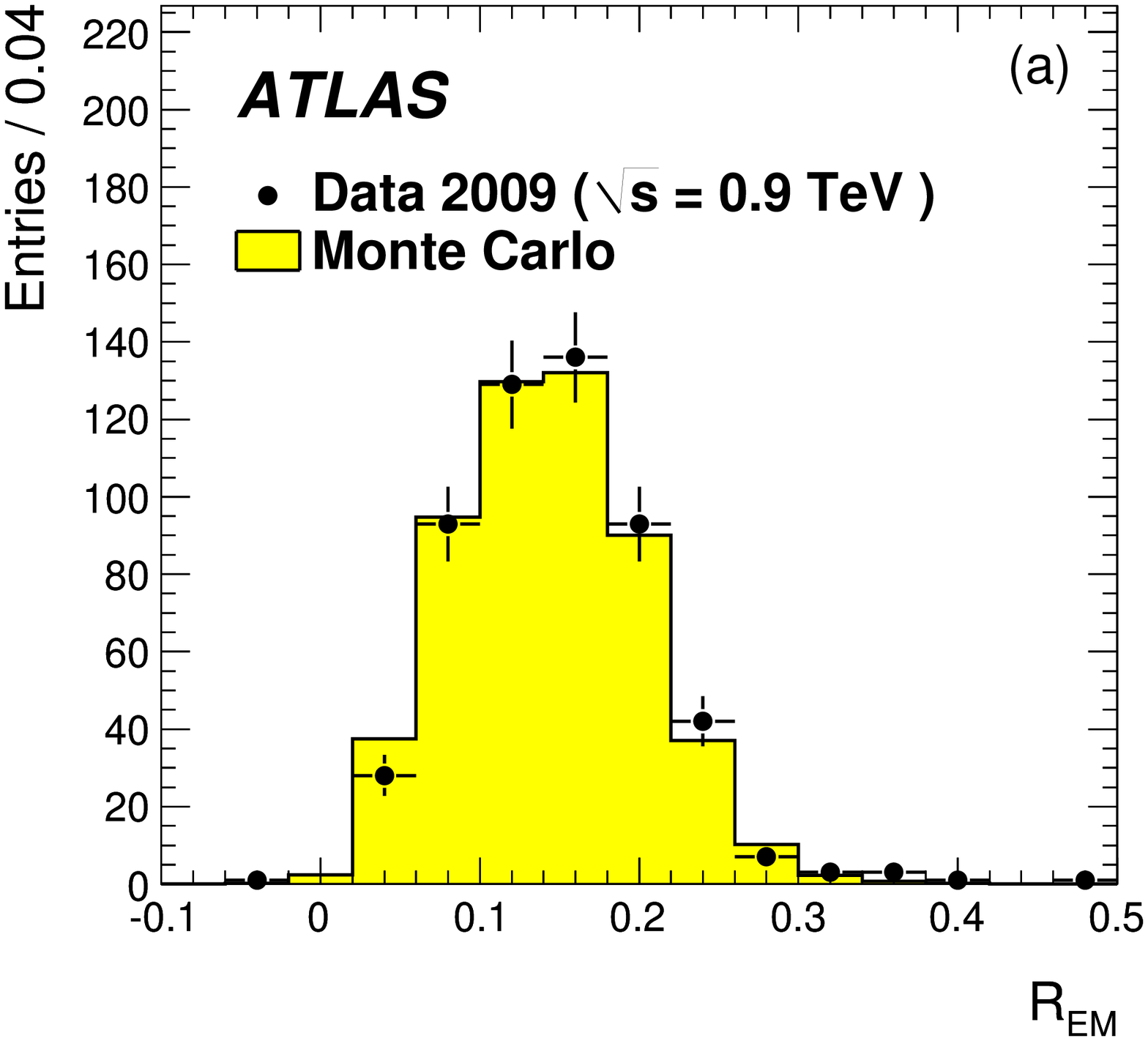}}
{\includegraphics[width=0.49\textwidth]{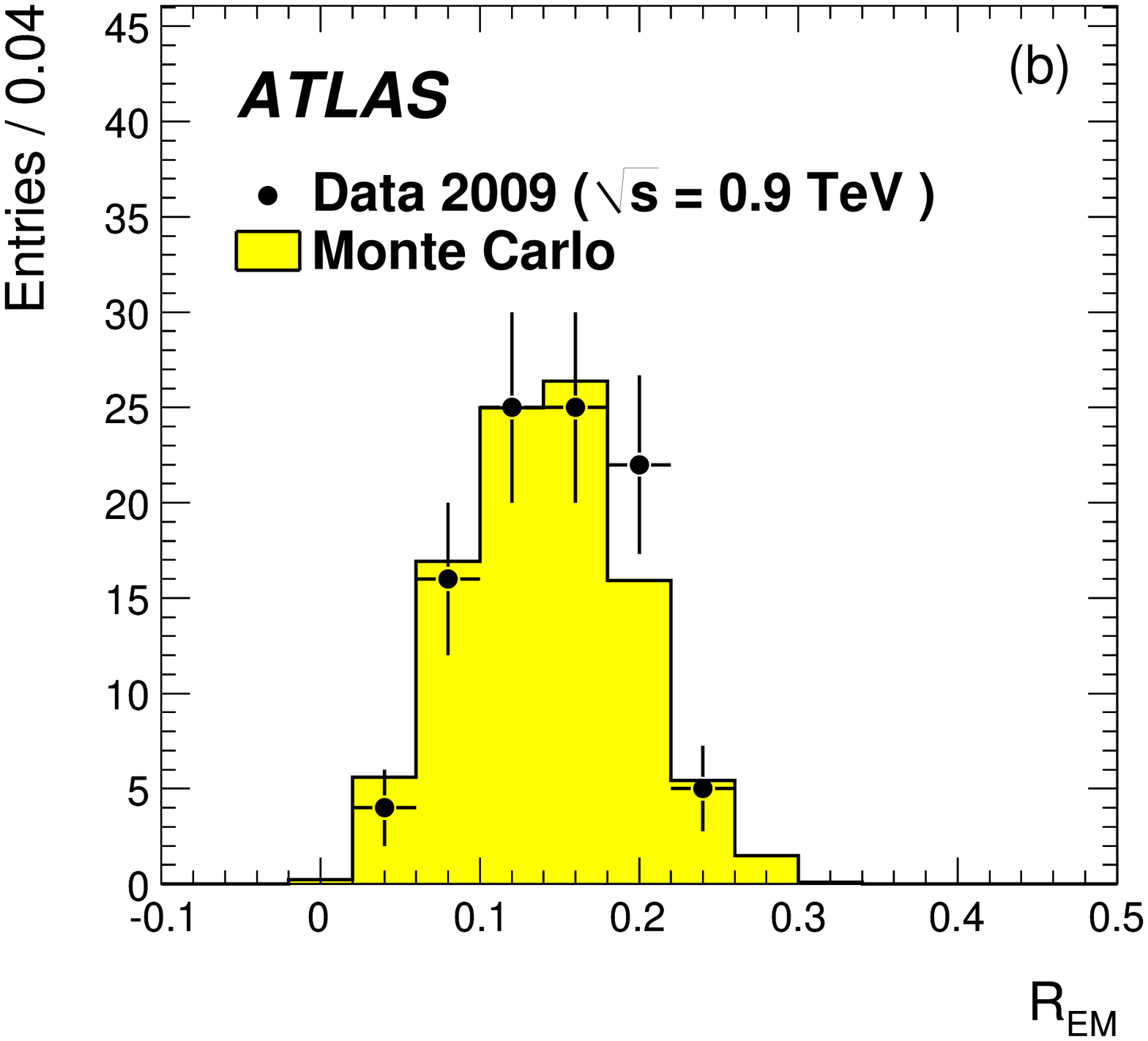}}
\caption{ (a) The electromagnetic radius\, $R_{\mathrm EM}$ (see text)  of
  the inclusive reconstructed tau candidates.  (b) The same variable
  with a tightened selection requiring di-jet events. The Monte Carlo is
  normalized to the same number of candidates as in the data.
}
\label{fig:JetTau}
\end{figure}

Jets from QCD processes form the largest background for the
reconstruction of hadronically decaying tau candidates. Therefore,
even though the actual number of tau leptons in the 2009 data is
expected to be negligible, basic track and cluster distributions have
been studied using jets with emphasis on the variables of importance
for  tau reconstruction and identification. 
In
Fig.~\ref{fig:JetTau}, the electromagnetic radius,  $R_{\mathrm EM}$, 
(the energy-weighted
mean $\Delta R$ of the jet components from the seed cell)  is shown for all tau
candidates with 
\Etjet$>$7\,\GeV\ on the left and for a subset with a
tight di-jet selection, which are more likely to be taken as $\tau$
candidates,
 on the right. Good agreement between data and
simulation is observed, which gives confidence in the results obtained
from earlier performance estimates based on Monte Carlo simulations.

\def\px {$E_{\rm x}^{\rm miss}$}
\def\py {$E_{\rm y}^{\rm miss}$}
\def\phimiss{$\varphi^{\rm miss}$}
\def\sumet{$\sum E_{\rm T}$}

\section{Missing Transverse Energy}
\label{met}

\label{met-intro}
A reliable measurement of the missing transverse energy, \MET, is a key
ingredient for many important analyses.
This
study considers \MET\  reconstructed from calorimeter information
only. As this measurement involves summing calorimeter cells over the
whole detector, it is sensitive to detector and reconstruction effects. In
particular events with rare unexpected noise contributions tend to
appear in the  tail of the \MET\ spectrum.

\MET\ reconstructed with the calorimeter is derived from the vector sum of the transverse energies of the cells.
Because of the high granularity of the calorimeter (about 187\,000 cells), it is
crucial to suppress noise contributions to \MET, i.e. to limit the number of cells
used in the sum. This is done by only using cells belonging to
three-dimensional topological clusters defined in
Section~\ref{jetscalo}. 
About 800 and 2\,500 cells on average are included in such clusters
 in randomly triggered and collision events, respectively.  
The 0.1\% of the cells classified as noisy are removed. 

\begin{figure}[htp]
\begin{center}
\includegraphics[width=0.70\textwidth]{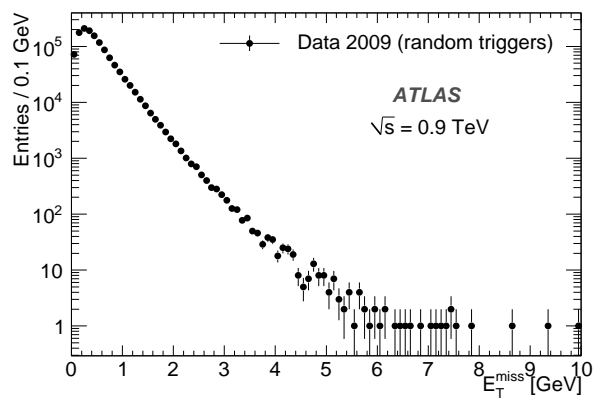}
\caption{Distribution of \MET~as measured in data from randomly
 triggered events. Only cells belonging to  topological
 clusters are included in the calculation; their  energies are
 calibrated at the EM scale.
}
\vspace*{-0.2cm}
\label{fig:MET_RNDM}
\end{center}
\end{figure}

The sensitivity to noise can be best studied in randomly triggered
events, where minimal energy is deposited in the calorimeters.
 The
 \MET~distribution of these events is shown in Fig.~\ref{fig:MET_RNDM},
 demonstrating  the level of tails in randomly triggered events.

\begin{figure}[htb]
\begin{center}
{\includegraphics[width=.49\textwidth]{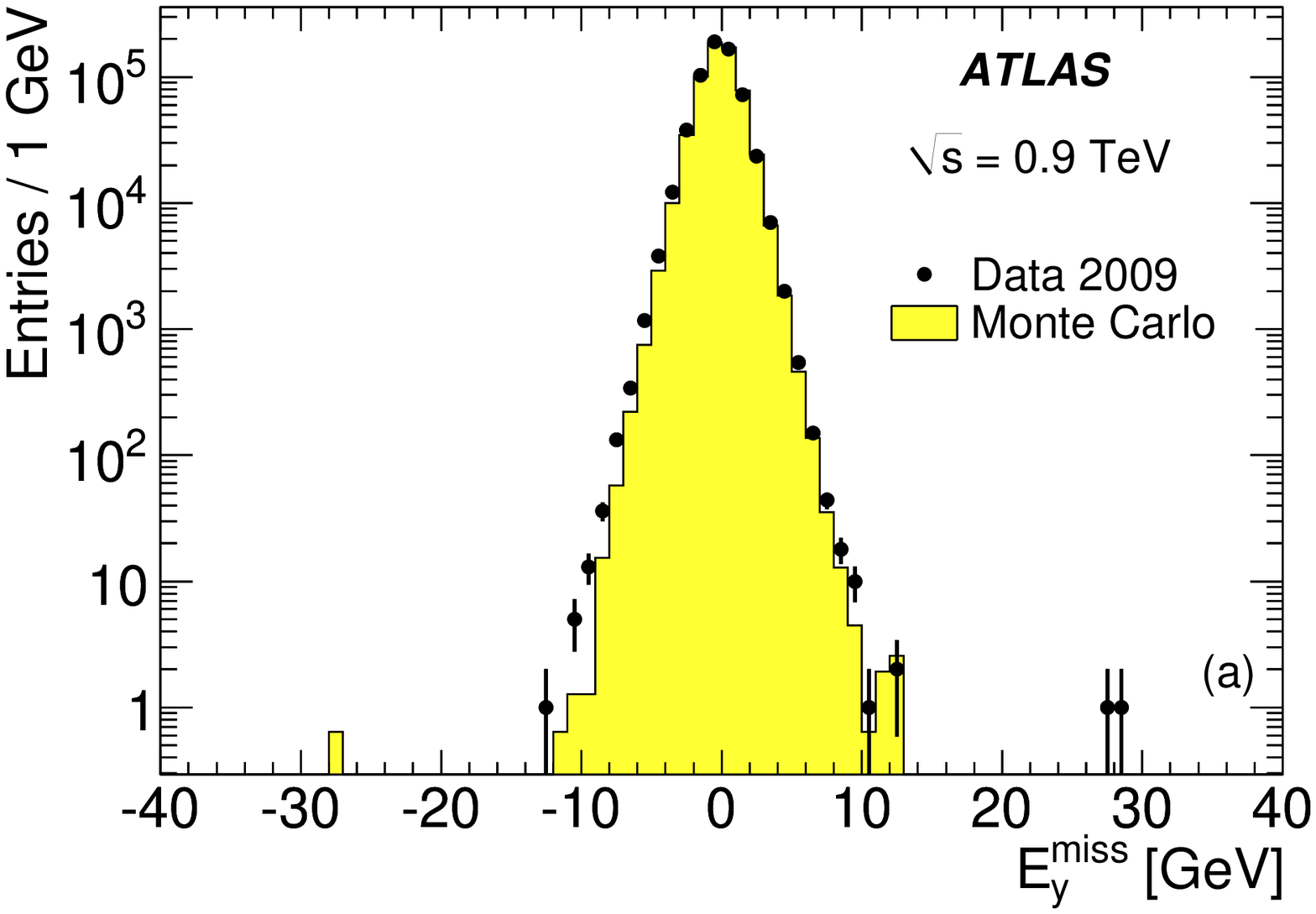}}
{\includegraphics[width=.49\textwidth]{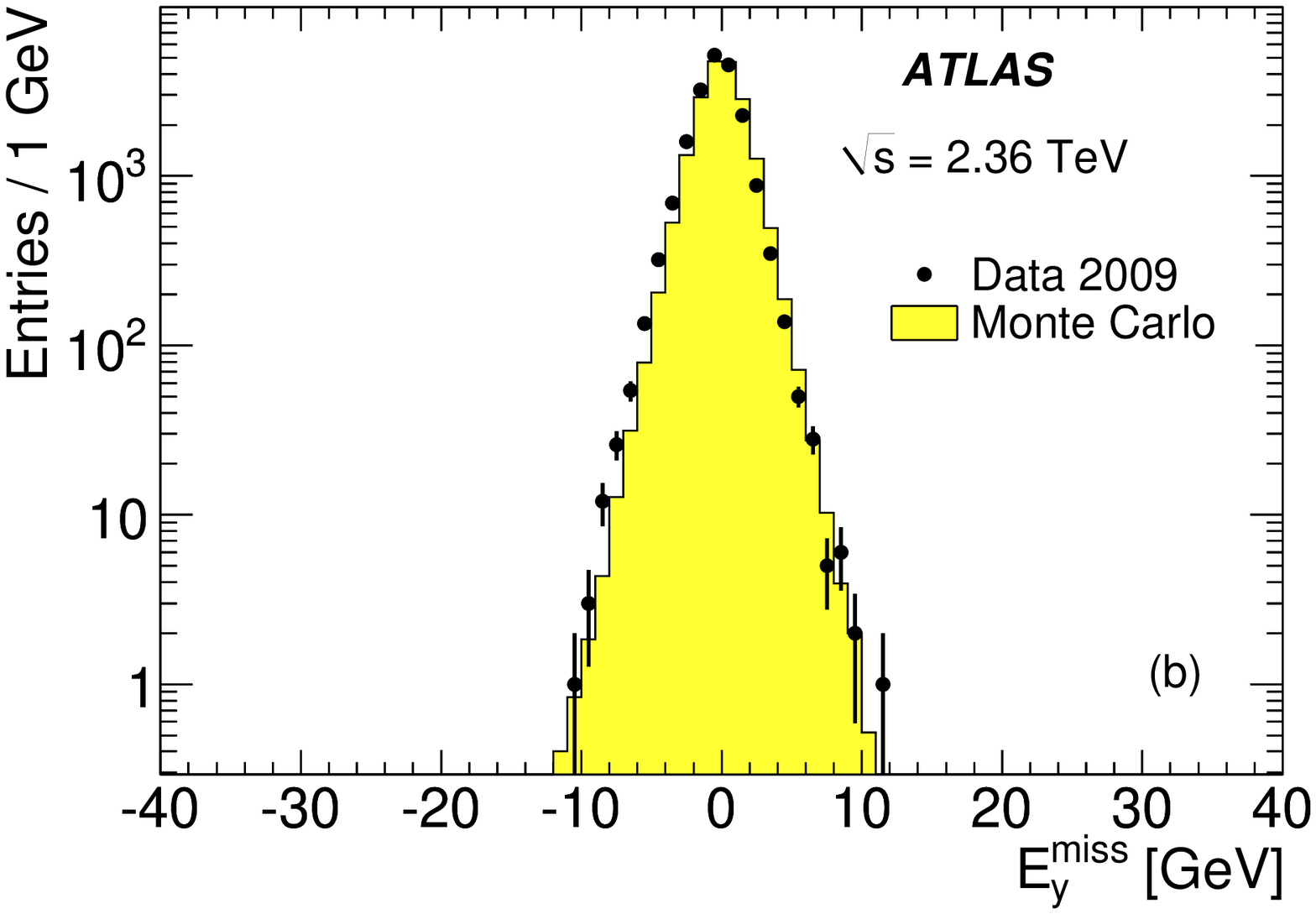}}
{\includegraphics[width=.49\textwidth]{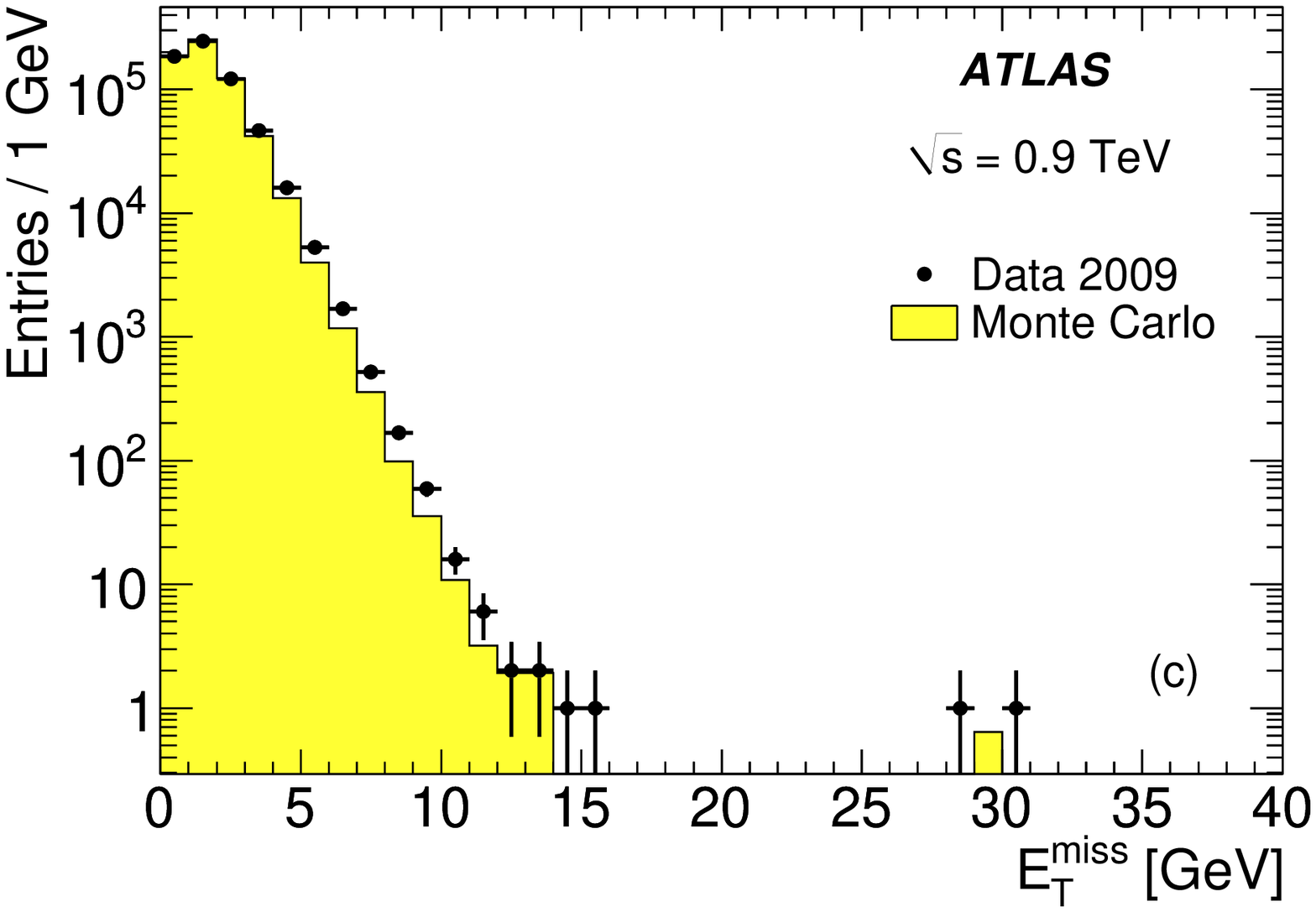}}
{\includegraphics[width=.49\textwidth]{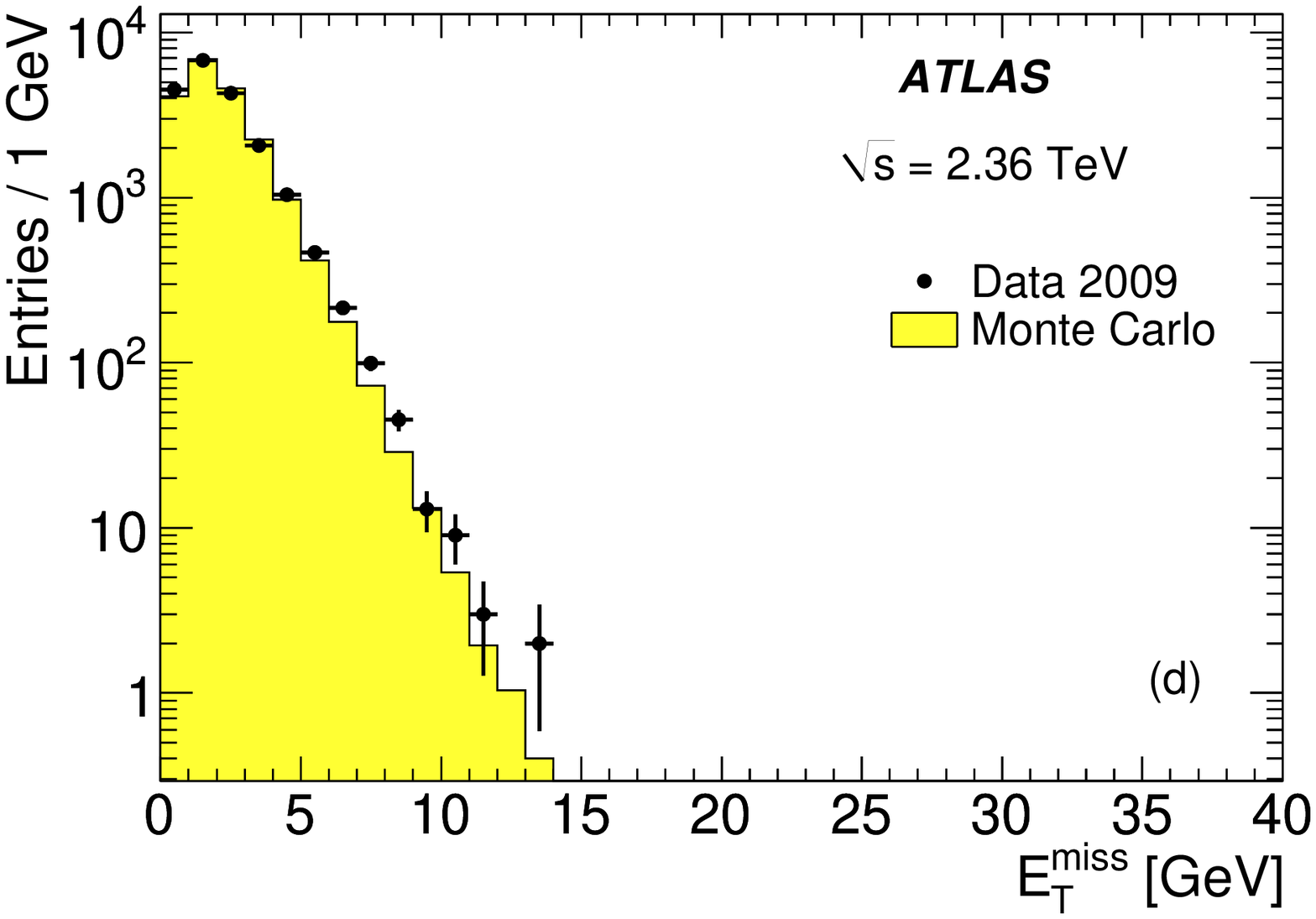}}
\caption{Distribution of \py~(a,b)~and \MET~(c,d) as measured in data
from minimum-bias events (dots) at 0.9~\TeV\ (a,c) and 2.36~\TeV\
(b,d) centre-of-mass energies. In the calculation only topological cluster
cells are used, with energies calibrated at the EM
scale. The expectations from Monte Carlo simulation are superimposed
(histograms) and normalized to the number of events in data.
 \vspace{-0.5cm}}
\label{fig:METxy}
\end{center}
\end{figure}

In soft proton-proton collisions, no true \MET~is expected. This is
confirmed by the Monte Carlo simulation.
Unlike in randomly triggered events, total transverse energies
(\sumet) up to 100~\GeV\ are deposited in the calorimeter for
minimum-bias
 events in the present data set.  Figure~\ref{fig:METxy} shows
 the measured  \py\ distributions as an example.
For both \px~and \py, the RMS  values are about 1.4~\GeV\
and 1.8~\GeV\ for 0.9~\TeV\ and 2.36~TeV centre-of-mass energies,
respectively.  These values are higher than in randomly triggered
events because the finite resolution in the presence of real energy
contributes
 to the width.

The \MET~distribution, also shown in Fig.~\ref{fig:METxy}, is found
to be satisfactory at this early stage.
 There are no  events outside the borders of the figure. 
The two data events with large \MET\ are consistent with arising
from out-of-time energy in the detector.  At least one of these
appears to be a cosmic ray event.  Such events are not included in the
Monte Carlo simulation sample.

\begin{figure}[hbtp]
\begin{center}
\includegraphics[width=0.75\textwidth]{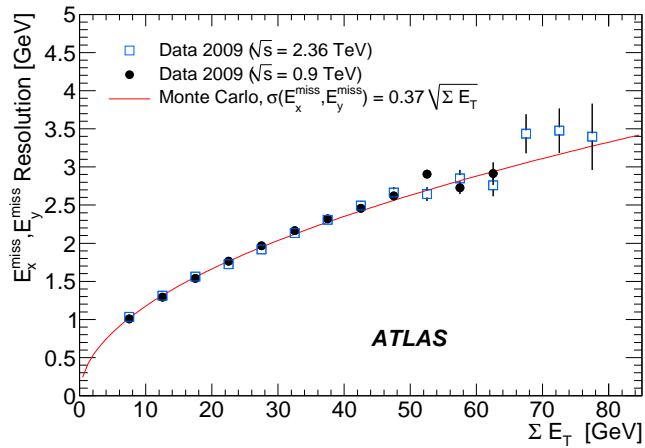}
\vspace*{0cm}
\caption{\MET~resolution as a function of the total transverse energy (\sumet)
for minimum-bias events. The line shows a fit to the resolution
expected from the 
Monte Carlo simulation  and the full dots (open squares) represents
the results with data at 0.9 (2.36)~\TeV. \px, \py, \sumet~are computed with
topological cluster cells at EM scale.
}
\label{fig:MET_Perf}
\end{center}
\end{figure}

A more quantitative evaluation of the \MET~performance can be obtained
from a study of the \px~and \py~resolutions as 
a function of the total transverse energy \sumet\ in the event. The
resolutions are expected to be proportional to 
$\sqrt{\sum E_{\rm T}}$. The resolutions observed in the ATLAS data at
both centre-of-mass energies are presented as a 
function of $\sqrt{\sum E_{\rm T}}$~in Fig.~\ref{fig:MET_Perf}. A very
good agreement between data and Monte Carlo simulation is
obtained at both centre-of-mass energies. The \MET~resolution can be
parameterized as $\sigma$(\px,\py)$=0.37 \times \sqrt{\sum E_{\rm T}}$
at the EM scale\footnote{$\sigma$(\px,\py) and $E_{\rm T}$ in \GeV}, with a negligible statistical error.

\section{Muons}
\label{sec:muons}

The calorimeters are surrounded by the muon
spectrometer~\cite{muonpaper}, which was designed to provide a trigger
and accurate standalone reconstruction for muons 
with \pt\ from several \GeV\ up to a few \TeV.  In contrast, the muons
studied in the sample collected at $\sqrt{s}$=0.9 \TeV\ are of
relatively low \pt. 
The air-core toroid system, with a long barrel and two inserted
end-cap magnets, generates an average field of 0.5~T (1~T), in the
barrel (end-caps)
over a large volume.
 Multiple scattering effects are thereby minimized, and
excellent muon momentum resolution is achieved with three layers of
high precision tracking chambers. Over most of the $\eta$ range,
tracks are measured by Monitored Drift Tubes (MDT). For $2 < |\eta| <
2.7$, Cathode Strip Chambers (CSC) with higher granularity are used in
the innermost layer, to withstand the demanding rate and background
conditions. In addition, the muon spectrometer includes dedicated
trigger chambers with nominal timing resolutions between
$1.5$ and $4$~ns. They are composed of Resistive Plate Chambers
(RPC) in the barrel and Thin-Gap Chambers (TGC) in the end-cap
regions. Besides  providing trigger signals they also measure the muon
coordinate in the direction orthogonal to that determined by the
precision-tracking chambers.

\begin{figure}[t]
\begin{center}
\includegraphics[width=.49\textwidth]{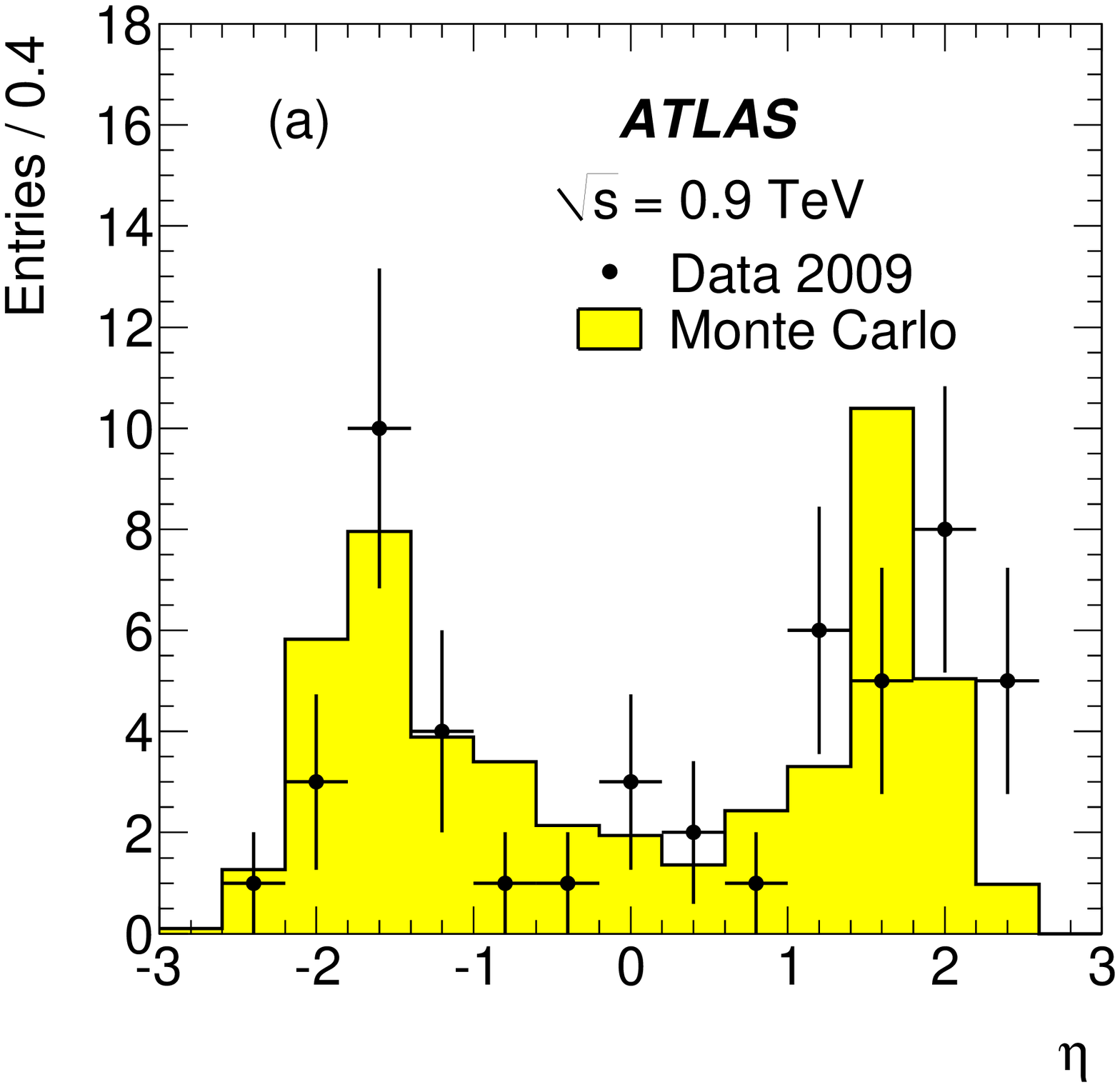}
\includegraphics[width=.49\textwidth]{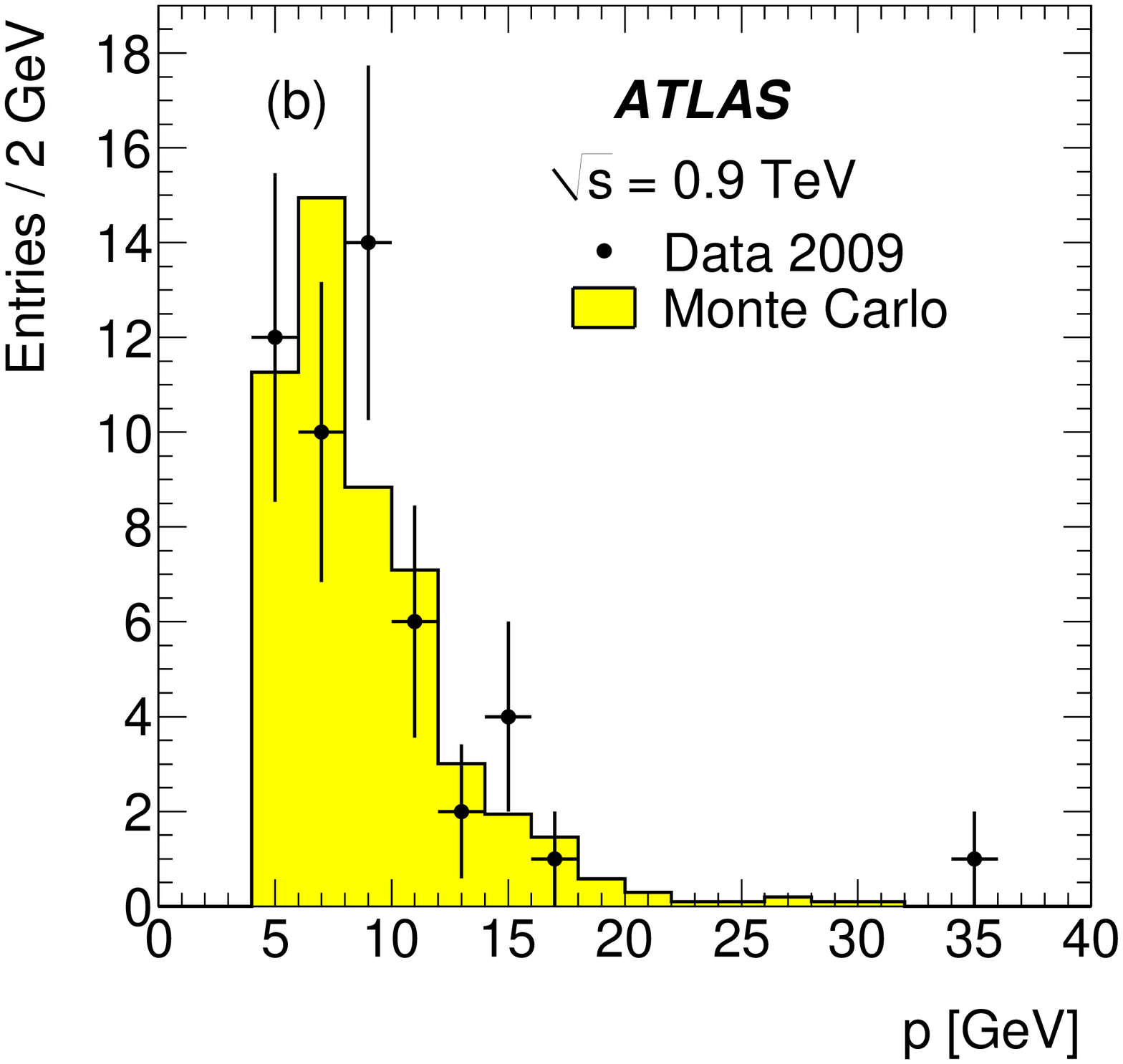}
\caption{The distribution of $\eta$ and $p$ of reconstructed muons in the $0.9$~\TeV\ data. The simulation
distributions are normalized to the number of entries in data.
 \vspace{-0.5cm}}
\label{muonfig}
\end{center}
\end{figure}

The muon analysis uses a somewhat smaller data set than other analyses
as the toroid system was not always operational. In addition, it was
checked that the MDTs,
TGCs and RPCs were all operating normally.
The algorithms used for muon reconstruction combine tracks from the muon
systems and the inner detector, and are developed from those
described in Ref.~\cite{CSC}. For the results presented here, two
independent algorithms are used and 
only candidates selected by both are accepted.  This selects a total
of 50 muon candidates. Raw kinematic distributions for these
candidates are presented in Fig.~\ref{muonfig}.
The muon spectrum observed is soft and strongly peaked in the forward
direction, where the momentum of the muons more often exceeds the 3.2~GeV needed
to penetrate through the forward calorimeter.
The
kinematic distributions are compared to the predictions from
minimum-bias simulation with the Monte Carlo normalized to the number
of muons found in data. Within the large statistical uncertainties
good agreement is found indicating a reasonable understanding of the
initial performance of the ATLAS muon spectrometer.

The muon trigger, designed to
select high-\pt\ muons, has a limited acceptance for the muon tracks
reconstructed offline in 2009. 
Of the 38 muons in the end-cap regions, 13 were triggered at L1 by the
TGC. Of the 12 muons in the barrel, one was triggered at L1 by the RPC with
the correct timing, another 9 were triggered with +1 bunch crossing
misalignment during the timing adjustment phase, while the other 2 muons
were outside the trigger acceptance.
Only one muon
passed the full trigger chain up to the EF combined trigger after
applying the \pt\ $> 4$~\GeV\ selection. 
The muon momenta and directions measured by the L2
and EF are in good agreement with the offline measurement.

\section{Conclusions}
\label{sec:conclusion}

The overall performance of the ATLAS detector at the LHC was established in 
first collision data at centre-of-mass energies of 0.9 \TeV\ and 2.36 \TeV. 
Although the detector has not been optimized for the low energy
particles studied,  
its performance was found to be remarkably good,  particularly
in view of the early stage of data taking.

 The overall data-taking efficiency was about
90\% and the subdetectors were typically  99\% operational. The entire
computing infrastructure of Trigger/DAQ was immediately effective. Collision 
candidates were selected with a negligible background level.

The tracking detectors and  electromagnetic calorimeters have, by
the nature of the data set, been the  most extensively
tested components and they perform well.  
The  
hit efficiencies, resolutions and particle identification capabilities of the tracking detector 
are well modelled by Monte Carlo simulations. 
The discrimination between electrons and pions using transition radiation 
was demonstrated and outperforms any previous colliding beam experiment. 
The results of various studies suggest that the material distribution in 
the inner part of the tracker  is well understood, at a level of
a few percent of 
the total in the barrel silicon systems.

The momentum scale linked to the
\kshort\ mass  is known at the per mille level and the calibration of
the electromagnetic calorimetry 
in the region of the \pizero\  mass was checked at the level of 1\%. 
Electron and photon reconstruction was extensively tested and
performs well. 
The candidates typically have 
transverse momenta well below those for which
reconstruction and identification algorithms were optimized, but their
properties are shown to be in good agreement with simulation.
Good calorimeter performance was also demonstrated by the
measurement of the  
resolution of the missing
transverse energy, which follows closely the expectations
from Monte Carlo  
over the entire energy range probed. 

The muon system was not extensively tested with this data set, but
performs as expected within the precision available.
It was well tested in cosmic ray data taken in
2008~\cite{muonpaper}. 


\section{Acknowledgements}

We are greatly indebted to all CERN's departments and to the LHC project for their immense efforts not only in building the LHC, but also for their direct contributions to the construction and installation of the ATLAS detector and its infrastructure. 
All our congratulations go to the LHC operation team for the superb
performance during this initial data-taking period.
We acknowledge equally warmly all our technical colleagues in the collaborating Institutions without whom the ATLAS detector could not have been built. Furthermore we are grateful to all the funding agencies which supported generously the construction and the commissioning of the ATLAS detector and also provided the computing infrastructure.

The ATLAS detector design and construction has taken about fifteen years, and our thoughts are with all our colleagues who sadly could not see its final realisation.

We acknowledge the support of ANPCyT, Argentina; Yerevan Physics Institute, Armenia; ARC and DEST, Australia; Bundesministerium f\"ur Wissenschaft und Forschung, Austria; National Academy of Sciences of Azerbaijan; State Committee on Science \& Technologies of the Republic of Belarus; CNPq and FINEP, Brazil; NSERC, NRC, and CFI, Canada;
CERN; CONICYT, Chile; NSFC, China; COLCIENCIAS, Colombia; Ministry of
Education, Youth and Sports of the Czech Republic, Ministry of
Industry and Trade of the Czech Republic, and Committee for
Collaboration of the Czech Republic with CERN; Danish Natural Science
Research Council and the Lundbeck Foundation; European Commission,
through the ARTEMIS Research Training Network; IN2P3-CNRS and
Dapnia-CEA, France; Georgian Academy of Sciences; BMBF, DFG, HGF and
MPG, Germany; Ministry of Education and Religion, through the EPEAEK
program PYTHAGORAS II and GSRT, Greece; ISF, MINERVA, GIF, DIP, and
Benoziyo Center, Israel; INFN, Italy; MEXT, Japan; CNRST, Morocco; FOM
and NWO, Netherlands; The Research Council of Norway; Ministry of
Science and Higher Education, Poland; GRICES and FCT, Portugal;
Ministry of Education and Research, Romania; Ministry of Education and
Science of the Russian Federation and State Atomic Energy Corporation
ROSATOM; JINR; Ministry of Science, Serbia; Department of
International Science and Technology Cooperation, Ministry of
Education of the Slovak Republic; Slovenian Research Agency, Ministry
of Higher Education, Science and Technology, Slovenia; Ministerio de
Educaci\'{o}n y Ciencia, Spain; The Swedish Research Council, The Knut
and Alice Wallenberg Foundation, Sweden; State Secretariat for
Education and Science, Swiss National Science Foundation, and Cantons
of Bern and Geneva, Switzerland; National Science Council, Taiwan;
TAEK, Turkey; The Science and Technology Facilities Council and The
Leverhulme Trust, United Kingdom; DOE and NSF, United States of
America.

\bibliographystyle{atlasnote}
\bibliography{perfpaper}

\nolinenumbers
\newpage
\begin{flushleft}
{\Large The ATLAS Collaboration}

\bigskip

G.~Aad$^{\rm 48}$,
E.~Abat$^{\rm 18a}$$^{,*}$,
B.~Abbott$^{\rm 111}$,
J.~Abdallah$^{\rm 11}$,
A.A.~Abdelalim$^{\rm 49}$,
A.~Abdesselam$^{\rm 118}$,
O.~Abdinov$^{\rm 10}$,
B.~Abi$^{\rm 112}$,
M.~Abolins$^{\rm 88}$,
H.~Abramowicz$^{\rm 152}$
% ...
\bigskip

% ...
$^{10}$ Institute of Physics, Azerbaijan Academy of Sciences, H. Javid Avenue 33, AZ 143 Baku, Azerbaijan\\
$^{11}$ Institut de F\'isica d'Altes Energies, IFAE, Edifici Cn, Universitat Aut\`onoma  de Barcelona,  ES - 08193 Bellaterra (Barcelona), Spain\\
% ...
$^{18}$ Bogazici University$^{(a)}$, Faculty of Sciences, Department of Physics, TR - 80815 Bebek-Istanbul; Dogus University$^{(b)}$, Faculty of Arts and Sciences, Department of Physics, 34722, Kadikoy, Istanbul; $^{(c)}$Gaziantep University, Faculty of Engineering, Department of Physics Engineering, 27310, Sehitkamil, Gaziantep, Turkey; Istanbul Technical University$^{(d)}$, Faculty of Arts and Sciences, Department of Physics, 34469, Maslak, Istanbul, Turkey\\
% ...
$^{48}$ Albert-Ludwigs-Universit\"{a}t, Fakult\"{a}t f\"{u}r Mathematik und Physik, Hermann-Herder Str. 3, D - 79104 Freiburg i.Br., Germany\\
$^{49}$ Universit\'e de Gen\`eve, Section de Physique, 24 rue Ernest Ansermet, CH - 1211 Geneve 4, Switzerland\\
% ...
$^{88}$ Michigan State University, Department of Physics and Astronomy, High Energy Physics Group, East Lansing, MI 48824-2320, United States of America\\
% ...
$^{111}$ University of Oklahoma, Homer L. Dodge Department of Physics and Astronomy, 440 West Brooks, Room 100, Norman, OK 73019-0225, United States of America\\
$^{112}$ Oklahoma State University, Department of Physics, 145 Physical Sciences Building, Stillwater, OK 74078-3072, United States of America\\
% ...
$^{118}$ Oxford University, Department of Physics, Denys Wilkinson Building, Keble Road, Oxford OX1 3RH, United Kingdom\\
% ...
$^{152}$ Tel Aviv University, Raymond and Beverly Sackler School of Physics and Astronomy, Ramat Aviv, IL - Tel Aviv 69978, Israel\\
% ...
$^{a}$ Present address FermiLab, USA\\
$^{b}$ Also at CPPM, Marseille, France.\\
$^{c}$ Also at TRIUMF, 4004 Wesbrook Mall, Vancouver, B.C. V6T 2A3, Canada\\
$^{d}$ Also at Faculty of Physics and Applied Computer Science of the AGH-University of Science and Technology, (FPACS, AGH-UST), al. Mickiewicza 30, PL-30059 Cracow, Poland\\
% ...
$^{*}$ Deceased\end{flushleft}


\end{document}